\documentclass[a4paper,11pt]{article}
\pdfoutput=1
\usepackage[no-natbib-sort]{jheppub} 
\usepackage{amsmath, amssymb}
\usepackage{graphicx}
\usepackage{float}
\usepackage{soul}
\usepackage{arydshln}
\usepackage[export]{adjustbox}
\usepackage[utf8]{inputenc}
\usepackage{slashed}
\usepackage{bbold}
\usepackage{subfig}
\usepackage{subfiles}
\usepackage{amsfonts}
\usepackage{pifont}
\usepackage{xcolor}
\usepackage{physics}
\usepackage{enumitem}
\usepackage{ifthen}
\usepackage{mathtools}
\usepackage{import}
\numberwithin{equation}{section}
\usepackage{standalone}
\usepackage{tikz}
\usepackage{comment}
\usepackage[compat=1.1.0]{tikz-feynman}

\usetikzlibrary{decorations.markings,arrows,arrows.meta,shapes}
\usetikzlibrary{decorations.text}
\usetikzlibrary{bending} 
\DeclareUnicodeCharacter{2212}{-}

\newcommand{\vk}{\vec{k}}
\newcommand{\vkp}{\vec{q}}

\newcommand{\vq}{\vec{q}}
\newcommand{\vp}{\vec{p}}

\newcommand{\vx}{\vec{x}}

\newcommand{\knl}{k_{\rm NL}}

\newcommand{\qp}{\vec{q}\cdot\vec{p}}

\newcommand{\peak}{\text{peak}}
\newcommand{\um}{h/\text{Mpc}}
\newcommand{\UV}{\text{UV}}
\newcommand{\td}{\tilde{\delta}}
\newcommand{\Ffive}{F_5 ( \vk, \vq,-\vq,\vp,-\vp )}
\newcommand{\fscr}[1]{f_{\rm screen} \left( #1 \right)}
\newcommand{\be}{\begin{equation}}
\newcommand{\ee}{\end{equation}}
\newcommand{\bea}{\begin{eqnarray}}
\newcommand{\eea}{\end{eqnarray}}

\newcommand{\eqn}[1]{Eq.~(\ref{#1})}
\newcommand{\kvec}{\vec{k}}
\newcommand{\qvec}{\vec{q}}
\newcommand{\pvec}{\vec{p}}
\newcommand{\xvec}{\vec{x}}

\newcommand{\xfl}{\xvec_{\rm fl}}
\newcommand{\appref}[1]{App.~\ref{#1}}

\newcommand{\Om}{\Omega_{\rm m}}
\newcommand{\secref}[1]{Sec.~\ref{#1}}
\newcommand{\figref}[1]{Fig.~\ref{#1}}
\newcommand{\tabref}[1]{Tab.~\ref{#1}}

\def\l{\left(}
\def\r{\right)}

\newcommand{\eqs}[2][0.3]{\includegraphics[width=#1\linewidth, valign=c]{#2}}
\allowdisplaybreaks

\graphicspath{{./figures/}}

\definecolor{MattGreen}{rgb}{.2,0.7,0.2}

\newcommand{\imineq}[2]{\vcenter{\hbox{\includegraphics[height=#2ex]{#1}}}}

\def\Tcal{{\cal T}}%
\def\Rcal{{\cal R}}%
\def\Pcal{{\cal P}}%
\def\pcal{{\it p}}%
\def\DeltaP{{\Delta{\cal P}}}%
\def\Plin{{\cal P}_{\rm lin.}}%
\def\Plo{{\cal P}_{\rm LO}}
\def\Pnlo{{\cal P}_{\rm NLO}}
\def\Pnnlo{{\cal P}_{\rm NNLO}}
\def\Knlo{{\cal K}_{\rm NLO}}
\def\Knnlo{{\cal K}_{\rm NNLO}}
\author[a]{Charalampos Anastasiou}
\affiliation[a]{Institute for Theoretical Physics, ETH Zurich, 8093 Z\"urich, Switzerland}
\emailAdd{babis@phys.ethz.ch}
\author[a]{Andrea Favorito}
\emailAdd{afavorito@phys.ethz.ch}
\author[b]{Matthew Lewandowski}
\emailAdd{matthew.joseph.lewandowski@cern.ch}
\affiliation[b]{Theoretical Physics Department, CERN, 1211 Geneva 23, Switzerland}
\author[a]{Leonardo Senatore}
\emailAdd{lsenatore@phys.ethz.ch}
\author[c]{Henry Zheng}
\affiliation[c]{Stanford Institute for Theoretical Physics, Physics Department, Stanford University, Stanford, CA 94306}

\title{Efficient evaluation of the dark-matter two-loop power spectrum in the EFT of LSS}
\graphicspath{{images/}{images/}}

\abstract{

Rapid progress in cosmological Large Scale Structure (LSS) surveys motivates precise theoretical predictions. The Effective Field Theory of Large-Scale Structure (EFTofLSS) is routinely applied to data, and requires fast computation of its predictions when sampling the large space of cosmological parameters. Going beyond existing one-loop techniques, we present a method to rapidly evaluate the two-loop power spectrum. Our method decomposes the typically small difference between a given linear power spectrum and a reference power spectrum into a cosmology-independent basis of functions resembling massive scalar propagators in Quantum Field Theory. By taking the leading terms in such a small difference, we numerically evaluate the cosmology-independent loop integrals where in the integrand only the relevant combinations of basis functions appear. We achieve an efficient numerical evaluation via physically motivated local ultraviolet subtractions and by arranging the cancellation of infrared singularities locally in the integrands. Final predictions are obtained by contracting these precomputed integrals with the cosmology-dependent coordinates of the expansion in the fixed basis.  We present and publicly release the precomputed integrals for the renormalized two-loop dark-matter power spectrum in the EFTofLSS. These require eight EFT counterterms, which include the effect of generated vorticity, and are sufficient to analyze the lensing galaxy signal in LSS surveys at this order.

}

\makeatletter
\gdef\@fpheader{}
\makeatother

\begin{document}
\vspace*{-25mm}

\begin{flushright}
{\small CERN-TH-2025-171}
\end{flushright}

\maketitle

\section{Introduction}

In the weakly non-linear regime, correlations of density fluctuations measured in the Large Scale Structure (LSS) of the universe can be predicted perturbatively, within the framework of the 
Effective Field Theory of the Large-Scale Structure (EFTofLSS)~\cite{Baumann:2010tm,Carrasco:2012cv}. In such a framework, the dynamics of the long wavelength perturbations are reproduced using an expansion in the amplitude of the fluctuations and in the ratio of their derivatives with respect to the non-linear scale of the universe, which is about $10 \, \text{Mpc}$ at the present redshift. {The  effect} of short distance physics at long distances is encoded in counterterms that need to be fit to data or to simulations. 

After a long development, the EFTofLSS has allowed {analyses in which all the LSS data} below a certain wavenumber are {included}, removing the need to perform suboptimal {data selections} (`cutting', `slicing', etc.). Such an analysis {is known as} `full-shape analysis', and was first introduced and performed in~\cite{DAmico:2019fhj,Ivanov:2019pdj,Colas:2019ret}. It is now routinely used by the main observational collaborations to extract cosmological information from their data, {as in} the latest data analyses by the DESI collaboration~\cite{DESI:2024jis,DESI:2024hhd,DESI:2025ejh,DESI:2025wzd}. 

{Being a perturbative method, the EFTofLSS yields predictions which become more accurate as the order increases and as correlation functions are computed for a greater number of external legs.}
At present, the most advanced EFTofLSS-based completed data analysis has been the one-loop analysis of both the power spectrum and bispectrum~\cite{DAmico:2022gki,DAmico:2022ukl}. 
{A natural next research direction is to either move to the two-loop order or to higher $n$-point functions at tree level and one loop.\footnote{For recent computations of two-loop spectra of dark matter in redshift space and of galaxies in the EFTofLSS, see e.g.~\cite{Taule:2023izt,Bakx:2025cvu}.}
In this paper, we focus on the power spectrum at two-loop order, developing novel  techniques that will enable the analysis of data at this order.}
In particular, we compute the renormalized two-loop dark-matter power spectrum, an observable that is very close
 to what is directly observed in galaxy lensing surveys such as DES~\cite{DES:2025xii} and Euclid~\cite{Euclid:2019clj}.  The method presented in this work can also be directly applied to including baryonic effects in the lensing power spectrum at two loops \cite{Lewandowski:2014rca,Braganca:2020nhv}.

In Ref.~\cite{Anastasiou:2022udy}, a method was introduced which maps one-loop corrections of  correlators to a moderate number of scalar integrals,  familiar in the context of perturbation theory for QFT amplitudes.  
Based on reduction methods to master integrals, a novel computation of the one-loop bispectrum was carried out in Ref.~\cite{Anastasiou:2022udy} and a general algorithm was proposed for the computation of generic $n$-point  one-loop correlation functions. The method of Ref.~\cite{Anastasiou:2022udy} is very well suited for the computation of one-loop cosmological correlators for the following reasons. First, at one-loop, reduction identities to master integrals can be diagonalised analytically in a symbolic form and then implemented recursively in an efficient computer algorithm which memorizes and reuses earlier steps in the reduction. Second, all master integrals can be computed in a closed analytic form.

At two loops, reduction identities are not generally known in a diagonal form and the master integrals are also not available, as this is a field of active mathematical developments. 
In this article, we develop a purely numerical  technique for the computation of two-loop corrections to the power spectrum. We focus on the dark-matter power spectrum, but the technique easily generalizes to the two-loop power spectrum of other cosmological objects (galaxies, Ly-$\alpha$, etc.).

A major objective of this work is the fast evaluation of the power spectrum for comparison with high-precision observations and for the practical inference of cosmological parameters. To achieve both precision and speed when computing loop corrections across a large cosmological parameter space, we find it important to: (i) isolate the cosmological model dependence into coefficients of model-independent basis integrals and (ii) precompute these universal loop integrals with very high accuracy.

The disentanglement of cosmological model dependence and integrations is based on a faithful emulation of model-dependent linear power spectra in terms of a few basis functions. For this purpose, we decompose the linear power spectrum for a generic cosmological model into two components: a component common to all models, which corresponds to the linear power spectrum of a reference model chosen to be a good fit to CMB data, and a second, small, model-dependent remnant that we fit to our basis functions. This allows us to make a perturbative expansion around the reference model, which reduces the number of model-independent loop integrals to ${\cal O}(300)$.

While these integrals cannot be readily computed with state-of-the-art analytic methods, we can compute them using simple stochastic integration. This is possible through a systematic treatment of ultraviolet (UV) and infrared (IR) enhancements in the loop integrands. We render our basis integrals regular in all UV regions by subtracting UV approximations at the integrand level, which, upon integration, yield contributions degenerate with the EFTofLSS counterterms. Furthermore, we cancel all IR singularities by mapping them to a minimal set of common points or surfaces where they cancel out across all diagrams.

After these systematic manipulations, the loop integrands become smooth in the entire integration domain, allowing us to compute them numerically with very high accuracy using modest computational resources.

A few other fast-evaluation strategies already exist in the literature, notably the approach of Ref.~\cite{Simonovic:2017mhp}, 
which decomposes loop integrals into power-law kernels using the so-called FFTLog basis (see \cite{Lewandowski:2018ywf} for an analytic implementation of the IR resummation in that basis), and recently the COBRA method, which factorizes the linear power spectra by performing a truncated singular value decomposition (SVD) of a set of template spectra and precomputing the corresponding loop tensors \cite{Bakx:2024zgu}. Our method differs from those by the expansion around a reference cosmology, by the choice of basis functions, and by the treatment of UV and IR divergencies.

While this paper was in the final stages of completion, Ref.~\cite{Bakx:2025jwa} appeared, which implements, as our method, a fast numerical evaluation of the two-loop matter power spectrum in the context of the EFTofLSS for arbitrary cosmologies, based on the COBRA method, which we have previously described. The approaches differ at a theoretical level by the number of counterterms, and in the numerical implementation.

{The paper is organized as follows. In Sec.~\ref{sec:EFTofLSS_biased_UV}, we briefly review the EFTofLSS perturbative expansion, motivating the origin of the convolution integrals over momentum. In Sec.~\ref{sec:PerturbativeSeries} we describe the perturbative loop expansion of the dark-matter power spectrum. In Sec.~\ref{sec:UV} we describe the issue of UV dependence, regularization and renormalization of the two-loop dark-matter power spectrum. {In Sec.~\ref{sec:IR}, we discuss a method by which cancellations of IR enhancements that occur across diagrams are made manifest at the level of the integrand (locally)}. In Sec.~\ref{sec:NumericalMethod} we discuss the numerical integration. In Sec.~\ref{sec:perturbative_expansion} we describe how practically all the power spectra of interest can be described as small deviations from a reference one. In Sec.~\ref{sec:Pfit} we describe how to factor out the cosmology dependence from the integrals. In Sec.~\ref{sec:Results} we show the results of our numerical integration, establishing the precision of our procedure. We conclude in Sec.~\ref{sec:conclusion}.}

\section{Review of perturbation theory}
\label{sec:EFTofLSS_biased_UV}
{We start by reviewing some general aspects of the perturbative expansion associated to the EFTofLSS, presenting, at the end of this section, the actual original material. In the review part, we mainly follow Sec. 2 of~\cite{DAmico:2022ukl}.}
As usual we start from the Boltzmann equations for the dark-matter particle distribution assuming a collisionless fluid. Taking the first and second moments of the Boltzmann equation and evaluating the {{long wavelength limit}} of these equations, one obtains the equations of motion of the dark-matter fields in terms of its density $\rho$ and velocity $v^i$ fields~\cite{DAmico:2022ukl,Baumann:2010tm,Carrasco:2013mua}. The background density field $\bar{\rho}(a)$ is position independent and evolves as,
\begin{align}
    \bar{\rho}(a) = \bar{\rho}_0\left(\frac{a_0}{a} \right)^3\, ,
    \label{eq:rhobar}
\end{align}
where $\bar{\rho}_0$ and $a_0$ are the current background density and scale factor respectively. We will denote $\dot{\mathcal{A}} = d\mathcal{A}/dt$.
Together with the Poisson equation, one obtains the following set of coupled equations,
\begin{align}
    \dot{\rho} + 3H\rho + \frac{\partial_i \pi^i}{a} &= 0 \ ,  \\
    \dot{\pi}^i + 4H\pi^i + \frac{\partial_j}{a}\left(\frac{\pi^i\pi^j}{\rho}\right) + \frac{\rho}{a}\partial_i\Phi &= -\frac{1}{a}\partial_j \tau^{ij} \ ,  \\
    \partial^2\Phi &= \frac{3}{2}\Omega_m H^2 a^2 \delta \ ,
\end{align}
where we have defined the overdensity $\delta(\vx, a) \equiv (\rho(\vx,a) - \bar{\rho}(a)) / \bar{\rho}(a)$
and momentum $\pi^i(\vx, a) \equiv \rho(\vx, a) v^i(\vx, a)$, $H \equiv \dot{a} / a $, $\Phi$ is the scalar perturbation to the metric, and $\Omega_m(a)$ is the total matter fraction. The no-counterterm  solutions (also known as SPT solutions) for the density and momentum fields are solutions to the above with the effective stress tensor $\tau^{ij} = 0$. Crucially, the presence of the $\tau^{ij}$ is required in the EFTofLSS as it encodes {the effect at long wavelengths of short distance fluctuations}.

The equations of motion for $\pi$ can be {simplified} by noting that we can express,
\begin{align}
    \rho \partial_i \Phi = \bar{\rho}(\delta \partial_i\Phi + \partial_i\Phi) = \bar{\rho}\partial_i \Phi + \frac{2}{3}\frac{\bar{\rho}}{\Omega_m \mathcal{H}^2}\partial_j\left(\partial_i\Phi\partial_j\Phi - \frac{1}{2}\delta_{ij}(\partial\Phi)^2\right) \ .
\end{align}
with $\mathcal{H} \equiv H \, a = \dot{a}$.
It is furthermore useful to decompose the momentum field into a scalar and vector,
\begin{align}
    \pi_S &\equiv \partial_i \pi^i \ , \\
    \pi^i_V &\equiv \epsilon^{ijk}\partial_j \pi^k\ , \\ 
    \pi^i &= \frac{\partial_i}{\partial^2}\pi_S - \epsilon^{ijk}\frac{\partial^j}{\partial^2}\pi^k_V  \ , \label{eq:combine_pi}
\end{align}
where $\epsilon^{ijk}$ is the Levi-Civita connection. Spatial indices are contracted with the Kronecker-Delta $\delta^K_{ij}$, and we generally do not distinguish between upper and lower spatial indices. 
Combining the above equations, the equations of motion are transformed to,
\begin{align}
    a\mathcal{H}\delta' + \frac{\pi_S}{\bar{\rho}} &= 0 \label{eq:EOM1}\ , \\
    \mathcal{H}\pi_S' + \frac{4\mathcal{H}}{a}\pi_S + \frac{3}{2a}\bar{\rho}\Omega_m \mathcal{H}^2\delta &= -\frac{\partial_i\partial_j}{a}\left(\frac{2}{3}\frac{\bar{\rho}}{\Omega_m \mathcal{H}^2}\left( \partial_i\Phi\partial_j\Phi - \frac{1}{2}\delta_{ij}(\partial \Phi)^2\right) \right. \nonumber \\
    &\left. \hspace{1in} + \frac{\pi^i\pi^j}{\rho}+ \tau^{ij}\right)\label{eq:EOM2}\ , \\
    \mathcal{H}\pi^{'i}_V + 4\frac{\mathcal{H}}{a}\pi^i_V &=-\frac{\epsilon^{ijk}}{a}\partial_j\partial_l\left(\frac{2}{3}\frac{\bar{\rho}}{\Omega_m \mathcal{H}^2}\partial_k\Phi\partial_l\Phi+ \frac{\pi^k\pi^l}{\rho} + \tau^{kl}\right)\ .\label{eq:EOM3}
\end{align}
where $\mathcal{A}' = \partial\mathcal{A}/\partial a$.
Combining \eqref{eq:EOM1} and \eqref{eq:EOM2}, the {equation} for $\delta$ can be written as,
\begin{align}
    a^2 \delta'' + \left(2 + \frac{a \mathcal{H}'}{\mathcal{H}} \right)a\delta' - \frac{3}{2}\Omega_m \delta &= \frac{\partial_i\partial_j}{\mathcal{H}^2\bar{\rho}}\left(\frac{2}{3}\frac{\bar{\rho}}{\Omega_m \mathcal{H}^2}\left( \partial_i\Phi\partial_j\Phi - \frac{1}{2}\delta_{ij}(\partial \Phi)^2\right)\right.\nonumber \\
    &\left. \hspace{1in} + \frac{\pi^i\pi^j}{\rho} + \tau^{ij}\right)\label{eq:EOM_delta}\  .
\end{align}

\subsection{{Factorizing the time dependence}} \label{solssec}
{Let us expand the solution to the non-linear equations as\footnote{{We will work with the following Fourier and integral conventions,
\begin{align}
& f ( \xvec ) = \int_{\kvec} e^{i \kvec \cdot \xvec} f ( \kvec)  \ , \quad     \int_{\vk_1,\dots,\vk_n} \equiv \int \frac{d^3\vk_1}{(2 \pi)^3}  \dots  \int \frac{d^3\vk_n}{(2 \pi)^3} , \quad \int^{\vk}_{\vk_1,\dots,\vk_n} \equiv \int_{\vk_1,\dots,\vk_n} ( 2 \pi)^3 \delta_D(\vk - \sum_{i=1}^n \vk_i)\   \ .
\end{align}}
}
\be
\delta(\vec k,a)=\sum_n \delta^{(n)}(\vec k,a) \ ,
\ee
where $\delta^{(n)}(\vec k,a)$ is of order $\delta^{(1)}(\vec k,a)^n$ and $\delta^{(1)}(\vec k,a)$ is the solution to the linearized equations. We make analogous definitions for the velocity and the momentum fields.
The solution to the linear equation, $ \delta^{(1)}(\vec k,a)$, factorizes into a time dependent and a space dependent part
\be
\delta^{(1)}(\vec k,a)=D(a)\tilde\delta^{(1)}(\vec k) \ ,
\ee
where $D(a)$ is known as the growth factor, and it satisfies
\be
 a^2 D''(a) + \left(2 + \frac{a \mathcal{H}'}{\mathcal{H}} \right)aD'(a) - \frac{3}{2}\Omega_m D(a)=0\ ,
\ee 
while $\tilde\delta^{(1)}(\vec k)$ is fixed by the initial conditions at some time $a_{\rm in}$, such that the linear power spectrum at $a_{\rm in}$, $\mathcal{P}_{\rm lin.} ( k)$, is defined by
\be
\langle \tilde \delta^{(1)} ( \kvec) \tilde \delta^{(1)} ( \kvec' ) \rangle = (2 \pi)^3 \delta_D ( \kvec + \kvec') \mathcal{P}_{\rm lin.} ( k) \ , 
\ee
and we have normalized the growth factor such that $D(a_{\rm in}) = 1$.

In the so-called EDS approximation, which works at per cent and even better at higher redshifts  (see for example~\cite{Donath:2020abv,Fasiello:2022lff,Garny:2022fsh}), we have $\Omega_m(a) \approx f(a)^2$, with 
$f (a) \equiv \frac{a D'(a)}{D(a)}$, and the $n$-th order solution  $\delta^{(n)}(\vec k,a)$ has a factorized time dependence
\be
\delta^{(n)}(\vec k,a)=D(a)^n \tilde\delta^{(n)}(\vec k)\ ,
\ee 
such that $\tilde \delta^{(n)}$ is time-independent. Analogously, we can define the time-independent momentum-density perturbations
\be \label{pitimeind}
\tilde{\pi}^{i}_{(n) } ( \kvec )  = \frac{-\pi^{i}_{(n)} ( \kvec , a ) }{aH\bar{\rho} fD(a)^n}   \ , 
\ee
such that
\be
\tilde \pi^i_{(n)} ( \xvec , a )  = \frac{1}{D(a)^n} \left[ (1 + \delta ( \xvec , a) ) \frac{\partial_i \theta ( \xvec , a )}{\partial^2} \right]^{(n)} \ . 
\ee 
We use the same scaling as \eqn{pitimeind} to define the time-independent parts $\tilde \pi_S$ and $\tilde \pi^i_{V}$. 

It is often convenient to use the velocity field, which, neglecting vorticity, at a given order can be rewritten in terms of its rescaled divergence  $\theta$, defined by 
\be
\theta(\vx, t') \equiv  - \partial_i v^i(\vx, t') / (a H f)  \ , 
\ee
as
\begin{align} \label{vastheta}
    v^{i}_{(n)} (\vx, t') = -a H f\frac{\partial_i}{\partial^2}\theta^{(n)}(\vx, t') = -\frac{a \dot{D}(t')}{D(t')}\frac{D(t')^n}{D(t)^n}\frac{\partial_i}{\partial^2}\theta^{(n)}(\vx, t)\  ,
\end{align}
where we note that $\theta$ also has a time dependence 
\be
\theta^{(n)}(\vx, t')  = \left( \frac{D(t')}{D(t)} \right)^n \theta^{(n)}(\vx, t) \ ,
\ee
such that we can also define the time-independent version $\tilde \theta$ as
\be
\tilde \theta^{(n)} ( \kvec) = D(a)^{-n} \theta^{(n)} ( \kvec , a ) \ . 
\ee
The fields are rescaled such that $\delta^{(1)} ( \kvec , a ) = \theta^{(1)} ( \kvec , a )$.  

Assuming $\tau^{ij} = 0$ (we will compute the contributions from $\tau^{ij}$ in \secref{dmctssec}), these scalings allow us to solve for the time-independent parts of the fields in the following way.  The $n$-th order part of~\eqref{eq:EOM_delta} becomes  
\begin{align}
    D(a)^n \frac{n-1}{2}\left(3\Omega_m + 2n f^2 \right)\tilde{\delta}^{(n)}(\kvec) = S^{(n)}_\delta (\kvec,a) \big|_{\tau = 0} \ ,
\end{align}
where we have denoted the $n$-th order part of the right hand side of \eqref{eq:EOM_delta} in Fourier space as a source term $S^{(n)}_\delta (\kvec,a) \big|_{\tau = 0} $. With the EDS-approximation $f(a)^2 = \Omega_m(a)$, we get 
\begin{align}
 \tilde{\delta}^{(n)}( \kvec ) = \frac{2}{(n-1)(3+2n)} \tilde{S}^{(n)}_\delta( \kvec) \big|_{\tau = 0} \ ,
\end{align}
where we have factorized the source term $S^{(n)}_\delta (\kvec, a) \big|_{\tau = 0} = D(a)^n\Omega_m \tilde{S}^{(n)}_\delta( \kvec) \big|_{\tau = 0} $.  This also means that 
\be
\tilde \pi_S^{(n)} ( \kvec ) =  n\, \tilde \delta^{(n)} ( \kvec ) \ . 
\ee
Similarly, denoting the right-hand side of \eqn{eq:EOM3} in Fourier space as ${S}^{i}_{\pi}(\kvec , a ) \big|_{\tau = 0}$, and the rescaled version as $\tilde{S}^{i}_{\pi, (n)}(\kvec) \big|_{\tau = 0} = {S}^{i}_{\pi, (n)}(\kvec,a) \big|_{\tau = 0} / (a H(a)^2 \bar \rho ( a ) \Om (a) )$, we have
\begin{align}
     \tilde \pi^{i}_{V,(n)}(\kvec) = \frac{-2}{3 + 2(n-1)} \tilde{S}^{i}_{\pi, (n)}(\kvec) \big|_{\tau = 0} \  .
\end{align}
Finally, the expression for $\pi^i_{(n)}$ is then computed with \eqn{eq:combine_pi}.

Explicitly, the expressions $\tilde{S}^{(n)}_\delta$ and $\tilde{S}^{i}_{\pi,(n)}$ in coordinate space are given by
\begin{align}
    \tilde{S}^{(n)}_\delta \big|_{\tau = 0} &= \left(\partial_i\partial_j\left(\frac{3}{2}\frac{\partial_i\tilde{\delta}}{\partial^2}\frac{\partial_j\tilde{\delta}}{\partial^2} - \frac{3}{4}\delta_{ij} \frac{\partial_k\tilde{\delta}}{\partial^2}\frac{\partial_k\tilde{\delta}}{\partial^2} + \frac{\tilde{\pi}^i\tilde{\pi}^j}{1+\tilde{\delta}}   \right)\right)^{(n)}\label{eq:d_response}\ ,\\
    \tilde{S}^{i,(n)}_{\pi} \big|_{\tau = 0}  &= - \left(\epsilon^{ijk}\partial_j\partial_l\left(\frac{3}{2}\frac{\partial_k\tilde{\delta}}{\partial^2}\frac{\partial_l\tilde{\delta}}{\partial^2} + \frac{\tilde{\pi}^k\tilde{\pi}^l}{1+\tilde{\delta}}+  \right)\right)^{(n)}\label{eq:pi_response}\ ,
\end{align}
where the notation $(T[\tilde \delta , \tilde \pi^i])^{(n)}$ indicates taking the $n$-th term of the expansion of $T$. For example $(\tilde{\pi}^{i}\tilde{\pi}^{j})^{(3)} = \tilde{\pi}^{i}_{(2)} \tilde{\pi}^{j}_{(1)}+\tilde{\pi}^{i}_{(1)} \tilde{\pi}^{j}_{(2)}$.  Note that the tilde fields are defined order by order in the perturbations. We write the tilde fields inside of the expression $(T[\tilde \delta , \tilde \pi^i])^{(n)}$ to remind the reader to use them once expanding in perturbations.

The solutions to the time-independent density and velocity fields, {assuming $\tau^{ij} =0$,} are well known and given by,
\begin{align}
\label{eq:SPTdellta}
    \tilde{\delta}^{(n)}(\vec{k}) &= \int_{\qvec_1 , \dots , \qvec_n}^{\kvec} F_n(\vec{q}_1,\dots,\vec{q}_n)\tilde{\delta}^{(1)}(\vec{q}_1)\dots\tilde{\delta}^{(1)}(\vec{q}_n)\  ,\\
    \label{eq:SPTvelo}
    \tilde \theta^{(n)} (\vec{k}) &= \int_{\qvec_1 , \dots , \qvec_n}^{\kvec} G_n(\vec{q}_1,\dots,\vec{q}_n)\tilde{\delta}^{(1)}(\vec{q}_1)\dots\tilde{\delta}^{(1)}(\vec{q}_n)\  ,
\end{align}
and assuming that the velocity field is irrotational, it is given by
\be
v^{i}_{(n)} ( \kvec , a ) = i a H f D(a)^n \frac{k^i}{k^2} \tilde \theta^{(n)} ( \kvec) \ .
\ee
Above, $F_n$ and $G_n$ are the symmetric kernels (in general, all perturbative kernels in this paper are assumed to be symmetric, unless otherwise stated) explicitly given by~\cite{Goroff:1986ep} 
\begin{align}
\begin{split} \label{symmfg}
    F_{n}(\vq_1,\ldots , \vq_n ) & =      \frac{1}{n!} \sum_{\rm perms.} \, \bar F_{n}(\vq_1,\ldots , \vq_n ) \ , \\ 
        G_{n}  (\vq_1,\ldots , \vq_n ) & =      \frac{1}{n!} \sum_{\rm perms.} \, \bar G_{n}(\vq_1,\ldots , \vq_n ) \ , 
\end{split}
\end{align}
and the recurrence relations,  
\begin{eqnarray}
\bar F_n(\vq_1,\dots,\vq_n) &=& 
\sum_{m=1}^{n-1} \frac{ \bar G_m(\vq_1,\dots,\vq_m)}{(2n+3)(n-1)}
\left[ (2n+1) \frac{\vk\cdot\vk_1}{k_1^2} \bar F_{n-m}(\vq_{m+1},\dots,\vq_n) \right. 
\nonumber \\
&&\qquad \left. +  \frac{k^2 (\vk_1\cdot\vk_2)}{k_1^2 k_2^2}
\bar G_{n-m}(\vq_{m+1},\dots,\vq_n) \right]  \ , \\
\nonumber \\
\bar G_n(\vq_1,\dots,\vq_n) &=& \sum_{m=1}^{n-1} \frac{\bar G_m(\vq_1,\dots,\vq_m)}{(2n+3)(n-1)}
\left[ 3 \frac{\vk\cdot\vk_1}{k_1^2} \bar F_{n-m}(\vq_{m+1},\dots,\vq_n) \right. \nonumber \\
&&\qquad\qquad\qquad\qquad \left. + n \frac{k^2 (\vk_1\cdot\vk_2)}{k_1^2 k_2^2}
\bar G_{n-m}(\vq_{m+1},\dots,\vq_n) \right] \ ,
\end{eqnarray}
with the starting conditions for the recurrence, 
\[
\bar F_1 = \bar G_1=1  \ . 
\]
The vectors in the above expressions are given by 
\[ 
\vk_1=\vq_1+\cdots+\vq_m \  , \;  
\vk_2=\vq_{m+1}+\cdots+\vq_n \ , \; 
\vk=\vk_1+\vk_2 \ .
\]
To obtain full and consistent solutions, the results of Eq. \eqref{eq:SPTdellta} should be combined with solutions sourced by the stress tensor, which we will denote with the subscript ${}_{\rm ct}$ and describe their construction next.

\subsection{Dark-matter counterterms} \label{dmctssec}

{Now that we have concluded the review of the perturbative solution without the contribution from counterterms, we will now proceed to derive the dark-matter counterterms up to third order in the EFTofLSS.}
Loop level corrections to cosmological $n$-point functions are generally UV sensitive as the integrands have support over momenta ranges well in the non-linear scale {within} which we do not have perturbative control. To remove this UV sensitivity, we require renormalizing the divergent loop integrals with EFT counterterms {obtained by expanding} the stress tensor $\tau^{ij}$ {in terms of the long-wavelength fields}. We will construct $\tau^{ij}$ at a given order in terms of rotational invariant and {diffeomorphism invariant (Galilean invariant in the non-relativistic limit)} tensors involving {{the long wavelength fields}}.

{The long wavelength fields at our disposal are $\delta$, $v^i$ and $\Phi$. Up to very high perturbative order~\cite{Carrasco:2013mua}, the velocity can be taken as the gradient of a scalar (we discuss vorticity in more detail at the end of this section). {To simplify the expressions further, it is more convenient to work with the density field $\delta$ instead of $\Phi$ and  the velocity field divergence $\theta$  instead of $v^{i}$. We will use the following {three} operators to expand~$\tau^{ij}$,
{\begin{equation}
    r_{ij} = \frac{\partial_i\partial_j \delta}{\partial^2},  \quad p_{ij} = \frac{\partial_i\partial_j \theta}{\partial^2}, \, \text{ and} \quad \delta^{K}_{ij}\  ,
\end{equation}
with $\delta^{K}_{ij}$ being the Kronecker-Delta. Note that $\delta = \delta^K_{ij}r_{ij}$ and $\theta = \delta^K_{ij}p_{ij}$.}}
{
The highest-order contribution to the two-loop power spectrum is fifth-order. This arises from the fifth-order kernel, ${\cal{P}}_{15}$, one of the terms that make up the two-loop power spectrum, which we will discuss in detail later.}
To renormalize the ${\cal{P}}_{15}$ diagram we will require the third order {counterterm} $\td^{(3)}_{\rm ct}$ which contains the third order stress tensor. {Therefore, 
in our expansion of $\tau^{ij}$, we can ignore terms of higher order than $\td^{(3)}_{\rm ct}$.}

We notice that the stress tensor is a local-in-space but non-local-in-time function of the long-wavelength fields and their derivatives. This is because the timescale of short modes, Hubble, is the same time scale as the one controlling long modes~\cite{Carrasco:2013mua,Senatore:2014eva}.\footnote{{This makes the EFTofLSS a somewhat unusual EFT, but we remind the reader that dielectric materials share the same property.}}  Therefore, the stress tensor at a spacetime point $x^\mu$ depends on the long wavelength fields evaluated on the full past trajectory of that point, the trajectory being the one associated to the one of the fluid element associated to $x^\mu$.  Let us briefly review this more in detail, following~\cite{Senatore:2014eva} except for a small correction that we will highlight.

The dark matter stress tensor (but the logic applies unaltered, with minor adjustments, to any UV sensitive quantity, such as the number density of galaxies, or the luminosity in Ly-$\alpha$ forest, etc.) is a very complicated function of many physical variables, evaluated on the past light cone of a point:
\bea
&&\tau_{ij}(\vec x,t)=\\ \nonumber
&&\quad f_{\rm very\, complicated}\left(\left\{H,\Omega_m,\dots,m_{\rm dm}, g_{\rm ew}, \ldots, \rho_{\rm dm},\partial_i v_{\rm dm}^j,\ldots\right\}_{{\rm past\, light\, cone\,} (\vx,t)}\right)_{ij}\ , \nonumber  
\eea
where $f_{\rm very\, complicated}(\ldots)_{ij}$ is a very complicated functional with the same transformation properties under spacetime diffeomorphisms as $\tau_{ij}$, $H(t)$ is the Hubble constant, $\Omega_m$ is the present dark matter abundance, $m_{\rm dm}$ is the mass of the dark matter, $g_{\rm ew}$ is the weak coupling constant, $\rho_{\rm dm}$, and $v^i_{\rm dm}$ are the dark matter density and velocities. The subscript ${}_{{\rm past\, light\, cone\,} (\vec x,t)}$ means that the variables need to be evaluated on the past light cone of the spacetime point $(\vx,t)$. In reality, since matter in the universe does not move more than about 10 Mpc, the non-linear scale, the dependence on the past light cone can be reduced to a dependence on the past tube around the fluid element that ended up at a given spacetime point, a tube of about 10 Mpc in spatial radius, and a length of about an Hubble time, as indicated by the subscript ${}_{{\rm past\, tube\,} (\vx,t)}$:
\bea
\tau_{ij}(\vec x,t)=f_{\rm very\, complicated}\left(\left\{H,\Omega_m,\dots,m_{\rm dm}, g_{\rm ew}, \ldots, \rho_{\rm dm},\partial_iv_{\rm dm}^j,\ldots\right\}_{{\rm past\, tube\,} (\vx,t)}\right)_{ij}\ . \nonumber \\ 
\eea
We have significantly divided the variables in three groups: the background cosmological parameters, the particle physics related parameters, and finally the fluctuating cosmological perturbations. Clearly, if one is interested in the value of $\tau_{ij}$ at a specific spacetime point, this formula is a formidable challenge, hence the name `very complicated.' For example, if we extend for a moment the concept to galaxies (but something similar happens also for dark matter), changing the electroweak constant will change the nuclear reactions, which in turn will change the number of stars, and so, perhaps, the mass of a galaxy. However, if we are interested only on $\tau_{ij}$ in the long wavelength limit (as in the EFTofLSS), several tremendously powerful but controlled approximations can be performed. 

First, since the long wavelength fields have small fluctuations, we can Taylor expand $f_{\rm very\, complicated}$ in powers of the long-wavelength fluctuating fields. Since the size of the fluctuations scales as powers of $k/\knl$, this corresponds to a controlled expansion. If we call~$\{{\cal{O}}\}$ the set of the long-wavelength fluctuating fields $\{{\cal{O}}_1,{\cal{O}}_2,\ldots\}$ (which can also be tensors under rotations, but we drop spatial indices here for brevity), with ${\cal{O}}_1$ being, for example,~$\delta$, we can write:
\bea\nonumber
&&\tau_{ij}(\vec x,t)=\sum_{n=0}^{\infty}\int_{{\rm past\, tube\,} (\vx,t)} d^4x_1\ldots \int_{{\rm past\, tube\,} (\vx,t)} d^4x_n \sum_{{\rm all\, subsets\,} {\left\{{\cal{O}}_1,\ldots, {\cal{O}}_n \right\}}\subset \{{\cal{O}}\}}\\ 
&&\qquad \qquad\left\{\left. \frac{\delta^n f_{\rm very\, complicated}(\left\{{\cal{O}}\right\}_{\rm past\, tube})}{\delta{\cal{O}}_1(x_1)\ldots \delta{\cal{O}}_n(x_n)}\right|_0 {\cal{O}}_1(x_1)\ldots {\cal{O}}_n(x_n) \right\}_{ij} \ ,
\eea
where $\left.\right|_0$ means that we set to zero all the long-wavelength fluctuating fields.  The right hand side should transform as $\tau_{ij}$, which is shown by the notation $\{\ldots\}_{ij}$. 

Up to this point, $\left.\frac{\delta^n f_{\rm very\, complicated}(\left\{{\cal{O}}\right\}_{\rm past\, tube})}{\delta{\cal{O}}_1(x_1)\ldots \delta{\cal{O}}_n(x_n)}\right|_0$ still depends on the short wavelength fields. Since short wavelength fields have short correlation functions, for long wavelength descriptions we can approximate this quantity in a controlled way with 
\bea
&&\left.\frac{\delta^n f_{\rm very\, complicated}(\left\{{\cal{O}}\right\}_{\rm past\, tube})}{\delta{\cal{O}}_1(x_1)\ldots \delta{\cal{O}}_n(x_n)}\right|_0\simeq\\\ \nonumber
&&\qquad\qquad\qquad \langle \left.\frac{\delta^n f_{\rm very\, complicated}(\left\{{\cal{O}}\right\}_{\rm past\, tube})}{\delta{\cal{O}}_1(x_1)\ldots \delta{\cal{O}}_n(x_n)}\right|_0\rangle+\epsilon_{1,\ldots,n}(x_1,\ldots, x_n)\ .
\eea
Here $\langle\ldots\rangle$ represents taking the expectation value over realizations of the universe, and $\epsilon_{1,\ldots,n}(x_1,\ldots, x_n)$ are fields, that are called stochastic counterterms, that accounts for the difference, in each realization and at long wavelengths, between the left-hand-side and its approximation with the expectation value. In particle physics, the step we have just performed would be described as `integrating out the short fields in a given long-wavelength configuration,' a small generalization of the background field method. Here, in this step we have just integrated out the dark matter halos and the galaxies, which are short objects with respect to the universe. Because only short-wavelength fields are present, $\epsilon_{1,\ldots,n}(x_1,\ldots, x_n)$ have correlation functions only among themselves, and they have support only at short distances, below the non-linear scale. Therefore, their correlation functions have spatial $\delta$-function support, $\sim \delta(\vec x-\vec x')$. The symbol $\simeq$ accounts for the fact that we are focusing on long wavelengths.

Since $\langle\left.\frac{\delta^n f_{\rm very\, complicated}(\left\{{\cal{O}}\right\}_{\rm past\, tube})}{\delta{\cal{O}}_1(x_1)\ldots \delta{\cal{O}}_n(x_n)}\right|_0\rangle$ can only depend on background quantities, its tensorial dependence can only be through a Kronecker delta (or $\epsilon_{ijk}$, but, if we assume parity as we do from now on, we can drop it).
The kernels $\langle\left. \frac{\delta^n f_{\rm very\, complicated}(\left\{{\cal{O}}\right\}_{\rm past\, tube})}{\delta{\cal{O}}_1(x_1)\ldots \delta{\cal{O}}_n(x_n)}\right|_0\rangle$ are still extremely complicated, with spatial dependence on scales shorter or comparable to the tube, {\it i.e. $1/\knl$}, but with time dependence of order Hubble. In fact, the orbiting time of two galaxies in a cluster is about an Hubble time, notwithstanding they are only about 10 Mpc far way.

Now, we can perform another controlled approximation. We can approximate the spatial dependence of the long wavelength fluctuating fields along the past light tube with a Taylor expansion around the spatial center of the tube, that we denote with $\vec{x}_{\rm fl}$. The fluid element $\vec{x}_{\rm fl}$ is defined iteratively as,
\begin{align}
    \vec{x}_{\rm fl}(\vx, t, t') = \vx  + \int^{t'}_t \frac{dt_1}{a(t_1)}\vec{v}(\vec{x}_{\rm fl}(\vx, t, t_1), t_1)\  .
\end{align}
After Taylor expansion, the spatial dependence within the tube is known, and one can perform the spatial integrals of each spatial slice of the tube, obtaining contributions of order $1/\knl$ for each spatial coordinate.\footnote{Doing this for the stochastic counterterms is justified by the fact that doing this is equivalent to Taylor expanding the distance between two far-away points that appears in the $\delta$-function in the correlation function among stochastic terms around the distance between the center of the respective past tubes.} Expanding the fields around the center of the past tube therefore corresponds to a perturbative expansion in $k/\knl$, $k$ being the wavenumber of the fluctuating fields, which is controlled at long wavelengths. This is the familiar derivative expansion in particle physics. We are therefore led to:
\bea\label{eq:final_stress}
&&\tau_{ij}(\vec x,t)\simeq\sum_{n=0}^{\infty}\int d t_1\ldots \int d t_n \sum_{\rm{all\, subsets\,} {\left\{ \mathcal{O}_1,\ldots, \mathcal{O}_n  \right\}}\subset \{{\cal{O}}\}}\\ \nonumber
&&\qquad\qquad\times  \left\{\left(K_{1,\dots,n}(t,t_1,\ldots, t_n)+\epsilon_{1,\ldots,n}(t, (\vec{x}_{\rm fl}(\vx, t, t_1),t_1) ,\ldots,  (\vec{x}_{\rm fl}(\vx, t, t_n),t_n )  )\right.\right.\\\nonumber
&&\qquad\qquad\quad\left.+\frac{\partial}{\knl}\epsilon_{1,\ldots,n,\partial}(t,(\vec{x}_{\rm fl}(\vx, t, t_1),t_1),\ldots, (\vec{x}_{\rm fl}(\vx, t, t_n),t_n))+\ldots\right)\;\\ \nonumber
&&\qquad\qquad\times \left.{\cal{O}}_1(\vec{x}_{\rm fl}(\vx, t, t_1),t_1)\ldots {\cal{O}}_n(\vec{x}_{\rm fl}(\vx, t, t_n),t_n)\right\}_{ij}  \ ,
\eea
where we have enlarged the set of $\{{\cal{O}}\}$ (without changing notation) to include fields obtained from the original fields by acting an arbitrary number of spatial derivatives, {evaluated at the fluid element}, acting on them. Notice that the fields are now evaluated on the past fluid trajectory corresponding to the final point~$(\vec x,t)$, denoted by $(\vec{x}_{\rm fl}(\vx, t, t_i),t_i)$, and are integrated only in $t_i$. $ K_{1,\dots,n}(t,t_1,\ldots, t_n)$ are the kernels resulting from the spatial integration along the spatial slices of the tube. We have used spatial translation invariance of the background to eliminate the dependence on~$\vec x$ in them. As discussed in~\cite{Senatore:2014eva}, it is unlikely that one can perform such a Taylor expansion also for the time dependence, because the time dependence of the kernels and of the long-wavelength fields are comparable. We have therefore reached the final formula for the stress tensor (and an analogous one holds for the number density of galaxies, etc.). Eq.~(\ref{eq:final_stress})  slightly differs from the original derivation of~\cite{Carrasco:2013mua,Senatore:2014eva} because there, when Taylor expanding $f_{\rm very\, complicated}$, {it was forgotten to insert the fields at different times along the fluid trajectory, but they were inserted always at the same time on the trajectory.}

After having reviewed the derivation of the counterterms, and made a small correction, we are ready to move forward. In this paper, we are interested in the stress tensor up to cubic order in the fluctuating fields. The stochastic terms contribute at a subleading order in $k/\knl$, starting at $(k/\knl)^4$ in the power spectrum, and can be neglected. We are therefore led to
\bea\label{eq:bias_exp}
&&\tau_{ij}(\vec{x},t) = {\frac{\Omega_m \mathcal{H}^2\bar{\rho}}{\knl^2}}\sum_l \int^t dt' H{(t')}  \left( c_{1,_{\mathcal{O}_l}}(t,t'){\mathcal{O}_{1,l}}(\vec{x}_{\rm fl}(\vec x, t,t'),t')_{ij}+\right.\\\nonumber
&&\quad+\int^t  dt'' H{(t'')} \left(c_{2,_{\mathcal{O}_l}}(t,t',t''){\mathcal{O}_{2,l}}((\vec{x}_{\rm fl}(\vec x, t,t'),t'),(\vec{x}_{\rm fl}(\vec x, t,t''),t''))_{ij}\right. \\ \nonumber
&&\quad+\int^t  dt''' H{(t''')} \left.\left.c_{3,_{\mathcal{O}_l}}(t,t',t'', t'''){\mathcal{O}_{3,l}}((\vec{x}_{\rm fl}(\vec x, t,t'),t'),(\vec{x}_{\rm fl}(\vec x, t,t''),t''),(\vec{x}_{\rm fl}(\vec x, t,t'''),t'''))_{ij}\right)\right) \ ,
\eea
with 
\bea \label{operators}
\mathcal{O}_{1,l} &&\in \left\{ r_{ij}, p_{ij},\delta^{K}_{ij}\delta, \delta^{K}_{ij}\theta,\delta^{K}_{ij} \frac{\partial^{2}}{\knl^{2}}\delta  \right\} \ , \\ \nonumber
\mathcal{O}_{2,l} &&\in \left\{ \delta^{K}_{ij}\delta^{2}, \delta^{K}_{ij}\theta^{2},\delta^{K}_{ij}\delta\theta, r_{ij}\delta, p_{ij}\delta, r_{ij}\theta, p_{ij}\theta, r_{ik}r_{kj}, p_{ik}p_{kj},  \right. \nonumber  \\ \nonumber
 &&\qquad\left. \frac{1}{2} ( r_{ik}p_{kj}+r_{jk}p_{ki}), \tr(r^{2})\delta^{K}_{ij}, \tr(p^{2})\delta^{K}_{ij}, \tr(rp)\delta^{K}_{ij}\right\} \ , \\ \nonumber 
 \mathcal{O}_{3,l} &&\in \left\{ r_{ik}r_{km}r_{mj}, \tr(r^{2})r_{ij}, \tr(r^{3})\delta^{K}_{ij}, r_{ik}r_{kj}\delta, \tr(r^{2})\delta\delta^{K}_{ij}, r_{ij}\delta^{2},\delta^{3}\delta^{K}_{ij}\right\} \ , 
\eea
where, again, $c_{i,\mathcal{O}}(t,t_1,\ldots,t_i)$ are unknowable time kernels with support of order one Hubble time {and time scale{, $\sim t-t_i$,} of order Hubble as well}. {At third order, there are more combinations of $r,p,\delta$ and $\theta$, but we can immediately remove some of the degeneracies since $\theta^{(1)} = \delta^{(1)}$ and $r^{(1)} = p^{(1)}$, and similarly for the higher derivative terms, that can be evaluated to linear order as they are suppressed by two powers of $k/\knl$.}
Each operator $\mathcal{O}$ appearing in~\eqref{eq:bias_exp} must then be further expanded, potentially up to second order along its fluid element, in the following way,
\begin{align}
    \mathcal{O}(\vec{x}_{\rm fl}(\vx, t, t'), t') &\approx \left[\mathcal{O}(\vx, t') + \partial_i \mathcal{O}(\vx, t')\int^{t'}_t\frac{dt_1}{a(t_1)}v^i(\vx, t_1)\right. \nonumber\\
    &+ \frac{\partial_i\partial_j}{2}\mathcal{O}(\vx, t')\int^{t'}_t\frac{dt_1}{a(t_1)}v^i(\vx, t_1)\int^{t'}_t\frac{dt_2}{a(t_2)}v^j(\vx, t_2) \nonumber \\
    &\left.+\partial_i\mathcal{O}(\vx, t')\int^{t'}_t\frac{dt_1}{a(t_1)}\partial_jv^i(\vx, t_1)\int^{t_1}_t\frac{dt_2}{a(t_2)}v^j(\vx, t_2)\right]\  ,\label{eq:fluid_exp}
\end{align}
where, using the expansion for $v^i$ in \eqn{vastheta}, each term in the expansion is evaluated {up to} third order.\footnote{{The time integrals in~\eqref{eq:fluid_exp} can be computed by noting that 
\be
\int^{t'}_tdt_1 \frac{\dot{D}(t_1)}{D(t_1)}\frac{D(t_1)^n}{D(t)^n}=\frac{1}{n}\left(\frac{D(t')^n}{D(t)^n}-1 \right) \ .
\ee }} 

In this way, we can generate all of the possible functions in the expansion of the stress tensor (see \appref{kernelsderiv} for systematic details on this construction).  Doing this, and then finding linear relations among the resulting functions, we find the following  basis of 17 independent operators, 
\begin{align} \label{Cbasis}
&\left\{\mathbb{C}^{ij(3)}_{r,1},\mathbb{C}^{ij(3)}_{r,2},\mathbb{C}^{ij(3)}_{r,3},\mathbb{C}^{ij(3)}_{r^2,1},\mathbb{C}^{ij(3)}_{r^2,2},\mathbb{C}^{ij(3)}_{r^3,1},\mathbb{C}^{ij(3)}_{(r^{2})r,1},\mathbb{C}^{ij(3)}_{\delta ,1},\mathbb{C}^{ij(3)}_{\delta ,2},\mathbb{C}^{ij(3)}_{\delta ,3},\right.\\ \nonumber
&\left. \hspace{.5in} \mathbb{C}^{ij(3)}_{\delta  r,1},\mathbb{C}^{ij(3)}_{\delta  r,2},\mathbb{C}^{ij(3)}_{\delta  r^2,1},\mathbb{C}^{ij(3)}_{\delta ^2,1},\mathbb{C}^{ij(3)}_{\delta ^2,2},\mathbb{C}^{ij(3)}_{\delta ^2 r,1},\mathbb{C}^{ij(3)}_{\delta ^3,1}\right\} \ , 
\end{align}
along with the higher-derivative kernel $\mathbb{C}^{ij(1)}_{\partial^{2}\delta,1}$. These functions, along with the explicit matrix structures for the labels $\mathcal{O}_m$ in $\mathbb{C}^{ij(n)}_{\mathcal{O}_m, \alpha}$ are explicitly given in \appref{app:counterterm expressions}, and are defined in terms of the multiple-time-insertion procedure in \appref{kernelsderiv}.  It turns out that this basis is equivalent to the single-time-insertion basis, so we have used single-time-insertion functions in \eqn{Cbasis} for simplicity.  We do not necessarily expect these two bases to be equivalent at all orders, however.

{The {kernel $\mathbb{C}^{ij(1)}_{\partial^{2}\delta,1}$} is a higher derivative term we include that will renormalize the double-UV integrand in $\mathcal{P}_{15}$. Since the stress tensor appears in the equation preceded by two spatial derivatives, up to two-loop order we need to include only linear terms with two spatial derivatives, or quadratic or cubic terms but with no derivatives. Similarly, the stochastic counterterms are also negligible, contributing as $(k/\knl)^4$.}

{We would like to also point out that if one performs the local in time expansion, i.e. setting
\be
c_{n , \mathcal{O}_l} ( t , t_1 , \dots , t_n) \propto \frac{\delta_D ( t - t_1)}{H(t_1)} \cdots \frac{\delta_D ( t - t_n)}{H(t_n)}  \ , 
\ee
 in Eq~\eqref{eq:bias_exp} and finds the minimally independent basis, one would find a set of 16 independent operators instead of 17~\cite{Ansari:2024efj}, so the non-locality in time of the EFTofLSS appears already at this order in the stress tensor.}

{By performing the described manipulations,} we can now provide the full expression of the first, second and third order stress tensor. {Redefining the $c_{\mathcal{O}_m, \alpha}$ by extracting useful factors of $D$, we can write}\footnote{{The assumed time dependence for the stress tensor holds only for the counterterm part, that is the part that absorbs the UV sensitive part of the loop. The finite part can have a different time dependence and it is treated separately.}}  
{\begin{align}\label{eq:stresstensor_summary_eq}
    \tau^{ij} = \frac{\Omega_m \mathcal{H}^2\bar{\rho}}{\knl^2}\left(D(a)^3{ \tilde \tau}^{ij}_{{\rm ct}, (1)} + D(a)^4{\tilde \tau}^{ij}_{{\rm ct}, (2)} + D(a)^5 \left( {\tilde \tau}^{ij}_{{\rm ct}, (3)} + {\tilde \tau}^{ij}_{{\partial^2 \rm ct}, (1)}  \right) \right)\ ,
\end{align}}
{we have} 
\begin{align}
\begin{split} \label{tauijexpansion}
    {\tilde \tau}^{ij}_{{\rm ct}, (1)}(\vx{,t}) &= c_{r,1}\mathbb{C}^{ij(1)}_{r,1} + c_{\delta,1}\mathbb{C}^{ij(1)}_{\delta,1}\  ,\\
    {\tilde \tau}^{ij}_{{\rm ct}, (2)}(\vx{,t}) &= c_{r,1}\mathbb{C}^{ij(2)}_{r,1} + c_{r,2}\mathbb{C}^{ij(2)}_{r,2} + c_{\delta,1}\mathbb{C}^{ij(2)}_{\delta,1} + c_{\delta, 2 }\mathbb{C}^{ij(2)}_{\delta,2} + c_{r^2,1}\mathbb{C}^{ij(2)}_{r^2,1} + c_{r\delta,1}\mathbb{C}^{ij(2)}_{r\delta,1}  \\
    &+ c_{\delta^2,1}\mathbb{C}^{ij(2)}_{\delta^2,1} \  ,\\
    {\tilde \tau}^{ij}_{{\rm ct}, (3)}(\vx{,t}) &= c_{r,1}\mathbb{C}^{ij(3)}_{r,1} + c_{r,2}\mathbb{C}^{ij(3)}_{r,2} + c_{r,3}\mathbb{C}^{ij(3)}_{r,3} + c_{\delta,1}\mathbb{C}^{ij(3)}_{\delta,1} + c_{\delta,2}\mathbb{C}^{ij(3)}_{\delta,2} + c_{\delta,3}\mathbb{C}^{ij(3)}_{\delta,3}    \\
    &+ c_{r^2,1}\mathbb{C}^{ij(3)}_{r^2,1} + c_{r^2,2}\mathbb{C}^{ij(3)}_{r^2,2} + c_{r\delta,1}\mathbb{C}^{ij(3)}_{r\delta,1} + c_{r\delta,2}\mathbb{C}^{ij(3)}_{r\delta,2} + c_{\delta^2,1}\mathbb{C}^{ij(3)}_{\delta^2,1}  \\
    &+ c_{\delta^2,2}\mathbb{C}^{ij(3)}_{\delta^2,2} + c_{r^3,1}\mathbb{C}^{ij(3)}_{r^3,1} + c_{r^2\delta,1}\mathbb{C}^{ij(3)}_{r^2\delta,1} + c_{r\delta^2,1}\mathbb{C}^{ij(3)}_{r\delta^2,1} + c_{\delta^3,1}\mathbb{C}^{ij(3)}_{\delta^3,1}  \\
    &+c_{(r^{2})r}\mathbb{C}^{ij(3)}_{(r^{2})r,1}  \ , \\
\tilde \tau^{ij}_{\partial^2 \rm ct, (1)} ( \xvec , t)  & = c_{\partial^2 \delta,1}\mathbb{C}^{ij(1)}_{\partial^2\delta,1} \ , 
\end{split}
\end{align}
{where the functions $\mathbb{C}^{ij(n)}_{\mathcal{O}_m, \alpha}$ are defined as:
\begin{align}
\begin{split}
    \mathbb{C}^{ij(1)}_{r,1} &= \frac{\partial_i\partial_j\tilde{\delta}^{(1)}}{\partial^2} \ , \quad \mathbb{C}^{ij(1)}_{\delta,1} = \delta^K_{ij}\tilde{\delta}^{(1)} \ , \quad \mathbb{C}^{ij(2)}_{r,1} = \frac{\partial_i\partial_j\partial_k\td^{(1)}}{\partial^2}\frac{\partial_k\td^{(1)}}{\partial^2} \ , \quad  \mathbb{C}^{ij(2)}_{\delta,1} = \delta^K_{ij}\partial_k\td^{(1)}\frac{\partial_k\td^{(1)}}{\partial^2}\ , \\
    \mathbb{C}^{ij(2)}_{r,2} &= \frac{\partial_i\partial_j\td^{(2)}}{\partial^2} - \frac{\partial_i\partial_j\partial_k\td^{(1)}}{\partial^2}\frac{\partial_k\td^{(1)}}{\partial^2} \ , \quad \mathbb{C}^{ij(2)}_{\delta,2} = \delta^K_{ij}\left[\td^{(2)} - \partial_k\td^{(1)}\frac{\partial_k\td^{(1)}}{\partial^2}\right] \ , \dots \ . 
    \end{split}
\end{align}
The complete list of the $\mathbb{C}^{ij(n)}_{\mathcal{O}_m, \alpha}$ up to $n = 3$ is given in App.~\ref{app:counterterm expressions} and the explicit definitions of the $c_{\mathcal{O}_m , \alpha}$ coefficients in terms of the original time kernels are given in App.~\ref{kernelsderiv}.

Following \cite{DAmico:2022ukl}, using the above, we solve for the first, second, and third order expressions for the density field $\delta$ as induced by the counterterms, that we call `responses' and denote by $\delta_{\rm ct}$, and, analogously, the first and second responses for the momentum field $\pi^i_{\rm ct}$.  The time dependences needed to cancel the UV divergences are
\begin{align}
\begin{split} \label{cttimedep}
& \delta^{(n)}_{\rm ct} ( \kvec , a )= D(a)^{n+2}  \tilde \delta^{(n)}_{\rm ct} ( \kvec )  \ , \quad  \pi^i_{\text{ct},(n)} ( \kvec , a ) = - a H  \bar \rho   f  D(a)^{n+2}  \tilde \pi^i_{\text{ct},(n)} ( \kvec )  \ ,   \\
& \delta^{(1)}_{\partial^2 \rm ct } (\kvec ,  a ) = D(a)^5 \tilde \delta^{(1)}_{\partial^2 \rm ct } ( \kvec )  \ , 
\end{split}
\end{align}
which also defines the time-independent tilde fields.  We perform the same scaling as $\pi^i_{\text{ct},(n)}$ in \eqn{cttimedep} to obtain $\pi^i_{V,\text{ct}}$ and $\pi_{S, \text{ct}}$.  Using the rescalings in \eqn{cttimedep}, \eqn{eq:stresstensor_summary_eq}, and \secref{solssec}, we have the following solutions in the EdS approximation
\begin{align}
\begin{split}
\tilde \delta_{\rm ct}^{(n)} & = \frac{2}{(n +1)(7+2n)} \tilde{S}^{(n)}_\delta( \kvec) \big|_{\tau, \text{L.D}}  \ , \\ 
\tilde \pi_{S,\text{ct}}^{(n)} & = (n +2) \tilde \delta_{\rm ct}^{(n)}  \ , \\
 \tilde \pi^{i}_{V,\text{ct},(n)}(\kvec) & = \frac{-2}{3 + 2(n+1)} \tilde{S}^{i}_{\pi, (n)}(\kvec) \big|_{\tau, \text{L.D}} \ ,
\end{split}
\end{align}
where the source terms in position space are now given by 
\begin{align}
    \tilde{S}^{(n)}_\delta \big|_{\tau, \text{L.D}} &= \left(\partial_i\partial_j\left(\frac{3}{2}\frac{\partial_i\tilde{\delta}}{\partial^2}\frac{\partial_j\tilde{\delta}}{\partial^2} - \frac{3}{4}\delta_{ij} \frac{\partial_k\tilde{\delta}}{\partial^2}\frac{\partial_k\tilde{\delta}}{\partial^2} + \frac{\tilde{\pi}^i\tilde{\pi}^j}{1+\tilde{\delta}}  + \tilde \tau^{ij}  \right)\right)^{(n)}_{\tau , \text{L.D}} \label{eq:d_responsect}\ ,\\
    \tilde{S}^{i,(n)}_{\pi} \big|_{\tau, \text{L.D}}  &= - \left(\epsilon^{ijk}\partial_j\partial_l\left(\frac{3}{2}\frac{\partial_k\tilde{\delta}}{\partial^2}\frac{\partial_l\tilde{\delta}}{\partial^2} + \frac{\tilde{\pi}^k\tilde{\pi}^l}{1+\tilde{\delta}}+ \tilde \tau^{kl} \right)\right)^{(n)}_{\tau, \text{L.D}} \label{eq:pi_responsect}\ ,
\end{align}
and the notation $(T[\tilde \delta , \tilde \pi^i])^{(n)}_{\tau, \text{L.D}}$ is similar to the one defined under \eqn{eq:pi_response}, but now we take all of the terms that are sourced by $\tau^{ij}$ at leading order in derivatives (i.e. with $c_{\partial^2 \delta , 1} = 0$), including plugging lower order counterterm solutions back into $\tilde \delta$ and $\tilde \pi^i$.  The higher derivative term is given explicitly in \eqn{delta3ct} below.

Explicitly, we obtain
\begin{align}\label{eq:countersol}
    \delta^{(1)}_{\rm ct}(a) &= \frac{D(a)^3}{9k^2_{\rm NL}}\partial_i\partial_j \tilde{\tau}^{ij}_{\rm ct, (1)}\, ,\\
    \tilde{\pi}^i_{\rm ct,(1)} &= \frac{1}{3k^2_{\rm NL}}\frac{\partial_i\partial_j\partial_k}{\partial^2}\tilde{\tau}^{jk}_{\rm ct,(1)}\  ,\\
    \delta^{(2)}_{\rm ct}(a) &= \frac{2 D(a)^4\partial_i \partial_j}{33 \knl^2} \left[ \tilde \tau_{\text{ct},(2)}^{ij} + \frac{\partial_i \tilde \delta^{(1)}}{\partial^2} \partial_k \tilde \tau^{jk}_{\text{ct},(1)} - \frac{1}{6} \delta^K_{ij} \frac{\partial_k \tilde \delta^{(1)}}{\partial^2} \partial_l \tilde \tau^{kl}_{\text{ct},(1)}  \right] \ , \\
\tilde \pi^i_{\text{ct},(2)} & = \frac{2}{9 \knl^2} \Bigg[  \partial_j \tilde \tau^{ij}_{\text{ct},(2)} + \frac{1}{11} \frac{\partial_i \partial_j \partial_k}{\partial^2} \left(   \tilde \tau^{jk}_{\text{ct}, (2)} + \frac{\partial_j \tilde \delta^{(1)}}{\partial^2}  \partial_l \tilde \tau^{kl}_{\text{ct}, (1)}  \right)   - \frac{2}{11} \partial_i \left(  \frac{\partial_j \tilde \delta^{(1)}}{\partial^2}\partial_k \tilde \tau^{jk}_{\text{ct},(1)}   \right)    \nonumber \\
& \hspace{.6in}  + \frac{1}{2} \partial_l \left(   \frac{\partial_i \tilde \delta^{(1)}}{\partial^2} \partial_m  \tilde \tau^{lm}_{\text{ct}, (1)}  +    \frac{\partial_l \tilde \delta^{(1)}}{\partial^2} \partial_j \tilde \tau^{ij}_{\text{ct},(1)}   \right)  \Bigg]  \ , \\
    \delta^{(3)}_{\rm ct}(a) &= \frac{D(a)^5\partial_i\partial_j}{26 }\Bigg[3\left(\frac{\partial_i\tilde{\delta}^{(2)}_{\rm ct}}{\partial^2}\frac{\partial_j\tilde{\delta}^{(1)}}{\partial^2}+\frac{\partial_i\tilde{\delta}^{(1)}_{\rm ct}}{\partial^2}\frac{\partial_j\tilde{\delta}^{(2)}}{\partial^2}\right) \nonumber \\
    -&\frac{3}{2}\delta_{ij}\left(\frac{\partial_k\tilde{\delta}^{(2)}_{\rm ct}}{\partial^2}\frac{\partial_k\tilde{\delta}^{(1)}}{\partial^2}+\frac{\partial_k\tilde{\delta}^{(1)}_{\rm ct}}{\partial^2}\frac{\partial_k\tilde{\delta}^{(2)}}{\partial^2}\right) + 2\left(\tilde{\pi}^i_{\rm ct, (2)}\tilde{\pi}^j_{(1)}+\tilde{\pi}^i_{\rm ct, (1)}\tilde{\pi}^j_{(2)}\right) \nonumber \\
    & - 2\tilde{\pi}^i_{\rm ct, (1)}\tilde{\pi}^j_{(1)}\tilde{\delta}^{(1)} - \tilde \pi^i_{(1)} \tilde \pi^j_{(1)} \tilde \delta^{(1)}_{\text{ct}} + \frac{1}{\knl^2} \tilde{\tau}^{ij}_{\rm ct,(3)}\Bigg]\  , \\
    \delta^{(1)}_{\rm \partial^2 ct}(a) & = \frac{D(a)^5}{26 \knl^2}  \partial_i \partial_j \tilde \tau^{ij}_{\partial^2 \rm ct, (1)}  \ .
\label{delta3ct}
\end{align}
These counterterm solutions up to second order were found in \cite{DAmico:2022ukl}.

 Let us make a brief comment about the velocity vorticity, $\omega^i$.  As discussed in \cite{Carrasco:2013mua}, vorticity is sourced starting from terms in the second-order stress tensor, through $\omega_{(2)}^i \sim  \epsilon^{ijk}\partial_j \partial_l \tau^{lk}_{(2)}$.\footnote{ Note that another possible second-order contribution $\omega_{(2)}^i \sim  \epsilon^{ijk}\partial_j(  \delta^{(1)} \partial_l \tau^{lk}_{(1)})$ is zero for response counterterms.}  In \cite{DAmico:2022ukl}, it was found that in order to renormalize the one-loop bispectrum, vorticity indeed needs to be generated.  Since in the equations of motion \cite{Carrasco:2013mua} for $\delta$ and $\theta$, vorticity enters multiplied by another field, the contribution to the solutions for $\delta$ and $\theta$ starts at third order.  Above, we have computed all of the counterterm contributions from a generic $\tau^{ij}$ to $\delta$ up to third order, and since we used $\pi^i$ directly to solve for these contributions (i.e. we did not assume $v^i$ is irrotational), this means that we have consistently included the effect of vorticity on the solution for $\delta$.  Explicitly, we find that the EFT coefficients $c_{r,1}$, $c_{\delta,1}$, $c_{r^2 , 1}$, and $c_{r \delta , 1}$ in $\tau^{ij}_{(2)}$ contribute to the vorticity at second order, which in turn contributes to $\delta$ at third order. 

Now, in our expansion of the stress tensor \eqn{tauijexpansion}, we assumed that $v^i$ was irrotational, so let us explain why this is consistent.   First, $\omega^i$ does not contribute to $ \tilde \tau^{ij}_{\text{ct},(1)}$ because $\omega^i_{(1)} = 0$.  Then, as discussed above, we generate $\omega_{(2)}^i \sim  \epsilon^{ijk}\partial_j \partial_l \tilde \tau^{lk}_{\text{ct}, \text{irr.},(2)}$, where~$\tilde \tau^{lk}_{\text{ct}, \text{irr.}}$ is the stress tensor using only the irrotational~$v^i$.  Since this is a second-order counterterm, it renormalizes fourth-order fields (see \cite{DAmico:2022ukl}).  Now, if one tried to include this back in~$\tilde \tau^{ij}_{\text{ct},(2)}$ through a term like~$\partial_i v^j \delta$, it would be a countertem plugged into a counterterm, and in this case it would act to renormalize sixth-order fields.  Since we only go up to fifth-order fields in this work, this means that we can ignore vorticity in our expansion of~$\tau^{ij}$.

\section{{Perturbative expansion of the power spectrum}}
\label{sec:PerturbativeSeries}

We are now ready to set up the computation of a particular, perhaps the most important, correlation function: the power spectrum.
The equal-time power spectrum is the Fourier transform of the two-point correlation function of the density contrast, 
 \begin{align}
\begin{split}
\label{eq:PS-definition}
 {\Pcal}\left(k , a \right) & \equiv \int d^3 \vec r  \, e^{- i \vec k \cdot \vec r} \langle \delta\left(\vec x , a \right) \, 
\delta\left(\vec x + \vec r , a \right)
  \rangle \  ,
\end{split}
\end{align}
where the fields are taken at the equal time $a$.  Furthermore, this means
\be
\langle \delta ( \kvec , a ) \delta ( \kvec' , a ) \rangle = ( 2 \pi)^3 \delta_D ( \kvec + \kvec ') \mathcal{P} ( k , a ) \ , 
\ee
and we sometimes use the notation $\langle \cdot \rangle '$ to mean that we strip off the $( 2 \pi)^3 \delta_D ( \kvec + \kvec ') $ from the correlation function, i.e.
\be
\langle \delta ( \kvec, a  ) \delta ( \kvec' , a) \rangle' = \mathcal{P} ( k, a ) \ . 
\ee
For small density constrasts, the power spectrum can be meaningfully computed perturbatively. 
In the EFTofLSS, the perturbative expansion is given by  
\begin{eqnarray}
\label{eq:noCT+EFT}
\Pcal(k, a) = \Pcal_{{\rm no}-{\rm CT}}(k, a) + \Pcal_{\rm CT}(k, a) \  .
\end{eqnarray} 
The first term on the right-hand side, labelled ``${\rm no}-{\rm CT}$,'' comprises a loop expansion with diagrams which emerge from neglecting the effective stress tensor, the so-called SPT terms,   \begin{eqnarray}
\label{eq:PS-loopexpansion}
    \Pcal_{{\rm no}-{\rm CT}}(k, a) = \Plin(k, a) + \Pcal_{\rm 1-loop}(k, a) + \Pcal_{\rm 2-loop}(k, a)+\ldots   \ . 
\end{eqnarray} 
The second term, $\Pcal_{\rm CT}(k , a)$, of the right-hand side in Eq.~\eqref{eq:noCT+EFT} comprises effective field theory counterterms which account for integrating out non-linear effects at large wavenumbers, and which were derived in the former section for the two-loop power spectrum.     In this work, we focus on the EdS approximation, such that the time dependence of the above terms is given by
 \begin{align}
 \begin{split}
 \Plin(k, a) & = D(a)^2 \, \Plin(k) \ , \\
 \Pcal_{\rm 1-loop}(k, a) & = D(a)^4 \, \Pcal_{\rm 1-loop}(k ) \ , \\
 \Pcal_{\rm 2-loop}(k, a)& = D(a)^6 \,  \Pcal_{\rm 2-loop}(k ) \ , 
 \end{split}
 \end{align}
 and the expressions for $\Pcal_{\rm CT}(k , a)$ are given explicitly in \secref{sec:check_renormalization}.  
 
The leading term is the linear solution, $\Plin(k)$. We represent it diagrammatically as, 
\begin{eqnarray}
\Plin(k) \equiv  \eqs[0.15]{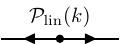} \ .
\end{eqnarray}
The linear power spectrum appears as a propagator in loop diagrams. Two diagrams contribute at one-loop,  
\begin{eqnarray}
\label{eq:P1loop-Diags}
\Pcal_{1-\rm loop}(k) =   \Pcal_{13}(k) + \Pcal_{22}(k)  \ , 
\end{eqnarray}
with 
\begin{eqnarray}
\Pcal_{13}(k) \equiv  \eqs[0.20]{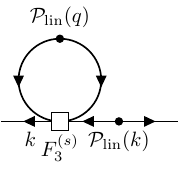}
\ , \quad      
\Pcal_{22}(k) \equiv  \eqs[0.20]{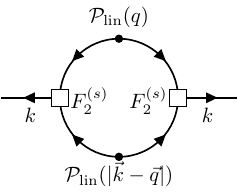} \ . 
\end{eqnarray}
In the above and later diagrams, we use the notation $F^{(s)}_{n}$ to remind the reader that we use the symmetric kernels defined in \eqn{symmfg}.  Here, lines with arrows pointing away from the vertex represent retarded Greens functions (due to the fact that we have factored out the time dependence, they represent just the unit number).

The two-loop correction consists of four diagrams, 
\begin{eqnarray}
\label{eq:P2loop-Diags}
\Pcal_{\rm 2-loop}(k) = \Pcal_{15}(k) + \Pcal_{42}(k) + \Pcal_{33}^{\rm (I)}(k) +\Pcal_{33}^{\rm (II)}(k) \ ,  \end{eqnarray}
with 
\begin{align}
\Pcal_{15}(k) \equiv \imineq{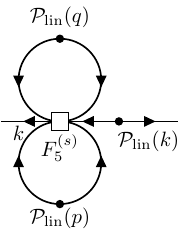}{15} \  ,
\quad & \quad 
\Pcal_{42}(k) \equiv \imineq{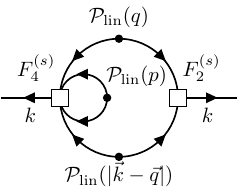}{14}  \  ,
\nonumber \\ 
\Pcal_{33}^{\rm (I)}(k) \equiv \imineq{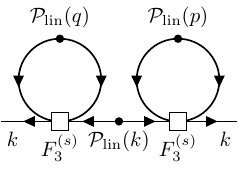}{12} \ ,
\quad & \quad
\Pcal_{33}^{\rm (II)}(k) \equiv \imineq{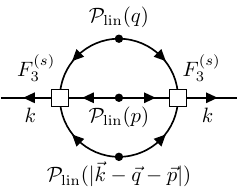}{14} \ . \label{p33feyn}
\end{align}

For every loop, an integration over the components of a three-dimensional  loop-momentum vector is required.  We will denote the integrand of a loop diagram $\Pcal_{ij}$ as $\pcal_{ij}$, so that 
\begin{eqnarray}
\label{eq:P1loop_integ}
 \Pcal_{ij}(k) &\equiv& \int_{\qvec} \pcal_{ij}(\kvec, \qvec) \, , \mbox{ at one loop, and, }  \\
 \label{eq:P2loop_integ}
 \Pcal_{ij}(k) &\equiv& \int_{\pvec , \qvec}   p_{ij}(\kvec, \pvec ,\qvec) \, ,\mbox{ at two loops.}  
\end{eqnarray}
The one-loop integrands read 
\begin{eqnarray}
\label{eq:P22}
\pcal_{22}(\kvec , \qvec) &=& 2 \, \Plin(q) \, \Plin (| \vk -\vq | ) \,   \left[F_2 (\vq,\vk-\vq)\right]^2 \ , 
\\
\label{eq:P13}
    \pcal_{13}(\kvec , \qvec) &=& 6 \Plin(k) \, \Plin(q) \,  F_3 (\vq,-\vq,\vk) \ . 
\end{eqnarray}
The two-loop integrands are
\begin{align}
\label{eq:P15}
& \pcal_{15}(\kvec, \pvec ,\qvec) =  30 \Plin(k) \, \Plin(q) \, \Plin(p) \, 
F_5 (\vk,\vkp,-\vkp,\vp,-\vp)  \ , \\
\label{eq:P42}
& \pcal_{42}(\kvec, \pvec ,\qvec) = 24 \, \Plin(q) \, \Plin(p) \, 
\Plin(|\vk-\vkp|) 
\nonumber \\
& \hspace{2.3cm} \qquad \times
F_2(\vkp,\vk-\vkp) F_4(-\vkp,\vkp-\vk, \vp,-\vp) \ , 
 \\
\label{eq:P33I}
& \pcal_{33}^{\rm (I)}(\kvec, \pvec ,\qvec) =  9 \,\Plin(k) \,\Plin(q) \, \Plin(p) \,  
 F_3 (\vk,\vq,-\vq) F_3 (-\vk,\vp,-\vp) \  , \\
 \label{eq:P33II}
& \pcal_{33}^{\rm (II)}(\kvec, \pvec ,\qvec) = 6 \, 
\Plin(q)\Plin(p)\Plin(|\vk-\vkp-\vp|)\,\nonumber \\
& \hspace{2.3cm} \qquad \times
  F_3 (\vkp,\vp,\vk-\vkp-\vp) F_3 (-\vkp,-\vp,-\vk+\vkp+\vp) \ . 
\end{align}

\section{Leading ultraviolet singularities and UV regularized diagrams}
\label{sec:UV}

The integrals of the one-loop $\Pcal_{\rm 1-loop}$ and two-loop $\Pcal_{\rm 2-loop}$ contributions in Eq.~\eqref{eq:noCT+EFT}, contain ultraviolet modes, for which a correct physical description requires incorporating non-linear/non-perturbative effects. However, to correctly encode the effect of these modes at long wavelengths, it suffices to include in the calculation the appropriate number of counterterms.
To quantify the fraction of the one- and two-loop integrals arising from the UV region, we define, at a given redshift
\begin{align}
    \label{eq:Delta_1loop}
    \Delta_{1\rm-loop}(k) &=
    \frac{\displaystyle \Pcal_{1\rm-loop}[\Plin](k)
    - \int_{|\vq| < q_{UV}} \frac{d^3\vq}{(2 \pi)^3} \;\,
    p_{1\rm-loop}[\Plin](\vq, \vk)}
    {\displaystyle \Pcal_{1\rm-loop}[\Plin](k)} \ , \\
    \label{eq:Delta_2loop}
    \Delta_{2\rm-loop}(k) &=
    \frac{\displaystyle \Pcal_{2\rm-loop}[\Plin](k)
    - \int_{|\vq| < q_{UV}} \frac{d^3\vq }{(2 \pi)^3}\int_{|\vp| < q_{UV}} \frac{d^3\vp}{( 2 \pi)^3} \;\,
    p_{2\rm-loop}[\Plin](\vq, \vp, \vk)}
    {\displaystyle \Pcal_{2\rm-loop}[\Plin](k)} \ ,
\end{align}
and $q_{UV}$ is an indicative  boundary of the ultraviolet region, set very conservatively to $q_{UV}=1.3$ $h/$Mpc. 
As indicated in square brackets, the loop integrals are functionals of the linear power spectrum at the given redshift, which is   the propagator entering in the corresponding Feynman diagrams and depends on the cosmological model. 
Here, we will consider a $\Lambda$CDM model  with parameters 
\begin{equation}
    \label{eq:Planck_cosmological_parameters}
    \begin{split}
        h &= 0.673 \ ,  \\
        \omega_b &= 0.02237 \ ,  \\
        \omega_{\rm cdm} &= 0.1203 \ ,  \\
        \ln ( 10^{10} A_s ) &= 3.044 \ ,  \\
        n_s&= 0.965 \ , \\
        \sum m_{\nu_i} &= 0.06\; {\rm eV} \ .
    \end{split}
\end{equation}
In the rest of this work, we refer to this linear power spectrum as the ``Planck" linear power spectrum, since it provides a favored fit to the CMB measurements by Planck~\cite{Planck:2018vyg}. We generate $\Plin(k)$ using \texttt{CLASS}~\cite{Blas:2011rf} for wavenumbers up to $k=10^4\,h/\mathrm{Mpc}$. Beyond this wavenumber, we extrapolate to higher values of $k$ assuming a $k^{-3}$ power-law behavior.

In \figref{fig:Delta_noCT} we show $\Delta^{\rm no-CT}_{\rm 1-loop}$ and $\Delta^{\rm no-CT}_{\rm 2-loop}$, where the superscript ``no$-$CT" highlights that we are only including diagrams of standard perturbation theory without any ultraviolet counterterms. 
We observe that contributions from the ultraviolet region are significant, especially at two loops. 
We note that the sharp peaks around $k \sim 0.082 \, \um$ at one loop and $k\sim 0.48 \, \um$ at two-loop are due to the loop corrections entering the denominators of Eqs.~\eqref{eq:Delta_1loop} and \eqref{eq:Delta_2loop} approaching zero. 
\begin{figure}[t!]
    \centering
    \begin{minipage}[c][5in][c]{0.55\textwidth} 
        \centering
        \includegraphics[]{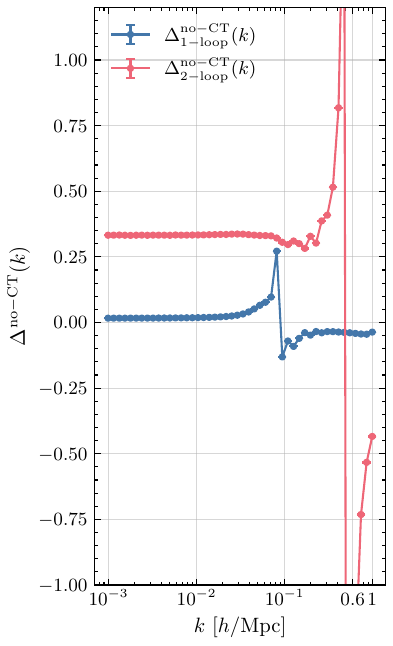}
        \caption{ \small The fractions $\Delta^{\rm no-CT}_{\rm 1-loop}$ (blue) and $\Delta^{\rm no-CT}_{\rm 2-loop}$ (red), computed using no$-$CT (often called SPT) kernels and the Planck linear power spectrum up to $k=10^4$~$h$/Mpc. The UV contribution to the loop corrections is significant, especially at two-loops. The sharp peaks at $k \sim 0.082 \, \um$ and $k\sim 0.48 \, \um$ are due to the loop correction vanishing. }
        \label{fig:Delta_noCT}
    \end{minipage}
    \hfill
    \begin{minipage}[c][7in][c]{0.4\textwidth}
        \centering
        \subfloat[\label{fig:Delta_1L_noCT_nu}]{%
            \includegraphics[width=\textwidth]{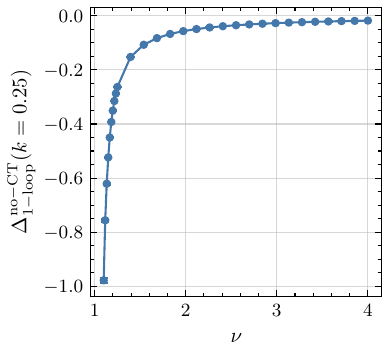}%
        }\\[5pt]
        \subfloat[\label{fig:Delta_2L_noCT_nu}]{%
            \includegraphics[width=\textwidth]{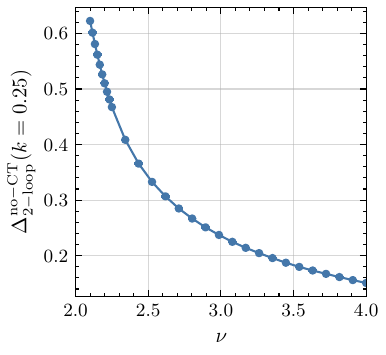}%
        }
        \caption{\small $\Delta^{{\rm no}-{\rm CT}}_{\rm 1-loop}(\nu)$ (\ref{fig:Delta_1L_noCT_nu}) and $\Delta^{{\rm no}-{\rm CT}}_{\rm 2-loop}(\nu)$ (\ref{fig:Delta_2L_noCT_nu}) at $k=0.25 \, \um$. We show that the infrared part of the loop corrections, without the counterterms, are sensitive to the tilt $\nu$ of the linear power spectrum in the UV.}
        \label{fig:Delta_noCT_nu}
    \end{minipage}
\end{figure}

The sensitivity of the loop corrections to the ultraviolet region depends on the asymptotic behavior of the linear power spectrum at large wavenumbers.            
To study this  behavior,  we  can modify the linear power spectrum so that, 
\begin{equation}
    \label{eq:Plin_nu}
    \Plin(k; \nu) = \Plin(k) \, \theta(q_{UV} - k) + \Plin(q_{UV}) \left(\frac{q_{UV}}{k}\right)^\nu \theta(k - q_{UV}) \ . 
\end{equation}
Below the ultraviolet cutoff $k<q_{UV}$ the linear power spectrum is the Planck power spectrum (which sometimes we refer to as `physical'), while above the cut-off we enforce a inverse power law behavior with a generic  power $\nu$ .   
Using this modified linear power spectrum as the kernel for the loop integrals,  the fractions defined in Eq.~\eqref{eq:Delta_1loop} and Eq.~\eqref{eq:Delta_2loop}, which indicate the relative size of the one-loop and two-loop corrections 
originating from ultraviolet regions,
can be studied as functions of the power $\nu$, 
\begin{equation}
\label{eq:Delta_nu}
    \Delta_{r\rm-loop}\left(\nu; k\right) = \Delta_{r\rm-loop} \left[ \Plin(k; \nu)  \right] (k)  \ . 
\end{equation} 
In \figref{fig:Delta_noCT_nu}  we show the fraction $\Delta^{\rm no-CT}_{\rm 1-loop}(\nu; k = 0.25)$ and $\Delta^{\rm no-CT}_{\rm 2-loop}(\nu; k = 0.25)$ respectively. We observe that the UV contribution grows rapidly as $\nu$ approaches $1$ at one loop and as $\nu$ approaches $2$ at two loops. 
This is anticipated by ultraviolet power-counting which we will detail in the next section. Indeed, it can be shown that the one-loop corrections to the power spectrum diverge for $\nu \leq 1$, while the two-loop corrections  diverge for $\nu \leq 2$.

As the ultraviolet contributions of the one- and two-loop corrections are meant to be ultimately replaced by physically determined EFTofLSS counterterms, it will be beneficial to rid the integrands of the corresponding loop diagrams 
from their ultraviolet sensitivity at the very start of our computation, 
before we carry out the integrations. This, on one hand,  will result in better stability for numerical integration, and on the other hand, the integrations will be dominated by physically legitimate and desired wavenumbers for which non-linearities are weak and perturbative.

Explicitly, we will separate the integrand of the loop diagrams with ultraviolet sensitivity into a simpler approximation  $p_{ij}^{\rm UV}$, which only captures the dominant (leading) ultraviolet regions of the full integrand $p_{ij}$, and a remainder $p_{ij}^{\rm UV-reg}$, 
\begin{eqnarray}
\label{eq:UV+regular}
    p_{ij} =  p_{ij}^{\rm UV} + p_{ij}^{\rm UV-reg}  \ . 
\end{eqnarray}
Removing the leading ultraviolet singularities will mathematically regularize  the divergences which are seen in the one- and two-loop corrections of Figs.~\ref{fig:Delta_1L_noCT_nu} and \ref{fig:Delta_2L_noCT_nu} for linear power spectra in Eq.~\eqref{eq:Plin_nu} with all $\nu \geq 1$ values. We will then observe  that the significance of the ultraviolet regions will be drastically reduced for the loop corrections of physical linear power-spectra too, which are the ones we are ultimately interested in. 
This is because the ultraviolet behavior of the linear power spectrum in viable cosmological models is typically more convergent than $\nu=1$. In fact, for realistic cosmologies all diagrams through two loops of no-CT solutions are mathematically integrable without  ultraviolet subtractions, having, asymptotically, a $\nu$ close to 3. The effect of our ultraviolet 
subtractions will reduce {these} finite ultraviolet  contributions to fractions of the total 
which are typically smaller than anticipated experimental uncertainties~\cite{Braganca:2023pcp}. 
{This procedure will therefore render the uncontrolled UV contribution from the loop diagrams negligible for observations.}

Operationally, we will build the $p_{ij}^{\rm UV}$ approximation from Taylor expansions of $p_{ij}$ in ultraviolet limits.  
We will then simplify the integrals over the ultraviolet approximations $\int p_{ij}^{\rm UV}$ analytically, using tensor reduction (cf. Ref.~\cite{Anastasiou:2023koq}). Tensor reduction will allow us to show that, as expected,  the ultraviolet approximations match the functional forms of the ultraviolet counterterm diagrams in $\Pcal_{\rm CT}$ of Eq.~\eqref{eq:noCT+EFT} which are furnished by the EFTofLSS. It will then be straightforward to absorb the  $\int p_{ij}^{\rm UV}$ integrals into the EFT counterterms of $\Pcal_{\rm CT}$. 
For the $p_{ij}^{\rm UV-reg}$ regular remainder, which will be insensitive to ultraviolet modes, we envisage integrating it numerically, after we address its infrared singularities. 

An important ingredient of ultraviolet power counting, necessary in the construction of $p_{ij}^{\rm UV}$, 
is the asymptotic ultraviolet behavior of 
the linear power spectrum $ \Plin(q) $, which becomes a function of loop momenta in the kernels of loop integrals. 
We will techically approach the ultraviolet regions by 
rescaling loop momenta with a $1/\delta$, where $\delta$ is a small parameter, 
sending $q \to q/\delta \to \infty$  as $\delta \to 0$. 
In this limit, as mentioned, we will assume a linear power spectrum scaling of the form, 
\begin{eqnarray}
    \label{eq:PlinUVscaling}
    \Plin\left(\frac{q}{\delta} \right) = \frac{C_{\rm lin.}}{q^\nu}   \delta^\nu  + {\cal O}(\delta ^{\nu +1})\ .  
\end{eqnarray}
The power-counting analysis of UV singularities and their subtraction will follow standard procedures (see, for example, Ref.~\cite{Herzog:2017bjx}), which are based on BPHZ renormalization.  In general, in this work, we will regulate any loop integrals that have a logarithmic or larger UV divergence for $\nu \geq 1$.
 If it is desired, in the future, 
 our analysis can be extended easily to remove subleading ultraviolet 
 singularities from the integrals, corresponding to divergencies appearing for $\nu<1$, following analogous steps.  
 
 \subsection{UV-regulated one-loop diagrams\label{sec:UV_1loop}}
 
In \tabref{tab:UV_1Loops} we display the ultraviolet regions which are singular in the one-loop diagrams of standard perturbation theory, for linear power-spectra with asymptotic ultraviolet behavior corresponding to  $\nu=1$ in Eq.~\eqref{eq:PlinUVscaling}. 
\begin{table}[ht]
\centering
    \begin{tabular}{|r|c|}
    \hline
               One-loop diagram       & UV singular region\\
              \hline
       $\Pcal_{13}$ & $\{\vq \to \infty \}$  \\ \hdashline
         $\Pcal_{22}$ &  --- \\
         \hline
    \end{tabular}
    \caption{\small Singular ultraviolet regions for the one-loop diagrams of standard perturbation theory in $\Pcal_{\rm 1-loop}$, assuming $\Plin(q) \sim 1/q$ in the ultraviolet.}
    \label{tab:UV_1Loops}
\end{table}
We find only a leading ultraviolet singularity in the diagram $\Pcal_{13}$. As a pedagogical example, we will describe in some detail the steps that we follow in order to separate the ultraviolet singular and regular part for this one-loop diagram. 

In the $\pcal_{13}$ integrand of Eq.~\eqref{eq:P13}, we encounter the vertex function 
\begin{eqnarray}
    &&F_3 (\vec{q},-\vec{q},\vec{k}) =\frac{1}{
   126 \, k^2 \, (q^2)^2 \, |\vec k - \vec q |^2  |\vec k + \vec q|^2 
    } \times 
 \Bigg\{
    28 (\vk\cdot \vq)^4 q^2 
    + 76 (\vk\cdot \vq)^4 k^2
    \nonumber  \\ 
&&   \hspace{2cm}
   -59 (\vk\cdot \vq)^2 (q^2)^2 k^2 
    -44 (\vk\cdot \vq)^2 q^2 (k^2)^2
    -21 (\vk\cdot \vq)^2  (k^2)^3
\nonumber  \\ 
&&   \hspace{2.5cm}
    +10 (k^2)^2 (q^2)^3 +10 (k^2)^3 (q^2)^2 
    \Bigg\}
   \ . 
\end{eqnarray}
For large loop-momentum $\vq \to \infty$, we can expand 
\begin{eqnarray} \label{tq0f3}
F_3 \left(\frac{\vec{q}}{\delta},-\frac{\vec{q}}{\delta},\vec{k}\right)  
= \delta^2 \, 
\Tcal^{(0)}_{q \to \infty} \left[ 
F_3 (\vec{q},-\vec{q},\vec{k}) \right] 
+{\cal O}(\delta^3) \ ,
\end{eqnarray}
where the coefficient of the leading term in the $\delta$ expansion reads
\begin{equation} \label{f3uvlimit}
\Tcal^{(0)}_{q \to \infty} \left[F_3 (\vec{q},-\vec{q},\vec{k}) \right] 
    = 
    \frac{1}{126 \, k^2}
    \frac{28 (\vk\cdot \vq)^4 - 59 k^2 (\vk \cdot \vq)^2 
   q^2 + 10 (k^2)^2 (q^2)^2}{(q^2)^3} \  . 
\end{equation}
In the  UV, the integrand of the  $\Pcal_{13}$ diagram combined with the integration measure scale as
\begin{eqnarray}
\label{eq:p13UVlimit}
    d^3  q \; \pcal_{13} &=& 
    d^3  q \; 6 \Plin(k) \, \Plin(q) \,  F_3 (\vq,-\vq,\vk) 
    \nonumber \\ 
    && 
    \to 
    \left( d^3  q 
    \, 6 \Plin(k) \, \frac{C_{\rm lin.}}{q^\nu}
    \,
    \Tcal^{(0)}_{q \to \infty} \left[ F_3 (\vec{q},-\vec{q},\vec{k}) \right]
    \right)
    \delta^{\nu+2-3} \ . 
\end{eqnarray}
Therefore, the degree of ultraviolet divergence in the $\vq \to \infty$ limit is 
\begin{equation}
\omega_{{\Pcal_{13}}} (\{\vq \to \infty\}; \nu) = 1-\nu \ . 
\end{equation}
For linear power-spectra with $\nu= 1$, the integral develops a logarithmic ultraviolet singularity.  Expanding \eqn{tq0f3} to the next order in $\delta$ would give an integral over $F_3 \sim k^4 / q^4$, which is much more convergent in the UV, so we do not need to regulate it.

 We will regulate the UV behavior by subtracting a suitable approximation of the integrand.  Our construction of the regulated integrand reads 
\begin{eqnarray}
\label{eq:p13-subtract}
    \pcal_{13}^{\rm UV-reg.}(\kvec, \qvec) = \pcal_{13}(\kvec, \qvec) 
    - \Rcal_{q \to \infty} \pcal_{13}(\kvec, \qvec) \ , 
\end{eqnarray}
with  
\begin{eqnarray}
\label{eq:P13_ct}
\Rcal_{q \to \infty } \, \pcal_{13}(\kvec, \qvec) 
= 
 6 \, \Plin(k) \, \Plin(q) \, 
    \Tcal^{(0)}_{q \to \infty} \left[ F_3 (\vec{q},-\vec{q},\vec{k}) \right] f_{\rm screen}(q^2)  \ , 
\end{eqnarray}
and
\begin{equation} \label{fscreeneq}
    f_{\rm screen}(q^2) = \theta (q^2 - M^2) \  .
\end{equation}
where we set $M = 0.6$ $h/$Mpc. The factor $f_{\rm screen}(q^2)$ is introduced to screen possible infrared singularities and it does not alter the ultraviolet behavior.
One can easily verify that the right-hand side of Eq.~\eqref{eq:P13_ct} has the same ultraviolet limit as the full integrand $p_{13}(\vk, \vq)$, shown in Eq.~\eqref{eq:p13UVlimit}.

 In the ultraviolet approximation of Eq.~\eqref{eq:P13_ct} for \(p_{13}(\vk,\vq)\), it is a natural choice not to approximate the linear power spectrum function with its asymptotic ultraviolet behavior $\Plin(q) \to C_{\rm lin}/q^\nu$. 
This choice is motivated by several reasons. First, in this way the ultraviolet components of the integrals will naturally match the form of the EFTofLSS counterterms. Second, from physical considerations, we would like to subtract the high wavenumber contributions as early as non-linearities settle in. This interval is larger than the range of values for which the asymptotic behavior of Eq.~\eqref{eq:PlinUVscaling} applies. 
Third, it is very convenient and elegant that the ultraviolet subtraction of Eq.~\eqref{eq:p13-subtract} acts only on the vertex of the diagram. This is a feature which will lead to an elegant formulation of ultraviolet regular integrands through two loops.

Indeed, we can implement the subtraction of Eq.~\eqref{eq:p13-subtract} by only decomposing the $F_3 $ vertex  into an ultraviolet, 
\begin{equation}
\label{eq:F3_UV-sing}
    F_3^{ \rm UV}(\vq, -\vq, \vk) 
    \equiv \eqs[0.15]{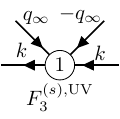}
    \equiv
\Tcal^{(0)}_{q \to \infty}[F_3(\vq, -\vq, \vk)]f_{\rm screen}(q^2) \ , 
\end{equation}
and a regular part 
\begin{equation}
\label{eq:F3_UV-reg}
    F_3^{ \rm UV-reg}(\vq, -\vq, \vk) 
    \equiv
\eqs[0.15]{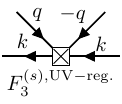}    
    \equiv F_3(\vq, -\vq, \vk) -
    F_3^{ \rm UV}(\vq, -\vq, \vk)\ .
\end{equation} 
Diagrammatically, we write
\begin{equation}
\label{eq:F3_UV-reg-diag}
 F_3 (\vq, -\vq, \vk) \equiv 
\includegraphics[width = 0.15\textwidth, valign=c]{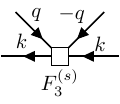}
=   \includegraphics[width = 0.15\textwidth, valign=c]{Figures/Diagrams/vertices/F3UV.pdf}
\, + \, 
\includegraphics[width = 0.15\textwidth, valign=c]{Figures/Diagrams/vertices/F3UVreg.pdf} \ . 
\end{equation} 

Then, the $\Pcal_{13}$ diagram is written as the sum of two integrals, 
\begin{eqnarray}
\label{eq:P13_reg+ct}
\includegraphics[width = 0.2\textwidth, valign=c]{Figures/Diagrams/1_loop/P13.pdf}
=    \includegraphics[width = 0.2\textwidth, valign=c]{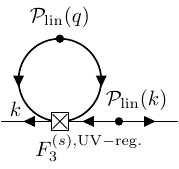}
\, + \, 
\includegraphics[width = 0.2\textwidth, valign=c]{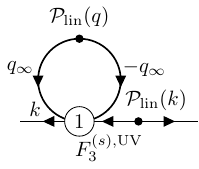} \ . 
\end{eqnarray}
The first integral in the right-hand side is ultraviolet finite and we can compute it with a direct numerical integration.  The second integral in the right-hand side contains the ultraviolet singularities.  One can easily see, for example using tensor reduction~\cite{Anastasiou:2023koq}, that this integral has a very simple form.  We find
\begin{eqnarray}\label{eq:P13UV}
 \includegraphics[width = 0.2\textwidth, valign=c]{Figures/Diagrams/1_loop/P13_UV.pdf}
 \, = \,  C_{13}(M)\, 
k^2 \,  \includegraphics[width = 0.15\textwidth, valign=c]{Figures/Diagrams/Plin.pdf} \ , 
\end{eqnarray}
with 
\begin{eqnarray}
\label{eq:P13_integrated_ct}
 C_{13}(M) =  -\frac{61}{315} \int_{\qvec} \frac{\Plin(q)}{q^2}f_{\rm screen}(q^2) \ .
\end{eqnarray}
As it is expected, the UV contribution is proportional to $k^2$ times the leading order power spectrum $\propto k^2 \Plin(k)$. 
The EFTofLSS formalism will add a counterterm of the same form. 

We can carry an analogous power-counting analysis of the ultraviolet behavior of the second one-loop diagram, $\Pcal_{22}$. We find that the ultraviolet degree of divergence is 
\begin{equation}
    \omega_{\Pcal_{22}} (\{\vq \to \infty \};\nu) = -1-2 \nu \ . 
\end{equation}
and the diagram  develops a singularity for $ \nu \leq -\frac 1 2$, outside the range of cosmological models that we consider here. 
This leading UV behavior is suppressed with respect to the $\Pcal_{13}$ diagram. As we are restricting our study to $\nu \geq 1$, we will consider this diagram regular and we will not introduce any ultraviolet subtractions for it.  Therefore, the $F_{2} $ vertex will not need a UV subtraction at one-loop.   Physically, this means that the UV contribution to $\Pcal_{22}(k)$, with $k$ in the IR, is much smaller than for $\Pcal_{13}(k)$.

In total, we can decompose the one-loop no-CT correction to the power spectrum 
in two parts, a part which is locally free of ultraviolet singularities and an ultraviolet ``counterterm'', 
\begin{equation}
\label{eq:P1-loop_UVreg+ct}
\begin{aligned}
    \Pcal_{\rm 1-loop}(k) &= C_{13}(M)\, k^2 \, 
    \eqs[0.15]{Figures/Diagrams/Plin.pdf}
    + \Pcal_{\rm 1-loop}^{\rm UV-reg.}(k)   \ ,
\end{aligned}
\end{equation}
where the first term is degenerate to an EFTofLSS counterterm and the second term 
is the sum of two one-loop diagrams which are free of ultraviolet singularities 
\begin{equation}
\label{eq:P1-loop_UVreg}
    \Pcal_{\rm 1-loop}^{\rm UV-reg.}(k) = \eqs[0.15]{Figures/Diagrams/1_loop/P13_UV_reg.pdf}
    \; + \; \eqs[0.15]{Figures/Diagrams/1_loop/P22.pdf} \  .  
\end{equation}

\subsection{Ultraviolet singularities at two loops} \label{uvsing2loopsec}

Power-counting analysis reveals several ultraviolet regions which yield leading singularities in the four diagrams of $\Pcal_{\rm 2-loop}$, at the next perturbative order. We summarize these leading singularities in \tabref{tab:UV_2Loops}.
\begin{table}[ht]
    \centering
    \begin{tabular}{|r|c|c|}
    \hline
                Two-loop diagram  & singular single UV regions & singular double UV region \\
              \hline
         $\Pcal_{15}$ & $\begin{array}{c} \{\vq \to \infty, \vp \sim {\rm fixed} \} \, \left[ 1- \nu \right] \\
         \{\vq \sim {\rm fixed}, \vp \to \infty \} \left[ 1- \nu \right] \end{array}$ & $\{ \vq,\vp \to \infty \} \, \left[ 2(2 - \nu) \right] $ \\ 
         \hdashline
         $\Pcal_{42}$ & $\{ \vq \sim {\rm fixed}, \vp \to \infty \} \, \, \left[ 1 - \nu \right] $ & --- \\ 
         \hdashline
         $\Pcal_{33}^{\rm (II)}$ & --- & ---  \\
         \hdashline
         $\Pcal_{33}^{\rm (I)}$ & 
         $\begin{array}{c} \{\vq \to \infty, \vp \sim {\rm fixed} \} \, \left[ 1 - \nu \right]  \\
         \{\vq \sim {\rm fixed}, \vp \to \infty \} \, \left[1 - \nu \right] \end{array}$
         & $\{\qvec , \pvec \to \infty \}\, \left[1 - \nu \right] $ \\
   \hline
    \end{tabular}
    \caption{ \small Singular ultraviolet regions for the two-loop diagrams of standard perturbation theory in $\Pcal_{\rm 2-loop}$, assuming $\Plin(q) \sim 1/q^\nu$ in the ultraviolet. In brackets, we include the degree of divergence of the integrals in the singular regions.}
    \label{tab:UV_2Loops}
\end{table}
The ``double UV region'' corresponds to the limit where both loop momenta are taken infinitely large.
To compute the degree of divergence, displayed in square brackets, we take the limit smoothly by scaling both loop-momenta 
$(\vp, \vq) \to (\vp/\delta, \vq/\delta) $ and computing the starting power of the $\delta \to 0^+$ expansion of the integrand, including the integration measure $d^3 q \, d^3 p \to d^3 q \, d^3 p / \delta^6$.   
``Single UV'' regions correspond to the limit in which one linear combination $c_p \, \vp + c_q \,  \vq$ of the loop momenta $\vp,\vq$ is taken infinitely large, while a second independent loop momentum is fixed.\footnote{{We check for the ultraviolet limit of linear combinations of internal momenta as they appear in the vertices of diagrams and in the arguments of the linear power spectrum functions (propagators) in the integrand: these are the only single UV limits  potentially giving rise to singularities.}}
We determine the degree of divergence with an analogous scaling and expansion procedure. 

For example, to compute the degree of divergence in a region $\{\vp+\vq \to \infty, \vq \sim {\rm fixed}\}$, we substitute  $\vp \to -\vq + \frac{\vp}{\delta}$ and expand the integrand and the integration measure, which scales as $d^3 p \,  d^3 p \to d^3 p \, d^3 p / \delta^3$, around $\delta \to 0^+$.  
Note that this specific region, $\{ \vp+\vq \to \infty , \vq \sim {\rm fixed}\}$, 
is part of our general study to identify all ultraviolet  singularities, 
although it does not yield a leading singularity. We would like to emphasize 
that for linear power spectra which fall less steeply than the assumed $\nu \geq 1$ power in the ultraviolet,  novel regions which do not appear in \tabref{tab:UV_2Loops}, such as   $\{ \vp+\vq \to \infty , \vq \sim {\rm fixed}\}$, also become singular and should not be neglected. 

We construct UV regulated diagrams $\pcal_{ij}^{\rm UV-reg.}$ at two loops by applying a sequence of subtractions, removing approximations of the two-loop diagram integrands $\pcal_{ij}$  in ultraviolet regions (cf.  Refs~\cite{Herzog:2017bjx,Anastasiou:2018rib,Anastasiou:2020sdt,Anastasiou:2022eym,Anastasiou:2024xvk}). In a first step, we subtract the singularities in single UV regions.  In a second step, we remove the double UV singularities of the remainder obtained after the subtractions of the first step. The final outcome of this sequence of subtractions is a UV regularized remainder,      
\begin{equation}
\label{eq:2loop_UVsub}
\begin{aligned}
    \pcal_{ij}^{\rm UV-reg.} &= \, \pcal_{ij} - \sum_{l \in \, {\rm single-UV}}\mathcal{R}_{ l} (\pcal_{ij})  \\
                             &\; - \mathcal{R}_{\rm double-UV} \l  \pcal_{ij} - \sum_{l \in \, {\rm single-UV}} \mathcal{R}_{ l } (\pcal_{ij}) \r  \  .
\end{aligned}
\end{equation}
In the above, we denote with $\Rcal_{\rm region}(f(\vp, \vq))$ an approximation of the loop momenta function $f(\vp, \vq)$ in a given singular region. 
 
As we described earlier, our UV approximations $\Rcal_{\rm region}(f(\vp, \vq))$ will not act on the $\Plin(q)$ kernel factors of the loop integrands, i.e., we take $\Rcal_{q \to \infty} \left(\Plin(q) \right) = \Plin(q)$. The $\Rcal_{\rm region}$ approximations will then be determined from taking ultraviolet limits only on the vertices in the diagrams. We will be able to implement the subtraction of Eq.~\eqref{eq:2loop_UVsub} by a simple decomposition of vertices into ultraviolet and regular parts, analogously to the decomposition of Eq.~\eqref{eq:F3_UV-reg-diag} for the $F_3 $ vertex which appeared already at one loop. 
Considering leading ultraviolet singularities at two loops and linear power spectra with $\nu=1$,  in addition to the decomposition of the $F_3 $ vertex,  we further need to decompose into regular and singular ultraviolet parts the $F_4$ and $F_5$ vertices. 

We decompose the $F_4$ vertex, appearing in diagram $\Pcal_{42}$ of Eq.~\eqref{eq:P42}, into an ultraviolet and a regular term,
\begin{eqnarray}
\label{eq:F4UV-reg.-Diag}
&& F_4(\vq - \vk, -\vq, \vp, - \vp) \equiv
 \includegraphics[width = 0.13\textwidth, valign=c]{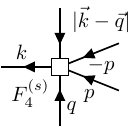}
= 
\includegraphics[width = 0.2\textwidth, valign=c]{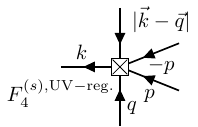}   
\, + \, 
\includegraphics[width = 0.15\textwidth, valign=c]{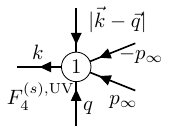}
\ . 
\nonumber \\
\end{eqnarray}
We define the  ultraviolet part as
\begin{eqnarray}
\eqs[0.15]{Figures/Diagrams/vertices/F4UV.pdf}
&\equiv&
\Tcal^{(0)}_{p\to\infty}[F_4(\vq - \vk, -\vq, \vp, - \vp)] f_{\rm screen}(p^2) \ . 
\end{eqnarray}
This expression corresponds to the limit of the loop momentum $\vp \to \infty$ while the $\vq$ loop momentum is kept fixed. We can identify the limit smoothly, from the expansion 
\begin{eqnarray}
\label{eq:F4UV}
  F_4\left(\vq - \vk, -\vq, \frac{\vp}{\delta}, - \frac{\vp}{\delta} \right)  =
  \Tcal^{(0)}_{p\to\infty}[F_4(\vq - \vk, -\vq, \vp, - \vp)] \, \delta^2 +{\cal O}\left(\delta^3 \right) \ .
\end{eqnarray}
Note that in the ultraviolet decomposition of the $F_4$ vertex in Eq.~\eqref{eq:F4UV-reg.-Diag} we need to isolate only one single ultraviolet region. The $F_2$ vertex factor in the integrand of the $\Pcal_{42}$ diagram, shown in Eq.~\eqref{eq:P42}, suppresses other ultraviolet regions. Specifically, the  single ultraviolet 
$\{\vq \to \infty, \vp \sim \, {\rm fixed}\}$ and double ultraviolet $\{\vp, \vq \to \infty\}$
regions do not yield singularities for $\nu \geq 1$, as their degrees of divergence, 
 $- 1 - 2\nu$ and $2 - 3\nu$ correspondingly, are negative. 

The ultraviolet decomposition of the $F_{5}$ vertex in the  diagram $\Pcal_{51}$ of Eq.~\eqref{eq:P15} takes the form 
\begin{eqnarray}
\label{eq:F5UV-reg.-Diag}
F_5(\vq, -\vq, \vp, -\vp, \vk) \equiv && 
\includegraphics[width = 0.13\textwidth, valign=c]{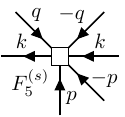}
=
\includegraphics[width = 0.2\textwidth, valign=c]{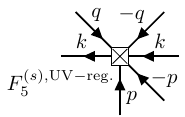}    
\nonumber + \\ 
&&
\hspace{-2cm}
+ 
\includegraphics[width = 0.15\textwidth, valign=c]{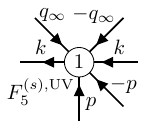} 
+ \includegraphics[width = 0.15\textwidth, valign=c]{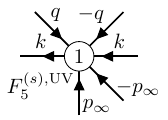}
+\includegraphics[width = 0.15\textwidth, valign=c]{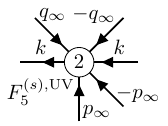} \ . 
\end{eqnarray}
The first graph of the right-hand side corresponds to the part of the $F_5$ vertex which is regular in all ultraviolet limits. 
The parts of the vertex which yield ultraviolet singularities in the $\Pcal_{51}$ integral are represented in the second line of Eq.~\eqref{eq:F5UV-reg.-Diag}. 

The first and second graphs of the second line originate from the $\sum_{l \in \, {\rm single-UV}}\mathcal{R}_{ l}$ term in the right-hand side of Eq.~\eqref{eq:2loop_UVsub} when it is applied to the diagram of  $p_{ij}=p_{15}$. 
The first graph of the second line contains the ultraviolet singularity in the region $\{\vq \to \infty, \vp \sim \, {\rm fixed}\}$ and we explicitly choose it to be
\begin{eqnarray}
\eqs[0.15]{Figures/Diagrams/vertices/F5UV1q.pdf} \equiv 
\Tcal^{(0)}_{q \to \infty}[F_5(\vq,-\vq, \vp, -\vp, \vk)] \,f_{\rm screen}(q^2)\  , 
\end{eqnarray}
where the leading term in this limit is obtained from the expansion, 
\begin{equation}
    F_5 \l \frac{\vq}{\delta}, -\frac{\vq}{\delta},\vp, -\vp, \vk \r = \delta^2 \Tcal^{(0)}_{q \to \infty}[F_5(\vq,-\vq, \vp, -\vp, \vk)] + \mathcal{O}(\delta^4) \ . 
\end{equation}
The second graph of the second line represents an analogous term for the symmetric ultraviolet region $\{\vp \to \infty, \vq \sim \, {\rm fixed}\}$.

The last graph of Eq.~\eqref{eq:F5UV-reg.-Diag} originates from the ${\rm {double}-UV}$ term in the right-hand side of Eq.~\eqref{eq:2loop_UVsub} and it contains the ultraviolet singularity in the region $\{\vq \to \infty, \vp \to \infty \}$. 
\begin{align}
    \label{eq:F5doubleUV}
    &\includegraphics[width = 0.15\textwidth, valign=c]{Figures/Diagrams/vertices/F5UV2.pdf}
     \equiv 
     \,  f_{\rm screen}(q^2) f_{\rm screen}(p^2) 
    \sum_{m = 0,2} \Tcal_{(q,p) \to \infty}^{(m)} \Bigg[
    F_5 (\vk, \vq,-\vq,\vp,-\vp) 
    \nonumber \\
    & -\Tcal^{(0)}_{q \to \infty} \left[F_5 ( \vk, \vq,-\vq,\vp,-\vp ) \right] 
    -\Tcal^{(0)}_{p \to \infty} \left[F_5 ( \vk, \vq,-\vq,\vp,-\vp ) \right]
    \Bigg] \ , 
    \end{align}
As the degree of divergence of this singularity in the $\Pcal_{51}$ integral is $4-2\nu$, it is necessary that 
Eq.~\eqref{eq:F5doubleUV} includes the first three terms in the Taylor expansion,   
\begin{align}
    \label{eq:F5_double_taylor}
    F_5 \l \vk, \frac{\vq}{\delta},-\frac{\vq}{\delta},\frac{\vp}{\delta},-\frac{\vp}{\delta} \r =& \sum_{m = 0}^{2} \delta^{2+m} \Tcal^{(m)}_{(q,p)\to \infty} [F_5 ( \vk, \vq,-\vq,\vp,-\vp ) ] + \mathcal{O}(\delta^5) \  , 
\end{align}
and not just the leading term. 
As expected from rotational and translation invariance, the $m=1$ power in the expansion around the double ultraviolet limit vanish.

\subsection{UV regulated two-loop corrections}

Let us now substitute the ultraviolet decompositions of Eqs.~\eqref{eq:F3_UV-reg-diag},~\eqref{eq:F4UV-reg.-Diag} and ~\eqref{eq:F5UV-reg.-Diag} for the $F_3$, $F_4$
and $F_5$ vertices into the ultraviolet singular two-loop diagrams ($\Pcal_{15}, \Pcal_{42}, \Pcal_{33}^{\rm (I)}$) 
of the no-CT solutions.  
The two-loop no-CT correction to the power spectrum, 
\begin{eqnarray} \label{p2loopdiagtot}
    \Pcal_{\rm 2-loop} &=& \eqs[0.15]{Figures/Diagrams/2_loop/P15.pdf}
    +\eqs[0.15]{Figures/Diagrams/2_loop/P42.pdf}
    +\eqs[0.15]{Figures/Diagrams/2_loop/P33I.pdf}
        +\eqs[0.15]{Figures/Diagrams/2_loop/P33II.pdf}\  , 
\end{eqnarray}
after these substitutions takes the form, 
\begin{eqnarray}
\label{eq:P2loop-UV-decomposition}
    \Pcal_{\rm 2-loop} &=& 
    \Pcal_{\rm 2-loop}^{\rm UV-reg.} + \Pcal_{\rm 2-loop}^{\rm UV-(1)} +\Pcal_{\rm 2-loop}^{\rm UV-(2)} \  ,
\end{eqnarray}
where the UV regulated contribution $\Pcal_{\rm 2-loop}^{\rm UV-reg.}$ is given in terms of the UV regulated vertices of \secref{uvsing2loopsec} as 
\begin{eqnarray}
\label{eq:P2loop-UV-reg}
    \Pcal_{\rm 2-loop}^{\rm UV-reg.} &=& \eqs[0.15]{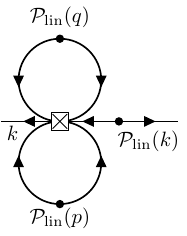}
    +\eqs[0.15]{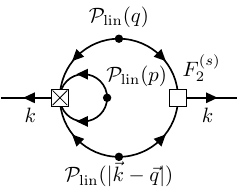}
    +\eqs[0.15]{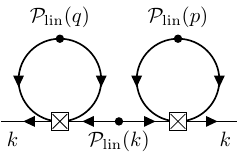}
        +\eqs[0.15]{Figures/Diagrams/2_loop/P33II.pdf}\  .
\end{eqnarray}
 Note that the last diagram in \eqn{eq:P2loop-UV-reg}, corresponding to $\mathcal{P}_{33}^{(\rm II)}$, is UV finite (see \tabref{tab:UV_2Loops}), and so is effectively already regulated.  All ultraviolet singularities are removed from the diagrams in \eqn{eq:P2loop-UV-reg}. We will then be able to numerically evaluate the integrals of $\Pcal_{\rm 2-loop}^{\rm UV-reg.}$. The remaining terms,  $\Pcal_{\rm 2-loop}^{\rm UV-(1)}$ and $\Pcal_{\rm 2-loop}^{\rm UV-(2)}$, on the right-hand side of \eqn{eq:P2loop-UV-decomposition}, correspond to single- and double-UV regions, respectively. In \secref{sec:check_renormalization}, we will demonstrate that all integrals in $\Pcal_{\rm 2-loop}^{\rm UV-(1)}$ and $\Pcal_{\rm 2-loop}^{\rm UV-(2)}$, after tensor reduction, become of the same form as EFT counterterms.

The two-loop contribution from singular single-UV regions reads 
\begin{eqnarray}
\Pcal_{\rm 2-loop}^{\rm UV-(1)} &=& 2 \, \eqs[0.15]{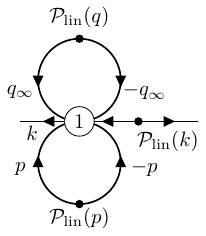}
+\eqs[0.15]{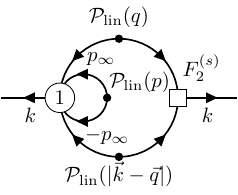}
   + 2 \,  \eqs[0.15]{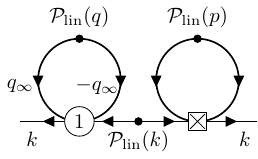} \ . 
\end{eqnarray}
After tensor reduction, we find that 
\begin{eqnarray}
\eqs[0.15]{Figures/Diagrams/2_loop/P33I_UV_q.pdf} = \frac{1}{2} k^2 C_{13}(M) \, \Pcal_{13}(k) 
\label{eq:P33_uv}\  ,     
\end{eqnarray}
\begin{align}
   & \eqs[0.15]{Figures/Diagrams/2_loop/P42_UV.pdf} = 
 \l \int_{\pvec}  \frac{\Plin(p)}{p^2}f_{\rm screen}(p^2)  \r \int_{\qvec}  \frac{\Plin(q) \Plin(|\kvec - \qvec|) F_2(\vq, \vk - \vq)}{169785 q^2 |\vk - \vq|^2} \times \nonumber \\
    & \quad \times (48096(\vk\cdot\vq)^3 + (16892 k^2-48096 q^2) (\vk\cdot\vq)^2 + \nonumber \\
    & \; \quad + (-32879 (k^2)^2+35744 q^2 k^2) (\vk \cdot \vq)-25933 q^2 (k^2)^2 + 6176 (q^2)^2 k^2)
    \  .
    \label{eq:P42_uv}\nonumber \\
    &
\end{align}
\begin{align} \label{eq:P51_uv_single}
   & \eqs[0.15]{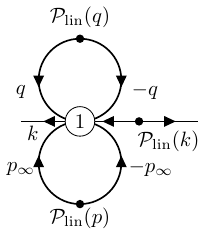}  = 30\Plin(k) \left( \int_{\qvec} \frac{\Plin(q)}{q^2} f_{\rm screen}(q^2) \right) \times \nonumber \\
   & \qquad \times \int_{\pvec}  \frac{\Plin(p)}{264864600 \, k^2 (p^2)^2 |\vk - \vp|^2 |\vk + \vp|^2} \times  \nonumber\\
    & \qquad \quad \times \biggl( -382844 (\vk \cdot \vp)^6 - 4 (k^2)^2 (p^2)^2 (k^2 + p^2) (79843 \, k^2 +32614 \, p^2) + \nonumber \\
    & \qquad \quad \quad \quad + (\vk \cdot \vp)^4 \l -1462930 (k^2)^2 + 6602262 \, k^2 p^2 + 173536 (p^2)^2 \r + \nonumber \\ 
    & \qquad \quad \quad \quad + k^2 (\vk \cdot \vp)^2 \Bigl( 427427 (k^2)^3 - 861688 (k^2)^2 p^2 - 1763891 (k^2) (p^2)^2 + \nonumber \\ 
    & \qquad \qquad \qquad \qquad \qquad \quad - 1832216 (p^2)^3 \Bigl) \biggl) \ . 
\end{align}
The two-loop contribution from double-UV regions reads 
\begin{eqnarray} \label{eq:P51_uv_double}
\Pcal_{\rm 2-loop}^{\rm UV-(2)} &=& \eqs[0.15]{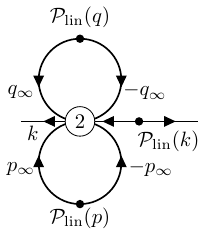}
   + \eqs[0.15]{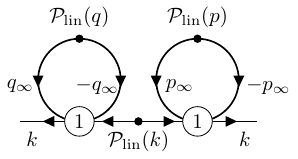} \ , 
\end{eqnarray}
where the first term on the right-hand side is defined in App.~\ref{sec:TRUVapproximations}, and the second term is
\begin{equation}\label{eq:P33_uv_double}
    \eqs[0.15]{Figures/Diagrams/2_loop/P33I_UV_qp.pdf} =
    \frac{1}{4} k^4 \Plin(k) (C_{13}(M))^2  \ . 
\end{equation}

\begin{figure}[t!]
    \centering
    \begin{minipage}[c][5in][c]{0.55\textwidth} 
        \centering
        \includegraphics[]{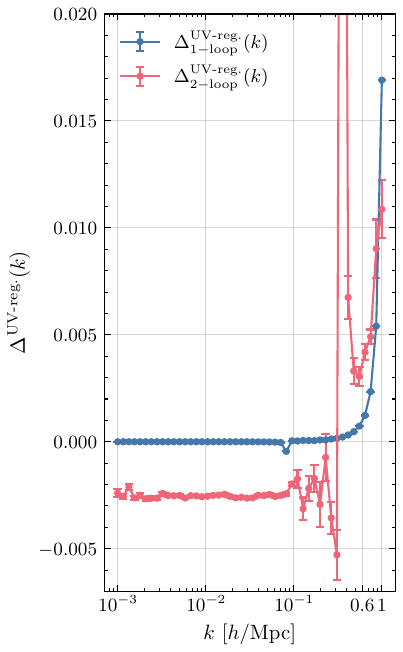}
        \caption{ \small The fractions $\Delta^{\rm UV-reg.}_{\rm 1-loop}$ (blue) and $\Delta^{\rm UV-reg.}_{\rm 2-loop}$ (red), computed using UV-reg. kernels and the Planck linear power spectrum up to $k=10^4$~$h$/Mpc. Compared to the no-CT 
        case, the UV contribution is significantly lower. The sharp peak for the two-loop at $k \sim 0.3 \, \um$ is due to the loop correction vanishing.}
        \label{fig:Delta_UVreg}
    \end{minipage}
    \hfill
    \begin{minipage}[c][8in][c]{0.42\textwidth}
        \centering
        \subfloat[\label{fig:Delta_1L_UVreg_nu}]{%
            \includegraphics[width=\textwidth]{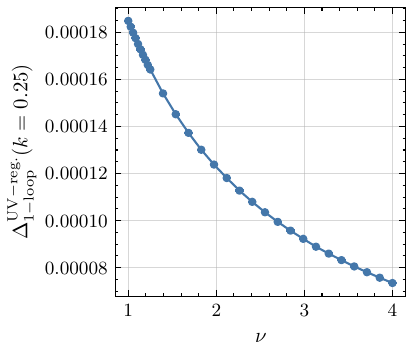}%
        }\\[5pt]
        \subfloat[\label{fig:Delta_2L_UVreg_nu}]{%
            \includegraphics[width=\textwidth]{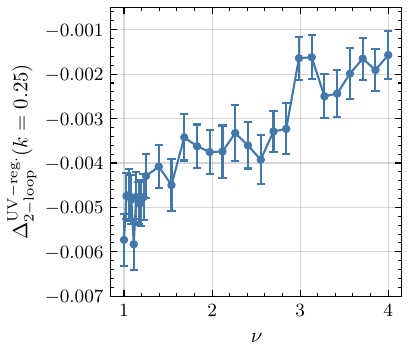}%
        }
        \caption{\small {$\Delta^{\rm UV-reg.}_{\rm 1-loop}(\nu)$~(\ref{fig:Delta_1L_UVreg_nu})~and} $\Delta^{\rm UV-reg.}_{\rm 2-loop}(\nu)$ (\ref{fig:Delta_2L_UVreg_nu}) at ${k=0.25~\um}$. We show that UV contribution to the loop corrections is now much less sensitive, compared to no-CT, to the tilt $\nu$ of the linear power spectrum. Furthermore, we are able to compute the loop integrands at $\nu=1$, where the no-CT expressions were singular.}
        \label{fig:Delta_UVreg_nu}
    \end{minipage}
    \vspace{-.5in}
\end{figure}

Having constructed the UV regulated integrands at one and two loops, we can now evaluate $\Delta^{\rm UV-reg.}_{r \rm{-loop}}$, as defined in Eqs.~\eqref{eq:Delta_1loop} and~\eqref{eq:Delta_2loop}, where the superscript ``UV-reg." denotes that we are using the UV-regulated kernels. \figref{fig:Delta_UVreg} shows that the UV dependence is now reduced to the permille level at two-loops, and is even smaller at one loop. As in the no-CT case, the sharp peak in the two-loop correction at $k \sim 0.35 \, \um$ occurs because the loop correction vanishes. A similar effect is present at one-loop around $k\sim 0.09~\um$, although it is less significant. We also observe a slight increase in both the one- and two-loop correction as $k$ approaches $1$, which is an artifact of our choice of the UV region starting at $q_{UV} = 1.3~\um$; since the integrals receive their dominant contributions in the region where loop momenta approach $k$, proximity to the UV boundary naturally leads to a more pronounced UV dependence at higher $k$. Furthermore, \figref{fig:Delta_UVreg_nu} presents the fraction $\Delta^{\rm UV-reg.}_{r \rm{-loop}}$ as a function of the tilt $\nu$, defined in~\eqref{eq:Delta_nu}, confirming that the UV dependence is removed up to a few permille for all $\nu \geq 1$.

\subsection{Check of renormalization}\label{sec:check_renormalization}

{So far, we have removed the UV contribution from the loop integrals, and accordingly defined ${ \Pcal_{n-{\rm loop}}^{\rm UV-reg.}}$, and $\Pcal_{n-{\rm loop}}^{{\rm UV-}(m)}$. In order for this procedure to be well defined, it needs to be that the $\Pcal_{n{\rm -loop}}^{{\rm UV-}(m)}$'s are degenerate with the contribution from the counterterms arising from the expansion of the effective stress tensor. This is what we are going to verify now.} 

{More explicitly, we will verify that proper choices of the counterterm coefficients cancel the contribution of UV modes to long wavelengths in the loop diagrams. We will perform this check in this subsection, by finding the value of the counterterms needed to cancel such a dependence. One could call these counterterms as the `bare' ones. Given that the time dependence of the loops in the EDS approximation is $D^4$ and $D^6$ at one and two loops respectively, the time dependence of the bare counterterms is known. In particular, given our choice in~(\ref{eq:stresstensor_summary_eq}), the bare counterterms are time independent.}\footnote{With the exception of the speed of sound counterterm $c_{\delta,1}$, which will take the role of absorbing an additional UV divergence arising from the one-loop counterterm diagram of ${\cal{P}}^{\rm ct}_{15}$ as will be shown later.}

Following the discussion in \secref{dmctssec}, we have found the following form of the counterterm contributions relevant to the two-loop power spectrum
\be
\delta_{\rm ct} ( a ) = D(a)^3 \tilde \delta^{(1)}_{\rm ct} + D(a)^4 \tilde \delta^{(2)}_{\rm ct} + D(a)^5 \tilde \delta^{(3)}_{\rm ct} + D(a)^5 \tilde \delta^{(1)}_{\partial^2 \rm ct} \ . 
\ee
We define the related perturbative kernels as
\begin{align}
\begin{split}
    \td^{(1)}_{\rm ct}(\vk)  & = F_{\rm ct, 1}(\vk)\td^{(1)}(\vk) \ , \quad \tilde \delta^{(1)}_{\partial^2 \rm ct}  ( \kvec ) = F_{\partial^2 \rm ct, 1} ( \kvec ) \tilde \delta^{(1)} ( \kvec )  \ , \\
    \td^{(2)}_{\rm ct}(\vk) &= \int^{\kvec}_{\vq_1, \vq_2}F_{\rm ct, 2}(\vq_1, \vq_2)\td^{(1)}(\vq_1)\td^{(1)}(\vq_2) \ , \\
     \td^{(3)}_{\rm ct}(\vk) &=  \int^{\kvec}_{\vq_1, \vq_2, \vq_3}F_{\rm ct, 3}(\vq_1, \vq_2, \vq_3)\td^{(1)}(\vq_1)\td^{(1)}(\vq_2)\td^{(1)}(\vq_3)\   .
\end{split} 
\end{align}
These give the following contributions to the dark-matter power spectrum 
\begin{align}
\begin{split}
\mathcal{P}_{13}^{\rm ct} ( k , a ) & = 2 D(a)^4  \langle \tilde \delta^{(1)}_{\rm ct} ( \kvec ) \tilde \delta^{(1)} ( \kvec ' ) \rangle '   \ , \\
\mathcal{P}_{15}^{\rm ct} ( k , a ) & = 2 D(a)^6 \left(  \langle \tilde \delta^{(3)}_{\rm ct} ( \kvec) \tilde \delta^{(1)} ( \kvec' ) \rangle ' +  \langle \tilde \delta^{(1)}_{\partial^2 \rm ct} ( \kvec) \tilde \delta^{(1)} ( \kvec' ) \rangle '  \right)  \ , \\
\mathcal{P}_{33}^{\rm ct} ( k , a ) & = D(a)^6 \left( 2 \langle \tilde \delta^{(1)}_{\rm ct} ( \kvec) \tilde \delta^{(3)} ( \kvec ' ) \rangle  ' +   \langle \tilde \delta^{(1)}_{\rm ct} ( \kvec ) \tilde \delta^{(1)}_{\rm ct} ( \kvec ' ) \rangle '  \right)  \ , \\
\mathcal{P}_{42}^{\rm ct} ( k , a ) & = 2 D(a)^6 \langle \tilde \delta^{(2)}_{\rm ct} ( \kvec ) \tilde \delta^{(2)} ( \kvec ' ) \rangle ' \ ,
\end{split}
\end{align}
and defining the time-independent contributions
\be
\mathcal{P}^{\rm ct}_{ij} ( k ) \equiv \mathcal{P}^{\rm ct}_{ij} ( k , a ) / D(a)^{i+j}\ ,
\ee  
we thus have the counterterm contributions in terms of the kernels defined above
\begin{align}
    \label{eq:2LEFTCT_diagrams}
\begin{split}
\mathcal{P}_{13}^{\rm ct} ( k )  & = 2 F_{\rm ct, 1} ( \kvec ) \mathcal{P}_{\rm lin.} ( k ) \ , \\
\mathcal{P}_{15}^{\rm ct} ( k ) & = 6 \mathcal{P}_{\rm lin.} ( k ) \int_{\qvec} F_{\rm ct , 3} ( \kvec , \qvec , - \qvec )  \mathcal{P}_{\rm lin.} ( q )  + 2 F_{\partial^2 \rm ct, 1} ( \kvec ) \mathcal{P}_{\rm lin.} ( k ) \ , \\
\mathcal{P}_{33}^{\rm ct} ( k ) & = 6 \mathcal{P}_{\rm lin.} ( k ) \int_{\qvec} F_{\rm ct , 1} ( \kvec ) F_3 ( \kvec , \qvec , - \qvec) \mathcal{P}_{\rm lin.} ( q )  + \left[ F_{\rm ct , 1} ( \kvec ) \right]^2 \mathcal{P}_{\rm lin.} ( k ) \ ,  \\
\mathcal{P}_{42}^{\rm ct} ( k ) & =  4 \int_{\qvec} F_{\rm ct , 2} ( \kvec - \qvec , \qvec) F_2 ( \kvec - \qvec , \qvec)  \Plin ( q ) \Plin ( | \kvec - \qvec | ) \ .
\end{split}
\end{align}

Naively, the linear diagram $\mathcal{P}_{13}^{\rm ct}$ and the three one-loop type contributions $\mathcal{P}_{15}^{\rm ct}$, $\mathcal{P}_{33}^{\rm ct}$, and $\mathcal{P}_{42}^{\rm ct}$, could contain the 17 EFT coefficients associated with the basis of $\tau^{ij}$ in \eqn{Cbasis}.  However, after explicitly doing the above contractions, the number of independent functional forms is decreased.  This is because $\tau^{ij}$ enters the counterterm diagrams with external derivatives (due to momentum conservation), and furthermore, the counterterms are integrated in the loop integrals in specific momentum configurations.  To see that, first we define the integrands $\mathbb{K}^{(3)}_{\mathcal{O}_m , \alpha }$ (which contain the factors of $\mathcal{P}_{\rm lin.}$) as,
\begin{align}
\begin{split}
\label{eq:EFT_ct_2L_sum}
	&2 \langle \td^{(2)}_{\rm ct} ( \kvec ) \td^{(2)} ( \kvec') \rangle' +2 \langle \td^{(3)}_{\rm ct} ( \kvec) \td^{(1)} ( \kvec') \rangle' + 2 \langle \td^{(1)}_{\rm ct} ( \kvec) \td^{(3)} ( \kvec') \rangle' \\
	& \hspace{.5in} = \int_{\qvec} \sum_{\mathcal{O}_m , \alpha } c_{\mathcal{O}_m , \alpha }\mathbb{K}^{(3)}_{\mathcal{O}_m , \alpha }(\vk, \vq)\  ,
\end{split}
\end{align}
i.e. we use Eqs. (\ref{eq:countersol}) - (\ref{delta3ct})  to solve for the counterterm solutions for $\delta$, we compute the above one-loop-type contributions to the power spectrum, and we organize the expression in terms of the coefficients $c_{\mathcal{O}_m , \alpha}$.  We then check for degeneracies (including after doing the angular integration) and we find the minimally independent basis 
\begin{align}
    \label{eq:counterKernels}
&\left\{\mathbb{K}^{(3)}_{\delta ,1},\mathbb{K}^{(3)}_{\delta ,2},\mathbb{K}^{(3)}_{\delta ,3}, \mathbb{K}^{(3)}_{\delta  r,1},\mathbb{K}^{(3)}_{\delta  r,2},\mathbb{K}^{(3)}_{\delta ^2,1},\mathbb{K}^{(3)}_{\delta ^2 r,1}\right\}\  .
\end{align}
 This minimal basis contains 7 operators (the degeneracy equations are given in App.~\ref{sec:counterDegens}).  We also find that we can set $c_{r,1} = 0$ in the linear counterterm, so that $c_{\delta , 1}$ is the only common EFT coefficient between the linear and one-loop type counterterms.  Along with these, we also have one higher-derivative counterterm (proportional to $c_{\partial^2 \delta , 1}$).\footnote{Previous studies of the dark matter two-loop power spectrum in the context of the EFTofLSS were performed in e.g.~\cite{Carrasco:2013mua, Foreman:2015lca, Fasiello:2022lff, Garny:2022fsh, Bakx:2025jwa} and at three loops in \cite{Konstandin:2019bay}, but the full renormalization, which includes vorticity and non-locality in time, was to our knowledge never performed before this work.  For example, in this work, we have found eight necessary non-stochastic counterterms relevant at two loops, while previous studies have used four or less.  In this work, we analytically check renormalization at two loops, while in previous works this test has only been done numerically.  We comment more on this in \secref{impeftctssec}. }

 The kernels in \eqn{eq:2LEFTCT_diagrams} specifically are
\begin{align}
\begin{split}
F_{\rm ct , 1} ( \kvec )  = - c_{\delta , 1}   \frac{k^2}{9 \knl^2} \ , \quad F_{\partial^2 \rm ct , 1} ( \kvec )  = c_{\partial^2 \delta , 1}   \frac{k^4}{26 \knl^4}  \ ,
\end{split}
\end{align}
and $F_{\rm ct, 2}$ and $F_{\rm ct, 3}$, which have the following dependence on the EFT parameters,
\begin{align}
\begin{split}
& F_{\rm ct, 2} ( \kvec - \qvec , \qvec ) [ c_{\delta , 1} , c_{\delta , 2} , c_{\delta^2 , 1} , c_{r \delta , 1} ] \ , \\
& F_{\rm ct , 3} ( \kvec ,  \qvec , - \qvec) [ c_{\delta , 1} , c_{\delta , 2},  c_{\delta^2 , 1}, c_{r \delta , 1}   , c_{\delta , 3} , c_{r \delta , 2} , c_{r \delta^2 , 1} ] \ ,
\end{split}
\end{align}
are given in \appref{sec:counterKernels}. 

We now move on to explicitly renormalize the loops.  For all EFT coefficients $c_n$ above, we will write
\be
c_n = c_n^{\rm fin.} + c_n^{[1]} + c_n^{[2]} \ ,
\ee
where the $c_n^{[1]}$ absorbs the UV parts of one-loop diagrams, $c_n^{[2]}$ absorbs the UV parts of two-loop diagrams, and the $c_n^{\rm fin.}$ are the finite, physical parameters that are fit to data.  As we will see, $c_{n}^{[1]} = 0 $ for all $n$ except $(\delta,1)$, which is the only EFT coefficient that enters at one and two loops.  Because of that, we will split up the $13$ counterterm contribution 
\be
\mathcal{P}_{13}^{\text{ct}, [L]} ( k )  \equiv 2 F_{\rm ct,1}^{[L]} ( \kvec ) P_{\rm lin.} ( k)  \ , 
\ee
with 
\be
F_{\rm ct , 1}^{[L]} ( \kvec )  = -  c_{\delta , 1}^{[L]}   \frac{k^2}{9 \knl^2}  \ ,
\ee
where $L$ stands for the loop at which the term contributes.

At one loop, ignoring stochastic terms which are small, only $\mathcal{P}_{13}$ has a UV divergence.  Using the regulated diagram in \secref{sec:UV_1loop} and the above counterterms, the full diagram is 
\be
\mathcal{P}_{13} + \mathcal{P}_{13}^{\rm ct} = \mathcal{P}_{13}^{\rm UV - reg.} + \mathcal{P}_{13}^{\rm UV} + \mathcal{P}_{13}^{\rm ct, [1]} \ ,
\ee
and letting 
\be \label{cdelta11sol}
c_{\delta,1}^{[1]} =  -\frac{61\knl^2}{70}\int_{\qvec} \frac{\Plin(q)}{q^2}f_{\rm screen}(q^2)  \ , 
\ee
the UV part $ \mathcal{P}_{13}^{\rm UV} $ cancels so that we have
\be \label{p13renormeq}
\mathcal{P}_{13} ( k ) + \mathcal{P}_{13}^{\rm ct, [1]} ( k )  =  \mathcal{P}_{13}^{\rm UV - reg.} ( k ) + 2 F_{\rm ct, 1}^{\rm fin.} ( \kvec )  \mathcal{P}_{\rm lin.} ( k ) \ . 
\ee
where
\be
F_{\rm ct, 1}^{\rm fin.} ( \kvec ) = F_{\rm ct, 1}(\kvec) \Big|_{c_n \rightarrow c_n^{\rm fin.}} \ . 
\ee

Next we move on to the two-loop diagrams.  First is the $42$ diagram
\be
\mathcal{P}_{42} + \mathcal{P}_{42}^{\rm ct}  = \mathcal{P}^{\rm UV - reg.}_{42} + \mathcal{P}^{\rm UV - (1) }_{42} + \mathcal{P}_{42}^{\rm ct} \ .
\ee
As summarized in \tabref{tab:UV_2Loops}, $\mathcal{P}_{42}$ only has a UV divergence when $\pvec \rightarrow \infty$.  We find that setting
\begin{align}
\begin{split}
c_{\delta , 2}^{[2]} & = -\frac{12409 \knl^2}{5880}\int_{\pvec} \frac{\Plin(p)}{p^2}f_{\rm screen}(p^2)   \ , \\
c_{r \delta , 1}^{[2]} & = -\frac{6997 \knl^2}{3430}\int_{\pvec} \frac{\Plin(p)}{p^2}f_{\rm screen}(p^2)  \ , \\
c_{\delta^2 , 1}^{[2]} & = -\frac{63149 \knl^2}{41160}\int_{\pvec} \frac{\Plin(p)}{p^2}f_{\rm screen}(p^2)  \ , 
\end{split}
\end{align}
along with using \eqn{cdelta11sol}, cancels the UV part $ \mathcal{P}^{\rm UV - (1) }_{42}$ giving
\be \label{p42ctrenorm}
\mathcal{P}_{42} ( k )  + \mathcal{P}_{42}^{\rm ct}  ( k )  = \mathcal{P}^{\rm UV - reg.}_{42} ( k )  +  4 \int_{\qvec} F^{\rm fin.}_{\rm ct , 2} ( \kvec - \qvec , \qvec) F_2 ( \kvec - \qvec , \qvec)  \Plin ( q ) \Plin ( | \kvec - \qvec | ) \ , 
\ee
where
\be
F^{\rm fin.}_{\rm ct , 2} ( \kvec - \qvec , \qvec) \equiv F_{\rm ct , 2} ( \kvec - \qvec , \qvec)\Big|_{c_n \rightarrow c_n^{\rm fin.}} \ ,
\ee
contains only the finite EFT parameters.  Note that the integral in \eqn{p42ctrenorm} has a negative degree of UV divergence, so we do not need to regulate that loop.  

Next, we have the $33$ diagram
\begin{align}
\begin{split}
\mathcal{P}_{33} + \mathcal{P}_{33}^{\rm ct} & = \mathcal{P}_{33}^{\rm (I)} + \mathcal{P}_{33}^{\rm (II)} + \mathcal{P}_{33}^{\rm ct} \ .
\end{split}
\end{align}
Looking at \tabref{tab:UV_2Loops}, we see that only $\mathcal{P}_{33}^{\rm (I)} $ has UV divergences.  From the diagram for $\mathcal{P}_{33}^{\rm (I)} $ in \eqn{p33feyn} and the expression in \eqn{eq:P33I}, we see that this is simply two copies of a $\mathcal{P}_{13}$ diagram, and so renormalization of this diagram will be automatic once we have renormalized $\mathcal{P}_{13}$.  To see that, we first write
\be
\mathcal{P}_{33}^{(\rm I)} ( k ) = \frac{\mathcal{P}_{13} ( k )^2}{4 \mathcal{P}_{\rm lin.} ( k ) }  \ ,
\ee
and
\be
\mathcal{P}_{33}^{\rm ct} ( k ) = \frac{\mathcal{P}_{13}^{\rm ct,[1]} ( k ) \mathcal{P}_{13} ( k ) }{2 \mathcal{P}_{\rm lin.} ( k ) } + \frac{\mathcal{P}_{13}^{\rm ct,[1]} ( k )^2 }{4 \mathcal{P}_{\rm lin.} ( k ) }  \ , 
\ee
so that
\be
\mathcal{P}_{33}^{(\rm I)} ( k )  + \mathcal{P}_{33}^{\rm ct} ( k )  = \frac{\left( \mathcal{P}_{13} (k) + \mathcal{P}_{13}^{\rm ct,[1]} ( k ) \right)^2  }{4 \mathcal{P}_{\rm lin.} ( k ) }  =  \frac{\left( \mathcal{P}_{13}^{\rm UV - reg.} ( k ) + 2 F_{\rm ct, 1}^{\rm fin.} ( \kvec )  \mathcal{P}_{\rm lin.} ( k ) \right)^2  }{4 \mathcal{P}_{\rm lin.} ( k ) }   \ ,
\ee
where we have used \eqn{p13renormeq} to replace $ \mathcal{P}_{13} (k) + \mathcal{P}_{13}^{\rm ct} ( k )$ in terms of finite quantities.  Thus, we finally have
\begin{align}
\begin{split}
   & \mathcal{P}_{33} ( k ) + \mathcal{P}_{33}^{\rm ct} ( k )  = \mathcal{P}_{33}^{(\rm II)} ( k )  +  \frac{\left( \mathcal{P}_{13}^{\rm UV - reg.} ( k ) + 2 F_{\rm ct, 1}^{\rm fin.} ( \kvec )  \mathcal{P}_{\rm lin.} ( k ) \right)^2  }{4 \mathcal{P}_{\rm lin.} ( k ) }   \\
    \label{eq:P33renorm}
   & \quad = \mathcal{P}_{33}^{(\rm II)} ( k ) + \mathcal{P}_{33}^{(\rm  I), UV-reg.} (k) + F_{{\rm ct},1}^{\rm fin.}(\vk) \mathcal{P}_{13}^{\rm UV-reg.}(k) + \Plin(k) \l  F_{{\rm ct},1}^{\rm fin.}(\vk) \r^2 \ , 
\end{split}
\end{align}
and we see that the UV parts automatically cancel, and the way in which the lower-order finite EFT parameters enter the renormalization of this diagram.  

Finally, we renormalize the $15$ diagram.  We have
\be
\mathcal{P}_{15} + \mathcal{P}_{15}^{\rm ct} + D^{-2} \mathcal{P}_{13}^{\rm ct , [2]} = \mathcal{P}_{15}^{\rm UV- reg.} + \mathcal{P}_{15}^{\rm UV- (1)} + \mathcal{P}_{15}^{\rm UV- (2)}+ \mathcal{P}_{15}^{\rm ct} +D^{-2} \mathcal{P}_{13}^{\rm ct , [2]}  \ .
\ee
First, we focus on the single UV limit $\mathcal{P}_{15}^{\rm UV- (1)}$, which will be cancelled by the term with $F_{\rm ct,3}$ in $\mathcal{P}_{15}^{\rm ct} $, giving us
\begin{align}
\begin{split}
 \mathcal{P}_{15}^{\rm UV- (1)} ( k ) + 6 \mathcal{P}_{\rm lin.} ( k ) \int_{\qvec} F_{\rm ct , 3} ( \kvec , \qvec , - \qvec )  \mathcal{P}_{\rm lin.} ( q )   =  6 \mathcal{P}_{\rm lin.} ( k ) \int_{\qvec} F^{\rm fin.}_{\rm ct , 3} ( \kvec , \qvec , - \qvec )  \mathcal{P}_{\rm lin.} ( q )  \ , 
\end{split}
\end{align}
where
\be
F^{\rm fin.}_{\rm ct , 3} ( \kvec , \qvec , - \qvec) \equiv F_{\rm ct , 3} ( \kvec , \qvec , -\qvec)\Big|_{c_n \rightarrow c_n^{\rm fin.}} \ ,
\ee
for
\begin{align}
\begin{split} \label{eighttermuvmatch}
c_{\delta , 3}^{[2]} & = -\frac{167099 \knl^2}{64680}\int_{\pvec} \frac{\Plin(p)}{p^2}f_{\rm screen}(p^2)   \ , \\
c_{r \delta , 2}^{[2]} & = -\frac{27241 \knl^2}{44100}\int_{\pvec} \frac{\Plin(p)}{p^2}f_{\rm screen}(p^2)  \ , \\
c_{r \delta^2 , 1}^{[2]} & = \frac{60145999 \knl^2}{169785000}\int_{\pvec} \frac{\Plin(p)}{p^2}f_{\rm screen}(p^2)  \ .
\end{split}
\end{align}

This means we are left with
\begin{align}
\begin{split} \label{p15step1_b}
\mathcal{P}_{15} ( k )  + \mathcal{P}_{15}^{\rm ct} ( k ) + D^{-2} \mathcal{P}_{13}^{\rm ct , [2]} ( k ) & = \mathcal{P}_{15}^{\rm UV- reg.}  (k)  + 2 F_{\partial^2 \rm ct,1} ( \kvec) \Plin ( k ) +D^{-2}  \mathcal{P}_{13}^{\rm ct , [2]} ( k ) \\
&  \quad + \mathcal{P}_{15}^{\rm UV- (2)}  ( k ) + 6 \mathcal{P}_{\rm lin.} ( k ) \int_{\qvec} F^{\rm fin.}_{\rm ct , 3} ( \kvec , \qvec , - \qvec )  \mathcal{P}_{\rm lin.} ( q )    \ .
\end{split}
\end{align}
At this point, both $\mathcal{P}_{15}^{\rm UV- (2)} $ and the integral over $F^{\rm fin.}_{\rm ct , 3} $ contain pieces proportional to $k^2 \Plin ( k)$ and $k^4 \Plin ( k )$, which can be cancelled by $ \mathcal{P}_{13}^{\rm ct , [2]} $ and $F_{\partial^2 \rm ct,1} $ respectively.  To see this, recall that 
\be
\mathcal{P}_{15}^{\rm UV- (2)}  ( k )  = \int_{\qvec, \pvec}  \mathcal{R}^{(0)}_{(q, p) \to \infty}(p_{15}( \kvec , \qvec , \pvec) ) + \int_{\qvec, \pvec}  \mathcal{R}^{(2)}_{(q, p) \to \infty}(p_{15} ( \kvec , \qvec , \pvec))  \ , 
\ee
where the first term on the right-hand side is proportional to $k^2 \Plin ( k)$, and the second term is proportional to $k^4 \Plin(k)$, and they are given explicitly in \appref{sec:TRUVapproximations}.  For later convenience, let us define the dimensionless numbers
\begin{align}
\begin{split}
\mathcal{R}^{(0)} & \equiv \frac{\knl^2}{k^2 \Plin(k)} \int_{\qvec, \pvec}  \mathcal{R}^{(0)}_{(q, p) \to \infty}(p_{15}( \kvec , \qvec , \pvec) )  \ , \\
\mathcal{R}^{(2)} & \equiv \frac{\knl^4}{k^4 \Plin(k)}  \int_{\qvec, \pvec}  \mathcal{R}^{(2)}_{(q, p) \to \infty}(p_{15} ( \kvec , \qvec , \pvec))  \ .
\end{split}
\end{align}
Furthermore, we can separate the UV part of the loop integral over $F^{\rm fin.}_{\rm ct , 3} $ by writing
\begin{align}
\begin{split} \label{f3ctfinuvreg}
 \int_{\qvec} F^{\rm fin.}_{\rm ct , 3} ( \kvec , \qvec , - \qvec )  \mathcal{P}_{\rm lin.} ( q ) & =  \int_{\qvec} F^{\rm fin., UV-reg.}_{\rm ct , 3} ( \kvec , \qvec , - \qvec )  \mathcal{P}_{\rm lin.} ( q ) \\ 
 & + \int_{\qvec} \frac{k^2}{\knl^2} F^{\text{fin.}, k^2}_{\rm ct , 3}   \mathcal{P}_{\rm lin.} ( q )  f_{\rm screen} ( q^2 ) \\
  & + \int_{\qvec} \frac{k^4}{q^2 \knl^2} F^{\text{fin.}, k^4}_{\rm ct , 3}   \mathcal{P}_{\rm lin.} ( q ) f_{\rm screen} ( q^2 ) \ , 
\end{split}
\end{align}
where
\begin{align}
\begin{split}
F^{\text{fin.}, k^2}_{\rm ct , 3}   & = -\frac{1}{810810 } \Big(  - 
    12460 c^{\rm fin.}_{\delta, 1} - 19392 c^{\rm fin.}_{\delta, 2}+ 
    23265 c^{\rm fin.}_{\delta, 3} - 16912 c^{\rm fin.}_{r \delta, 1}  \\
    & + 
    28908 c^{\rm fin.}_{r \delta, 2} - 23520 c^{\rm fin.}_{\delta^2, 1} + 
    17325 c^{\rm fin.}_{r \delta^2, 1}  \Big) \ , 
\end{split}
\end{align}
\begin{align}
\begin{split}
F^{\text{fin.}, k^4}_{\rm ct , 3}   & =  \frac{1}{5675670 } \Big(  14043 c^{\rm fin.}_{\delta, 1} +  7392 c^{\rm fin.}_{\delta, 3} + 2380 c^{\rm fin.}_{r \delta, 1} - 
 23760 c^{\rm fin.}_{r \delta, 2} \Big) \ , 
\end{split}
\end{align}
and 
\be
\label{eq:ct3_uv}
F^{\rm fin., UV-reg.}_{\rm ct , 3} ( \kvec , \qvec , - \qvec ) = F^{\rm fin.}_{\rm ct , 3} ( \kvec , \qvec , - \qvec ) - \left(  \frac{k^2}{\knl^2} F^{\text{fin.}, k^2}_{\rm ct , 3}   +  \frac{k^4}{q^2 \knl^2} F^{\text{fin.}, k^4}_{\rm ct , 3}   \right) f_{\rm screen} ( q^2 ) \ .
\ee

Going back to \eqn{p15step1_b}, the above implies that
\begin{align}
\begin{split}
\mathcal{P}_{15} ( k )  + \mathcal{P}_{15}^{\rm ct} ( k ) + D^{-2} \mathcal{P}_{13}^{\rm ct , [2]} ( k ) & = \mathcal{P}_{15}^{\rm UV- reg.}  (k) + 6 \Plin(k) \int_{\qvec} F^{\rm fin., UV-reg.}_{\rm ct , 3} ( \kvec , \qvec , - \qvec )  \mathcal{P}_{\rm lin.} ( q ) \\
& \hspace{-1in} +  \frac{k^2}{\knl^2} \Plin(k) \left( \mathcal{R}^{(0)} + 6 F^{\text{fin.}, k^2}_{\rm ct , 3} \int_{\qvec} \Plin(q) f_{\rm screen} ( q^2 )  -  \frac{2  c_{\delta , 1}^{[2]}  }{9 D^2} \right)  \\
& \hspace{-1in} + \frac{k^4}{\knl^4} \Plin(k )  \left( \mathcal{R}^{(2)} + 6 \knl^2 F^{\text{fin.}, k^4}_{\rm ct , 3} \int_{\qvec} \frac{\Plin(q) }{q^2} f_{\rm screen} ( q^2 ) + \frac{c_{\partial^2 \delta , 1}^{[2]} }{13} \right)   \ ,
\end{split}
\end{align}
and we see that the  UV sensitive terms inside of the parentheses cancel if
\begin{align}
\begin{split}
c_{\delta,1}^{[2]} & = \frac{9 D^2}{2} \mathcal{R}^{(0)} + 27 D^2  F^{\text{fin.}, k^2}_{\rm ct , 3} \int_{\qvec} \Plin(q) f_{\rm screen} ( q^2 ) +  D^2  \hat c_{\delta,1}^{[2]}  \ , \\
c_{\partial^2 \delta , 1}^{[2]} & = -13 \mathcal{R}^{(2)} - 78 \knl^2 F^{\text{fin.}, k^4}_{\rm ct , 3} \int_{\qvec} \frac{\Plin(q) }{q^2} f_{\rm screen} ( q^2 ) \ ,
\end{split}
\end{align}
where $\hat c_{\delta,1}^{[2]} $ is finite and determined by the renormalization conditions \cite{Carrasco:2013mua, Foreman:2015lca}.\footnote{The renormalization condition is, for example, ${\cal{P}}_{\rm 2-loop} + {\cal{P}}^{\rm ct}_{42} + {\cal{P}}^{\rm ct}_{33} + {\cal{P}}^{\rm ct}_{15} + D^{-2} \mathcal{P}_{13}^{\rm ct , [2]}  = 0 $ at some $k_{\rm ren}$ sufficiently small, which makes $\hat c_{\delta,1}^{[2]} $ not a free parameter. This renormalization condition effectively removes the terms in the full two-loop contribution, including terms proportional to two-loop counterterms, that scale like $k^2 \mathcal{P}_{\rm lin.}(k)$ for $k$ much smaller than $M = 0.6 \, h / \text{Mpc}$ in $f_{\rm screen}$ \eqn{fscreeneq}. This approximately means that the only $k^2\mathcal{P}_{\rm lin.}(k)$ contribution comes from $\mathcal{P}_{\rm 1-loop}  + \mathcal{P}_{13}^{\rm ct,[1]}$. } 
This leads to our final expression
\begin{align}
\begin{split}
\mathcal{P}_{15} + \mathcal{P}_{15}^{\rm ct} + D^{-2} \mathcal{P}_{13}^{\rm ct , [2]} & = \mathcal{P}_{15}^{\rm UV- reg.}  ( k ) + 6 \Plin(k)  \int_{\qvec} F^{\rm fin., UV-reg.}_{\rm ct , 3} ( \kvec , \qvec , - \qvec )  \mathcal{P}_{\rm lin.} ( q ) \\
& + c_{\partial^2 \delta , 1}^{\rm fin.} \frac{k^4}{13 \knl^4} \Plin ( k)  - \hat c^{[2]}_{\delta , 1} \frac{2 k^2}{9 \knl^2} \Plin(k)  \ . 
\end{split}
\end{align}
Notice that, although we expanded \eqn{f3ctfinuvreg} up to $k^4 \Plin ( k )$ for UV regularization, we only expanded $\mathcal{P}_{13}$ up to $k^2 \Plin ( k)$, see \eqn{tq0f3}.  These diagrams actually have the same level of UV divergence, logarithmic, for $\nu = 1$, and so we are being consistent.  The reason that \eqn{f3ctfinuvreg} goes up to $k^4 \Plin ( k )$ is because there is a factor of $1 / \knl^2$ due to the fact that it is a counterterm diagram, and so the whole kernel goes as $k^4 / ( q^2 \knl^2)$, and so the level of UV divergence is the same as for the $k^2 / q^2 $ term in \eqn{f3uvlimit}.

To summarize, the one-loop renormalized diagrams are given by
\begin{align} \label{finalp1loopren}
\mathcal{P}_{\rm 1-loop} ( k ) + \mathcal{P}_{13}^{\rm ct,[1]} ( k )  = \mathcal{P}_{22} ( k ) +  \mathcal{P}_{13}^{\rm UV - reg.} ( k ) + 2 F_{\rm ct, 1}^{\rm fin.} ( \kvec )  \mathcal{P}_{\rm lin.} ( k ) \ , 
\end{align}
and the two-loop renormalized diagrams are given by
\begin{align}
\begin{split} \label{finalp2loopren}
& {\cal{P}}_{\rm 2-loop} + {\cal{P}}^{\rm ct}_{42} + {\cal{P}}^{\rm ct}_{33} + {\cal{P}}^{\rm ct}_{15} + D^{-2} \mathcal{P}_{13}^{\rm ct , [2]}  \\
& \hspace{.5in}  = \mathcal{P}^{\rm UV - reg.}_{42} ( k )  +  4 \int_{\qvec} F^{\rm fin.}_{\rm ct , 2} ( \kvec - \qvec , \qvec) F_2 ( \kvec - \qvec , \qvec)  \Plin ( q ) \Plin ( | \kvec - \qvec | ) \\
& \hspace{.5in} +  \mathcal{P}_{33}^{(\rm II)} ( k )  +  \mathcal{P}_{33}^{(\rm I), UV-reg.} ( k )  + F_{{\rm ct},1}^{\rm fin.}(\vk) \mathcal{P}_{13}^{\rm UV-reg.}(k) + \Plin(k) \l  F_{{\rm ct},1}^{\rm fin.}(\vk) \r^2   \\
& \hspace{.5in} + \mathcal{P}_{15}^{\rm UV- reg.}  ( k ) + 6 \Plin ( k ) \int_{\qvec} F^{\rm fin., UV-reg.}_{\rm ct , 3} ( \kvec , \qvec , - \qvec )  \mathcal{P}_{\rm lin.} ( q ) \\
& \hspace{.5in} + c_{\partial^2 \delta , 1}^{\rm fin.} \frac{k^4}{13 \knl^4} \Plin ( k)   - \hat c^{[2]}_{\delta , 1} \frac{2 k^2}{9 \knl^2} \Plin(k)  \ , 
\end{split}
\end{align}
such that the final two-loop contributions contain only regularized loop diagrams and the response diagrams contain only finite counterterms.   Restoring the EdS time dependence, our final expression for the time-dependent renormalized power spectrum up to two loops is $D(a)^2 \Plin ( k)$ plus $D(a)^4$ times \eqn{finalp1loopren}, plus $D(a)^6$ times \eqn{finalp2loopren}. We explicitly show how to evaluate the one-loop-type counterterm contributions above in \appref{sec:2-loop_counterterms}.  Finally, we note that even though there is one more basis function in $\tau^{ij}$ up to third order in the non-local-in-time expansion, we find the non-local-in-time and local-in-time expansions to be equivalent for the two-loop power spectrum.  We have concluded the renormalization of the power spectrum at two loops, showing in particular that the UV subtractions of the kernels that we have made are matched by counterterms.

%
\subsection{Importance of EFT counterterms} \label{impeftctssec}

 Since all eight of our EFT counterterms are needed to fully match the UV of the two-loop integrals (for the degree of divergence that we consider), we conclude that all eight of these EFT counterterms are needed to fully renormalize the theory.  Previous works \cite{Carrasco:2013mua, Foreman:2015lca, Konstandin:2019bay, Fasiello:2022lff, Garny:2022fsh, Taule:2023izt, Bakx:2025jwa} have numerically found that only four (or less) EFT coefficients are necessary to make the theory sufficiently UV insensitive with respect to some given error budget over a reasonable range of wavenumbers.  This seems to us to be due to the fact that the no-counterterm loops are actually UV convergent for a realistic $\Plin$;  if one were to use a $\Plin$ that leads to bona fide UV divergences in the loop integrals, all eight EFT counterterms found in this work would be needed to make the theory finite.

A separate question from regularizing the no-counterterm loops is how important each of the eight EFT counterterms will be in a realistic data analysis.  To assess this, we can look at the Fisher matrix
\begin{align}
F_{ij} & = \frac{1}{2} \frac{\partial^2}{\partial \theta^i \partial \theta^j}  \sum_{n,m}   \mathcal{P} ( k_n)  C^{-1}(k_n , k_m)  \mathcal{P} ( k_m)   \Big|_{\theta = 0} \\
& = \sum_{n,m} \left(  \frac{\partial \mathcal{P} ( k_n) }{\partial \theta^i } C^{-1}(k_n , k_m)\frac{\partial \mathcal{P} ( k_m) }{\partial \theta^j } +  \delta^K_{i1} \delta^K_{j1}  \sum_{n,m}  \frac{\partial^2 \mathcal{P} ( k_n)}{\partial c_{\delta,1}^2}   C^{-1}(k_n , k_m)  \mathcal{P} ( k_m)   \right) \Big|_{\theta = 0}  \ , \nonumber
\end{align}
where $n$ and $m$ label the data bin, $C$ is the covariance matrix, and $\theta^i$ is the EFT coefficient vector
\be
\theta^i = \{   c_{\delta , 1} , c_{\delta , 2},  c_{\delta^2 , 1}, c_{r \delta , 1}   , c_{\delta , 3} , c_{r \delta , 2} , c_{r \delta^2 , 1}, c_{\partial^2 \delta , 1} \}  \ .
\ee 

To get a rough sense of the constraints on the EFT parameters, we imagine comparing to a dark-matter distribution, and we approximate the covariance as diagonal, taking
\be
C(k_n, k_n) \equiv \sigma_{\rm data}^2 ( k_n ) \ .
\ee 
For $\sigma_{\rm data}^2$, in order to get an idea of constraints from current, realistic-sized surveys, we will use the covariance of a dark-matter distribution with DESI-like volume. Specifically, we take 
\be \label{sigmadata}
\sigma_{\rm data}^2 ( k_n ) = \frac{4 \pi^2 \mathcal{P}_{\rm lin.} ( k_n , z_{\rm eff} )^2}{V_{\rm eff}  k_n^2 \delta k} \ , 
\ee
at $z_{\rm eff} = 1.23$, with $V_{\rm eff} = 20 \times 10^9 ( \text{Mpc}/h)^3$.\footnote{ \label{covfn} To find this, we first constructed the DESI galaxy covariance using the parameters for Emission Line Galaxies from App. A of \cite{Braganca:2023pcp} using the redshifts $z_i = 1.15, 1.25, 1.35, 1.45, 1.55, 1.65$.  Recall that a good approximation to the Gaussian galaxy power spectrum covariance is
\be
C(k_n, k_n; z , V(z) , b_1(z), \bar n ( z) ) = \frac{4 \pi^2}{V(z) k_n^2 \delta k} \left( b_{1}(z)^2 \mathcal{P}_{\rm lin.} ( k_n , z ) + \frac{1}{\bar n ( z)} \right)^2  \ ,
\ee 
and the inverse covariance of multiple redshift bins is obtained by summing over the inverse covariances of each bin
\be \label{invcovsum}
C^{-1} (k_n, k_n; z_{\rm eff} ) = \sum_i C^{-1} (k_n, k_n; z_i , V(z_i) , b_1(z_i), \bar n ( z_i ) ) \ , 
\ee
where
\be
z_{\rm eff} = \frac{\sum_i z_i V(z_i) \bar n (z_i)^2 }{\sum_i  V(z_i) \bar n (z_i)^2} \ , 
\ee
(we ignore FKP and systematic weights for simplicity).   Using the values for $V(z)$, $b_1 ( z)$, and $\bar n(z)$ from \cite{Braganca:2023pcp}, we find for the six redshift bins mentioned above, that $z_{\rm eff} = 1.23$ and that 
\be
\sigma^2_{\rm DESI} ( k_n) \equiv C(k_n, k_n; z_{\rm eff} , V_{\rm eff} , b_{1,\text{eff}}, \bar n_{\rm eff} ) \ , 
\ee
with $V_{\rm eff} = 20 \times 10^9 ( \text{Mpc}/h)^3$, $b_{1,\text{eff}} = 1.5$, and $\bar n_{\rm eff} = 2.8 \times 10^{-4} ( h / \text{Mpc})^3$ is approximately $C (k_n, k_n; z_{\rm eff} ) $ in \eqn{invcovsum}.  Then we take $\sigma_{\rm data}^2 ( k_n)$ in \eqn{sigmadata} to be $\sigma^2_{\rm DESI} ( k_n)$ in the dark-matter limit $b_{1,\text{eff}} = 1$ and $ \bar n_{\rm eff} \rightarrow \infty$, i.e. $\sigma_{\rm data}^2 ( k_n ) = C(k_n, k_n; z_{\rm eff} , V_{\rm eff} , 1, \infty)$.}
Taking $\knl = 1 \, h / \text{Mpc}$,  imposing the renormalization condition ${\cal{P}}_{\rm 2-loop} + {\cal{P}}^{\rm ct}_{42} + {\cal{P}}^{\rm ct}_{33} + {\cal{P}}^{\rm ct}_{15} + D^{-2} \mathcal{P}_{13}^{\rm ct , [2]}  = 0 $ at $k_{\rm ren} = 0.005 \, h / \text{Mpc}$, binning with $\delta k = 0.005 \, h / \text{Mpc}$, taking $k_{\rm min} = 0.001 \, h / \text{Mpc}$ and $k_{\rm max} = 0.601 \, h / \text{Mpc}$ in the sum over $k$-bins, and then diagonalizing the Fisher matrix such that
\be
F = U D U^{\rm t} \ , 
\ee
we find 
\be \label{diagfisher}
D^{-1/2} \approx   \{0.0034, 0.013, 0.57, 12, 140 , 1.1 \times 10^3, 2.4 \times 10^3,  5.1 \times 10^3 \} , 
\ee
as the estimated size of the error bars of the linear combinations of the $\theta^i$ corresponding to the eigenvectors of $F_{ij}$.  We should compare these errors to the typical size of the priors that would be put on the EFT parameters, which is typically $\mathcal{O}(2)$.  From this, we see that three parameters are constrained significantly better than the priors, and so are determined well by the data.   

Reasoning similar to that above can be done to determine how many EFT parameters could be kept in any given data analysis.  We would like to make the following points about this observation, though.  As long as there are reasonable priors on all of the EFT parameters, the constraints on the cosmological parameters of interest should not change much if one includes extra EFT parameters (i.e. combinations that are not well constrained by the data), because indeed they do not play a role in the prediction.  That said, we can give a rough indication here about which parameters can be dropped from the hypothetical dark-matter analysis mentioned above.  Keeping only the parameters $\{   c_{\delta , 1} ,  c_{\delta^2 , 1} \} $, we can estimate the following errors for the corresponding eigenvectors
\begin{align}
\begin{split}
\sigma & \approx 0.0035 \quad \text{for} \quad 0.968 \, c_{\delta,1} + 0.249 \, c_{\delta^2,1}  \ ,   \\
\sigma & \approx 0.018 \quad \text{for} \quad - 0.249 \, c_{\delta,1} + 0.968\, c_{\delta^2,1}   \ .
\end{split}
\end{align}
These errors are quite close to the smallest errors in \eqn{diagfisher}, which means that these parameters should describe the leading eigenvectors, and therefore the data, well.  Keeping additionally $c_{\partial^2\delta , 1}$, for example, we have

\begin{align}
\begin{split}
\sigma & \approx 0.0035 \quad \text{for} \quad 0.968 \, c_{\delta,1} + 0.249 \, c_{\delta^2,1} - 0.027 \, c_{\partial^2\delta , 1}  \ ,   \\
\sigma & \approx 0.017 \quad \text{for} \quad - 0.251\, c_{\delta,1} + 0.962\, c_{\delta^2,1} - 0.105\, c_{\partial^2\delta , 1}  \ , \\ 
\sigma& \approx 1.85 \quad \text{for} \quad -  0.0004\, c_{\delta,1} + 0.109\, c_{\delta^2,1}  + 0.994 \, c_{\partial^2\delta , 1}  \ ,
\end{split}
\end{align}
and further keeping $c_{r \delta  , 2}$, for example, we have
\begin{align}
\begin{split} \label{4params}
\sigma & \approx 0.0035 \quad \text{for} \quad 0.967\, c_{\delta,1} + 0.250 \, c_{\delta^2,1} + 0.053\, c_{r \delta , 2}   - 0.027 \, c_{\partial^2\delta , 1} \ ,  \\
\sigma & \approx 0.017 \quad \text{for} \quad -0.257 \, c_{\delta,1} + 0.941\, c_{\delta^2,1} + 0.195 \, c_{r \delta , 2}   - 0.103 \, c_{\partial^2\delta , 1}  \ ,  \\
\sigma & \approx 0.63 \quad \text{for} \quad -0.001\, c_{\delta,1} - 0.153\, c_{\delta^2,1} + 0.922\, c_{r \delta , 2}   + 0.355 \, c_{\partial^2\delta , 1}  \ ,  \\
\sigma & \approx 13 \quad \text{for} \quad   10^{-5} \, c_{\delta,1} + 0.170 \, c_{\delta^2,1} - 0.330 \, c_{r \delta , 2}   + 0.929 \, c_{\partial^2\delta , 1}  \ .
\end{split}
\end{align}
Thus, we see that we seem to have saturated the constraining power implied by \eqn{diagfisher}.  Consistent with \cite{Foreman:2015lca,Konstandin:2019bay}, we find that $\{   c_{\delta , 1} ,  c_{\delta^2 , 1},  c_{\partial^2\delta , 1} \} $ should lead to a good reproduction of this data.\footnote{We note that even though the comparisons in \cite{Foreman:2015lca,Konstandin:2019bay} were done on simulations with a much larger $V_{\rm eff}$ and at $z = 0$, the conclusion about the number of parameters needed to fit the data appears to be the same.  One can also see from \eqn{4params} that the combination of parameters $\{   c_{\delta , 1} ,  c_{\delta^2 , 1},  c_{r \delta , 2}  \} $ is actually better constrained than $\{   c_{\delta , 1} ,  c_{\delta^2 , 1},  c_{\partial^2\delta , 1} \} $ (because the eigenvector with $0.922 c_{r \delta , 2}$ has error $\sigma \approx 0.63$ and the eigenvector with $0.929 c_{\partial^2 \delta , 1}$ has error $\sigma \approx 13$).  However, as pointed out in \cite{Foreman:2015lca}, adding a parameter like $c_{r \delta , 2}$ leads to over-fitting, and so \cite{Foreman:2015lca} chose to use analogues of our $\{   c_{\delta , 1} ,  c_{\delta^2 , 1},  c_{\partial^2\delta , 1}  \} $. It is not a surprise, though, that we found that $c_{r \delta , 2}$ is well constrained, since \cite{Foreman:2015lca} found that an analogous parameter for them significantly affected the fit (although with an unnaturally large coefficient that signaled over-fitting).}  We would also like to stress that this kind of reduction in EFT parameters should be updated for other data sets or when comparing higher-order statistics to data.

We give these expressions just as an indication of how many and which parameters are expected to be important in a dark-matter analysis with DESI-like volume.  A similar analysis could be done with the lensing potential, which can be computed from the dark-matter power spectrum that we have computed here, and is relevant for the Euclid satellite, for example.  While the overall size of the errors for the Euclid survey will be different than the example we discussed above (in particular, the errors on the lensing potential will be larger), we do not necessarily expect the number of well-measured parameters to be much different from what we found.  We leave a further exploration of the lensing potential to future study.

\section{Infrared cancellations}
\label{sec:IR}

In the previous section, we constructed the UV-regulated expressions $\Pcal_{\rm 1-loop}^{\rm UV-reg.}(k)$ and $\Pcal_{\rm 2-loop}^{\rm UV-reg.}(k)$ in \eqn{eq:P1-loop_UVreg}  and \eqn{eq:P2loop-UV-reg} respectively, consisting of regulated diagrams which are locally, {\it i.e.} at the level of integrand, free of ultraviolet singularities. 
These diagrams can be, in principle, integrated numerically. 
However, a substantial practical problem still remains before numerical integration, as the diagrams of Eq.~\eqref{eq:P1-loop_UVreg} and Eq.~\eqref{eq:P2loop-UV-reg} exhibit integrable infrared singularities. 
All infrared singularities cancel in the sum of diagrams~\cite{Carrasco:2013sva}. However, these singularities develop at various regions in the integration domain, and the cancellations do not occur locally. This fact compromises the stability and convergence of stochastic numerical integration. 

Challenges due to IR singularities have been studied in Refs~\cite{Carrasco:2013sva, Lewandowski:2017kes}, which dealt with constructions of IR-safe loop integrands. In this article, we will construct an IR-safe integrand as well, with a method similar to the one of Ref.~\cite{Carrasco:2013sva}. 
A difference in our current procedure is that we will use partitions of unity (cf. Refs~\cite{Becker:2012aqa, Capatti:2019edf}) instead of step functions to disentangle infrared regions.

In order to chart the infrared singularities of the one- and two-loop diagrams, we will need to carry out an infrared power-counting analysis (cf. Ref~\cite{Anastasiou:2018rib}). For this purpose, we parametrize the soft limit by rescaling the loop momentum as $q \to q \delta$, with $q$ fixed and $\delta \to 0$. In this way, the linear power spectrum scales in the infrared as\footnote{Modes with wavelengths longer than $k_{\rm eq} \simeq 0.015 h/\text{Mpc} $ enter the horizon after matter-radiation equality, and thus their power spectrum is directly related to the inflationary power spectrum \cite{dodelson}.  For an inflationary power spectrum with scalar tilt $n_s$, this means that the matter power spectrum scales as $\Plin ( q ) \propto q^{n_s}$ for $ q \ll k_{\rm eq}$.  For simplicity, we take $n_s = 1$ in \eqn{eq:PlinIRscaling} because it is a close approximation and does not meaningfully change our results.}
\begin{eqnarray}
    \label{eq:PlinIRscaling}
    \Plin\left(q \, \delta \right) = C_{\rm lin.}^{\rm IR}\,  q \,  \delta  + {\cal O}(\delta^2)\  .
\end{eqnarray}
The constant $ C_{\rm lin.}^{\rm IR}$ is positive and depends on the cosmological model and cosmological parameters.
With the IR scaling of the linear power spectrum that we assumed in Eq.~\eqref{eq:PlinIRscaling}, some integrable singularities which were treated in Ref.~\cite{Carrasco:2013sva} are subleading and they will not concern us in this publication. 

\subsection{Infrared cancellations at one loop}
\label{sec:1L-IRnUVsafe}

\begin{table}[H]
    \centering
    \begin{tabular}{|r|c|}
    \hline
               Diagram       & Leading IR regions \\
              \hline
         $\Pcal_{13}$ & $ \{\vq \to 0\} \, [-1]$ \\ 
         \hdashline
         $\Pcal_{22}$ & $\begin{array}{c}\{\vq \to 0\} \, [-1] \\ \{\vq -\vk \to 0\} \, [-1] \end{array}$\\
         \hline
    \end{tabular}
    \caption{\small Locations of leading infrared singularities in one-loop diagrams. In brackets we include the infrared degree of divergence of the integrand \textit{excluding the integration measure} in the specified location.}
    \label{tab:IR-1L}
\end{table}
In \tabref{tab:IR-1L}, we report the infrared regions that yield the leading infrared singularities in one-loop diagrams. With the exception of one singularity at $\vq = \vk$ in the $\Pcal_{22}$ diagram, the remaining singularities originate from the $\vq \to 0$ limit. We will first disentangle the $\vq \to \vk$ singularity from the $\vq \to 0$ singularity. Then, with a shift of the integration variables to the contribution that exhibits the $\vq \to \vk$  singularity, we bring all the singularities to the same region $\vq \to 0$, where they will cancel locally.  

To disentangle the two infrared regions, we will  use a partition of unity to split the $\Pcal_{22}$ integral into two ``channels", each with only one singularity,
\begin{align}
\label{eq:P22_multi-channeling}
    \Pcal_{22}(k) =& \int_{\qvec} p_{22}(\vq, \vk) = \int_{\qvec} \frac{p_{22}(\vq, \vk) \, (|\vk - \vq|^2)^2}{(q^2)^2 + (|\vk - \vq|^2)^2} + \int_{\qvec} \frac{p_{22}(\vq, \vk) \, (q^2)^2}{(q^2)^2 + (|\vk - \vq|^2)^2} \ . 
\end{align}
We define the two channels as:
\begin{align}
    p^{(1)}_{22}(\vq, \vk) = \frac{p_{22}(\vq, \vk) \, (|\vk - \vq|^2)^2}{(q^2)^2 + (|\vk - \vq|^2)^2} \ , \\
    p^{(2)}_{22}(\vq, \vk) = \frac{p_{22}(\vq, \vk) \, (q^2)^2}{(q^2)^2 + (|\vk - \vq|^2)^2} \ .
\end{align}
As seen from these definitions, $p^{(1)}_{22}$ contains a singularity only at $\vq = 0$, while $p^{(2)}_{22}$ has a singularity only at $\vq = \vk$. 
By shifting $\vq \to \vk - \vq$ in $p^{(2)}_{22}$ and using the property $p^{(2)}_{22}(\vk - \vq, \vk) = p^{(1)}_{22}(\vq, \vk)$, which follows directly form the definition of $F_2$, we obtain
\begin{equation}
    \Pcal_{22}(k) = \int_{\qvec} 2 p^{(1)}_{22}(\vk, \vq)  \ . 
\end{equation}

We have now arrived at an integrand for the $ \Pcal_{22}$ diagram which has desired properties in the infrared. After the combination of partitions that we have just described, we can define the partitioned kernel   $F_{22}^{\rm part.}$ as 
\begin{equation}
\label{eq:F2_multi}
    F_{22}^{\rm part.} (\vq, \vk - \vq) \equiv \frac{2((|\vk - \vq|)^2)^2}{(q^2)^2 + ((|\vk - \vq|)^2)^2} \l F_2 (\vq, \vk - \vq) \r^2 \ ,
\end{equation}
so that the complete integrand reads
\begin{equation}
\label{eq:p22_multi}
    p_{22}^{\rm part.}(\vq, \vk) = 2\, F_{22}^{\rm part.} (\vq, \vk - \vq) \Plin(q) \Plin(k - q) \ .
\end{equation}
This integrand has only one singular denominator factor as $\vq \rightarrow 0$. The singularity 
is manifestly suppressed by casting the integration measure in spherical coordinates.  After applying this procedure, the IR singularities in all one-loop diagrams, namely 
the $p_{13}^{\rm UV-reg.}$ and $p_{22}^{\rm part.}$ diagrams, are located at $\vq = 0$. In the sum of diagrams, these singularities now completely cancel. 

In conclusion, we have constructed the following integrand for the one-loop power spectrum
\begin{equation}
\label{eq:1L_lin_IRUV_integrand}
    p^{\rm UV\&IR-reg.}_{\rm 1-loop}(\vk, \vq) = p^{\rm UV-reg.}_{13}(\vk, \vq) + p_{22}^{\rm part.}(\vk, \vq) \ .
\end{equation}
This integrand, when combined with the integration measure in spherical coordinates, is locally smooth in  both UV and IR limits. The integration can then be readily carried out numerically.

\subsection{Infrared cancellations at two loops}

\begin{table}[H]
    \centering
    \begin{tabular}{c|c|c}
                      & Single IR regions & Double IR regions \\
              \hline
         $\Pcal^{\rm UV-reg.}_{15}$ & $\begin{array}{c}\{\vq \to 0, \vp \neq 0\} \, [-1] \\ \{\vq \neq 0, \vp \to 0 \} \, [-1] \end{array}$ & $\{(\vq, \vp) \to (0,0)\} \, [-2]$ \\ 
            \hdashline
         $\Pcal^{\rm UV-reg.}_{42}$ & $\begin{array}{c}\{\vq \to 0,  \vp \neq 0\} \, [-1] \\ \{\vq \to \vk,  \vp \neq 0\} \, [-1]\\ \{\vq \neq 0 , \vp \to 0 \} \, [-1] \end{array}$ & $ \begin{array}{c} \{(\vq, \vp) \to (0,0)\} \, [-2] \\ \{(\vq, \vp) \to (\vk,0)\} \, [-2] \end{array}$ \\ 
            \hdashline
         $\Pcal_{33}^{\rm (II)}$ & $\begin{array}{c}\{\vq \to 0,  \vp \neq 0, \vq + \vp \neq \vk \} \, [-1] \\ \{ \vq \neq 0, \vp \to 0, \vq + \vp \neq \vk \} \, [-1] \\ \{\vq \neq 0 , \vp \neq 0, \vq + \vp \to \vk\} \, [-1] \end{array}$ & $ \begin{array}{c} \{(\vq, \vp) \to (0,0)\} \, [-2] \\ \{(\vq, \vp) \to (\vk,0)\} \, [-2] \\ \{(\vq, \vp) \to (0,\vk)\} \, [-2] \end{array}$ \\
            \hdashline
         $\Pcal_{33}^{(\rm I), \rm UV-reg.}$ & $\begin{array}{c}\{\vq \to 0,  \vp \neq 0\} \, [-1] \\ \{\vq \neq 0, \vp \to 0 \} \, [-1] \end{array}$ & $\{(\vq, \vp) \to (0,0)\} \, [-2]$ \\ 
    \end{tabular}
    \caption{ \small Relevant single and double IR limits for the two-loop diagrams. In brackets we include the infrared degree of divergence of the integrand \textit{excluding the integration measure} in the specified location.}
    \label{tab:IR_2L}
\end{table}

{We can construct an infrared-safe integrand at two-loops by following steps analogous to our one-loop construction.} First, we identify IR-divergent regions by a power-counting analysis, scaling the loop momenta. At two loops, singularities can arise from both ``single'' and ``double'' limits, where either one or both loop momenta (or combinations thereof) approach singular configurations. These are summarized in \tabref{tab:IR_2L}.
We can see that the diagrams $\Pcal_{42}$ and $\Pcal_{33}^{\rm (II)}$ have single infrared singularities that are not located at $\vq = 0$ or $\vp = 0$, and double singularities that are not located at the origin $(\vq, \vp) = (0,0)$.

The procedure for remapping all infrared singularities to the same locations is similar to the one-loop case. First, we decompose each integrand into channels using a partition of unity, isolating each singularity. Second, we apply linear transformations to shift all singularities to the same region. Then, we recombine all the channels into one integrand.
This process results in the partitioned kernels 
\begin{align}
\label{eq:F42_multi}
    F_{42}^{\rm UV-reg., part.} (\vq, \vp, \vk) &= \frac{ 2 (|\vk - \vq|^2)^3}{(q^2)^3 + (|\vk - \vq|^2)^3} F_4 (- \vq, \vq - \vk, \vp, - \vp) F_2(\vq, \vk - \vq) \ , \\
\label{eq:F33II_multi}
    F_{33}^{\rm (II), part.} (\vq, \vp, \vk) &= \frac{ 3 (|\kvec-\qvec-\pvec|^2)^2}{(q^2)^2 + (p^2)^2 + (| \kvec-\qvec-\pvec|^2)^2} \times \nonumber \\
    & \qquad \quad \times \, F_3 (\vq, \vp, \vk - \vq - \vp)  F_3 (- \vq, - \vp, -\vk + \vq + \vp) \ .
\end{align}
The corresponding integrands are
\begin{align}
    p_{42}^{\rm UV-reg., part.}(\vq, \vp, \vk) &= 24 \, F_{42}^{\rm UV-reg., part.} (\vq, \vp, \vk) \Plin(q)\Plin(p)\Plin(|\kvec - \qvec|) \ , \\
    p_{33}^{(\rm II), part.}(\vq, \vp, \vk) &= 6 \, F_{33}^{\rm (II), part.} (\vq, \vp, \vk) \Plin(q)\Plin(p)\Plin(|\kvec - \qvec - \pvec|) \ .
\end{align}
These reproduce the original diagrams upon integration, and all IR singularities are now mapped to the same region as those in $\Pcal_{15}$ and $\Pcal_{33}^{\rm (I)}$. 

Finally, we construct the full two-loop integrand
\begin{eqnarray}
\label{eq:2L_lin_IRUV_integrand}
   && q^2 p^2 p_{\rm 2-loop}^{\rm UV\&IR-reg.}(\vk,\vq,\vp) =q^2 p^2\left( p_{33}^{\rm (I), UV-reg.}(\vk,\vq,\vp) + p_{33}^{\rm (II), part.}(\vq, \vp, \vk) +\right. \nonumber\\
        &&\qquad\qquad \left.+ p_{42}^{\rm UV-reg., part.}(\vq, \vp, \vk) + p_{15}^{\rm UV-reg.}(\vk,\vq,\vp)\right)\ ,
\end{eqnarray}
The above integrand is locally free of both UV and IR divergences throughout the whole integration domain, making it fully suitable for stable numerical integration. 
Moreover, in $p_{\rm 2-loop}^{\rm UV\&IR-reg.}$, {\it i.e.}, after factoring out the $q^2p^2$ from the measure, the leading singularity in the double soft limit $(\vq, \vp) \to (0,0)$ is manifestly cancelled.  The partition of unity method used here is also applied to the integrals appearing in the two-loop EFT counterterms defined in Eq.~\eqref{eq:2LEFTCT_diagrams}. Details are provided in \appref{subapp:2-loop_count_IR}.

\section{Numerical evaluation of the power spectrum through two loops}
\label{sec:NumericalMethod}

We have now derived the integrands $p_{\rm 1-loop}^{\rm UV\&IR-reg.}$ in Eq.~\eqref{eq:1L_lin_IRUV_integrand} and $p_{\rm 2-loop}^{\rm UV\&IR-reg.}$ in Eq.~\eqref{eq:2L_lin_IRUV_integrand} that contribute to the no-counterterm part of the one-loop and two-loop corrections in the EFTofLSS.  The resulting integrands are locally free of infrared singularities and have their ultraviolet divergences subtracted, making them well-suited for stochastic numerical integration. The corresponding integrals, complemented with {finite remainders of} counterterms from the EFTofLSS, provide perturbative corrections through two-loop order for the dark-matter power spectrum. 
The numerical coefficients of the EFT counterterms can eventually be determined through a phenomenological comparison with data from surveys of large scale structure~\cite{DAmico:2022osl} or with a comparison against predictions from N body numerical simulations~\cite{Carrasco:2013mua}.  
As these investigations are beyond the scope of the current publication, we will for the moment focus on the no$-$CT contributions to the EFTofLSS one and two-loop corrections, the latter being numerically the most challenging, and we will evaluate the contribution from the counterterm diagrams later on.
With this assumption, in what follows, we will focus on obtaining numerical results for the perturbative corrections to the power spectrum originated solely from the 
UV\&IR-regulated integrals of Eqs.~\eqref{eq:1L_lin_IRUV_integrand} and~\eqref{eq:2L_lin_IRUV_integrand}.

We define the leading order (LO), next-to-leading-order (NLO) and next-to-next-to-leading-order (NNLO) no-CT power spectra as
\begin{eqnarray}
\label{eq:Plo}
    \Plo(k , a) &\equiv& \mathcal{D}(a)^2 \Plin(k) \ ,  \\ 
    \label{eq:Pnlo}
    \Pnlo(k , a) &\equiv&  \mathcal{D}(a)^2 \Plin(k) +  \mathcal{D}(a)^4 \Pcal^{\rm UV\&IR-reg.}_{\rm 1-loop}(k) \ ,  \\
\label{eq:Pnnlo}
    \Pnnlo(k , a) &\equiv& \mathcal{D}(a)^2 \Plin(k) + \mathcal{D}(a)^4 \Pcal^{\rm UV\&IR-reg.}_{\rm 1-loop}(k) + \mathcal{D}(a)^6 \Pcal^{\rm UV\&IR-reg.}_{\rm 2-loop}(k) \ , 
\end{eqnarray}
where we defined the UV\&IR$-$reg. integrals
\begin{align}
\begin{split}\label{uvirintdefs}
    \Pcal^{\rm UV\&IR-reg.}_{\rm 1-loop}(k) &= \int_{\qvec} p_{\rm 1-loop}^{\rm UV\&IR-reg.}(\vk, \vq) \ ,  \\
    \Pcal^{\rm UV\&IR-reg.}_{\rm 2-loop}(k) &= \int_{\qvec, \pvec}  p_{\rm 2-loop}^{\rm UV\&IR-reg.}(\vk, \vq, \vp) \ .  
\end{split}
\end{align}
We stress that for the correct prediction from the EFTofLSS, counterterms should be added to \eqn{eq:Pnlo} and \eqn{eq:Pnnlo} as in \secref{sec:check_renormalization}.  For later convenience, to assess the convergence of the perturbative series, we define the K-factors 
\begin{eqnarray}
    \Knlo(k , a) \equiv \frac{\Pnlo(k , a)}{\Plo(k, a)} \ , 
    \quad 
    \Knnlo(k , a) \equiv \frac{\Pnnlo(k , a)}{\Plo(k , a)} \ . \label{eq:KNNLO}
\end{eqnarray}

\subsection{Numerical integration method}
\label{sec:numerical_integration}

To integrate numerically the UV\&IR-regulated integrands, we will use the Monte Carlo integration method. Specifically, we will use the VEGAS+ algorithm~\cite{Lepage:2020tgj} as implemented in the \texttt{python} library \texttt{vegas}~\cite{peter_lepage_2025_14834979}. To ensure fast evaluation times, we write the integrands in \texttt{Cython}.
We set \texttt{vegas} to generate points in the $[0,1]^{3r}$ hypercube, where $r$ is the number of loops of the integral. Then, we use the parametrization in \cite[Sec.~6.2]{Capatti:2019edf} to map the integration {in $[0,1]^{3r}$} to { $r$-copies} of spherical coordinates, one for each loop momenta
\begin{equation}
    \int_{\qvec} = \frac{1}{(2 \pi)^{3}} \int_0^\infty q^2 dq \int_0^{2 \pi} d \phi \int_{-1}^{1} d \cos \theta \ , 
\end{equation}
where $\theta$ and $\phi$ are the polar and azimuthal angles, respectively.

The numerical evaluation of $F_5^{ \rm UV-reg.}$ can require a higher precision than the default \texttt{double} precision due to potential precision loss. This occurs when the loop momenta approach certain limits where denominators become small, while lengthy numerators exacerbate numerical instability. To circumvent this issue without resorting to higher precision arithmetic, which can be several orders of magnitude slower, we analytically expand the expression of $F_5^{ \rm UV-reg.}$ in these regions. This produces a significantly simplified form that maintains accuracy within \texttt{double} precision introducing only a small error, which remains negligible as long as the regions where expanded expressions are applied are sufficiently small. We use these expanded expressions only when the loop momenta exceed a certain threshold empirically determined.

The linear power spectra for any cosmological model can be obtained using CLASS~\cite{Blas:2011rf}, that provides $N$ discrete data points over the range:
\begin{equation}
\label{eq:class_range}
k_{{\rm min}} \leq k \leq k_{{\rm max}}       \ , 
\end{equation}
where we set
\begin{equation}
\label{eq:class_range_values}
k_{{\rm min}}=10^{-5} \, \um \ ,  \quad  k_{{\rm max}}=1.3 \, \um \,, \quad  N = 10^3 \ . 
\end{equation}
As we have discussed above, loop integrations extend from $k=0$ to $k = \infty$. To perform the integrations beyond $k_{{\rm max}}$, we impose the scaling:
\begin{equation}
\label{eq:Plin_scaling}
    \Plin(k) = \Plin(k_{{\rm max}}) \left(\frac{k_{{\rm max}}}{k}\right)^n
    \ ,  \quad k \in [ k_{{\rm max}},\infty)\ ,
\end{equation} 
where $n$ is a cosmology dependent exponent defined as $n \equiv -\frac{d\ln \mathcal{P}_{\rm lin.} (\ln k)}{d\ln k}\vert_{k = k_{{\rm max}}}$.  In the IR,  we extrapolate the integrand linearly,  
\begin{equation}
\label{eq:IRextrapolation}
\Plin(k) = \Plin(k_{{\rm min}}) \, \left( \frac{k}{k_{{\rm min}}} \right)\ , \quad  k \in [0, k_{{\rm min}}] \ .    
\end{equation}
These UV and IR scalings will be imposed for any cosmological model considered in the rest of this work.

\subsection{Magnitude of perturbative corrections}
\label{sec:K-factors}

\begin{figure}[t!]
    \centering
    \subfloat[]{
        \includegraphics[]{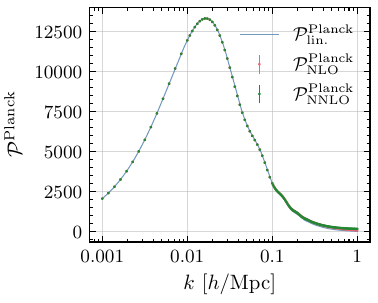}
    }
    \subfloat[]{
        \includegraphics[]{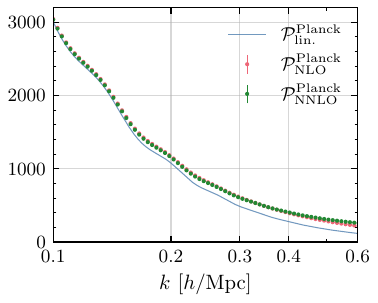}
    }
    \caption{\small The power spectrum (i) $\Plo^{\rm Planck}$ computed at leading order (linear contribution) (blue), (ii) 
    $\Pnlo^{\rm Planck}$ computed at  NLO which includes 
    the one-loop correction (red),  
    and (iii) $\Pnnlo^{\rm Planck}$ computed at NNLO which includes both the one-loop and two-loop corrections (green). In Fig.~(a) we show the scaling of the full power spectrum across the range $[10^{-3}, 1]$ $h/$Mpc, while in Fig.~(b) we focus on the $k$ range $[0.1, 0.6]$ $h/$Mpc, where the effect of the loop correction becomes more significant. Both figures are at redshift $z = 0.57$.
    }
    \label{fig:Plin0_lo+nlo+nnlo}
\end{figure}

We present here the result of the numerical evaluation for the case of the Planck linear power spectrum, corresponding to the blue curve of \figref{fig:Plin0_lo+nlo+nnlo}, which has parameters reported in Eq.~\eqref{eq:Planck_cosmological_parameters}.   In \figref{fig:Plin0_lo+nlo+nnlo}, in addition to the leading order linear power spectrum, we present the power spectrum at NLO $\Pnlo^{\rm Planck}$ and at NNLO $\Pnnlo^{\rm Planck}$. The power spectrum and its perturbative corrections span a wide range of values over several orders of magnitude. 

In \figref{fig:Plin0-k-factors}, we plot the NLO and NNLO K-factors that give a measure of the importance of the loop corrections.
We observe that the NLO K-factor becomes a few percent for wavenumbers larger than $k \sim 0.1 \, h/{\rm Mpc}$, and the effects of the two-loop correction over the one-loop correction start to becomes a few percent at wavenumbers $k\sim 0.4 \, h/{\rm Mpc}$. We see that around $k \sim 0.5 \, h/{\rm Mpc}$, the total perturbative correction is becoming uncomfortably large.  However, this statement should be taken at ${\cal{O}}(1)$, as the counterterms, which are not included here and whose numerical value should be taken from data, indeed typically change the total NLO and NNLO contributions of the EFTofLSS in these regimes by ${\cal{O}}(1)$.}  

\begin{figure}[t!]
    \centering
    \subfloat[]
    {
        \includegraphics[]{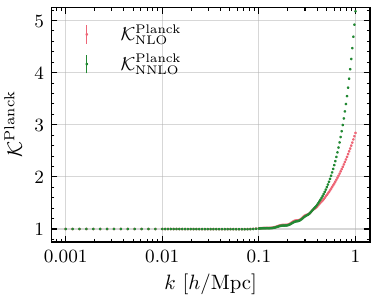}
        \label{subfig:Planck_k_factors_full}
    }
    \subfloat[]{
        \includegraphics[]{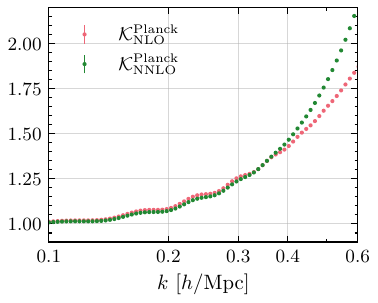}
        \label{subfig:Planck_k_factors_focus}
    }
    \caption{\small The K-factors for the power spectrum. The NLO K-factor $\Knlo^{\rm Planck} = \frac{\Pnlo^{\rm Planck}}{\Plo^{\rm Planck}}$  (red), and NNLO K-factor $\Knnlo^{\rm Planck} = \frac{\Pnnlo^{\rm Planck}}{\Plo^{\rm Planck}}$ (green) are shown. Both figures are at redshift $z = 0.57$.
    }
    \label{fig:Plin0-k-factors}
\end{figure}

\section{A perturbative expansion around a standard cosmology model}
\label{sec:perturbative_expansion}

In this work, we aim to efficiently compute the perturbative corrections for a potentially large number of cosmological models, each identified by a different combination of cosmological parameters. This is needed because typically many such combinations, yielding different linear power spectra, give a reasonable fit to data and therefore need to be evaluated. In Eqs.~\eqref{eq:Plo}-\eqref{eq:KNNLO}, the dependence of the power spectrum on cosmological parameters is introduced via the (leading order) linear solution $\Plin(k)$, which serves as an integration kernel for the loop corrections, plus the dependence of the counterterm coefficients, which depend on the cosmology. In this paper we focus on the loop corrections, and therefore focus on the cosmological dependence of $\Plin(k)$, as the one of the counterterm coefficients easily factors out. 

We label the cosmological models and the associated combination of parameters with a collective index $j$. The corresponding linear power spectrum is denoted as $\Plin^{[j]}{(k)}$ and the loop-corrected no-counterterm power spectrum is given by:
\begin{eqnarray}
\label{eq:PS-loopexpansion-model}
    \Pcal^{[j]}_{\rm no-CT}(k , a) =   \mathcal{D} (a)^2 \Plin^{[j]}(k) +  \mathcal{D} (a)^4 \Pcal_{\rm 1-loop}^{[j]}(k) +  \mathcal{D} (a)^6 \Pcal_{\rm 2-loop}^{[j]}(k)+ \ldots  \ ,  
\end{eqnarray}
where 
\begin{eqnarray} \label{eq:Pr-loop_jmodel}
\Pcal_{r\rm-loop}^{[j]}(k) \equiv \Pcal^{\rm UV\&IR-reg.}_{r-\text{loop}}   \left[ \Plin^{[j]} \right](k)      \ , 
\end{eqnarray}
and $\Pcal_{r\rm-loop}^{[j]}$ for $r = 1,2$ are defined in \eqn{uvirintdefs}.  We define the power spectrum $\Plo^{[j]}$, $\Pnlo^{[j]}$, $\Pnnlo^{[j]}$ at LO, NLO and NNLO  for each model $j$, as in Eqs.~\eqref{eq:Plo}-\eqref{eq:Pnnlo}.  Since physically acceptable models will need to be compatible with the same data, which by now are very precise, the shapes of the various $\Plin^{[j]}$ are not expected to vary significantly.  As such, the Planck linear power spectrum, that we will denote with $j=0$, is expected to be a good approximation of the linear power spectrum in other models (with $j \neq 0$), so we choose to expand around this.

We define the difference between the $j$-th and Planck linear power spectra as
\begin{equation}
\label{eq:DeltaP_definition}
\DeltaP_{\rm lin.}^{[j]}(k) = \Plin^{[j]}(k) - {\cal N}^{[j]}  \, \Plin^{[0]}(k) \ , 
\end{equation}
and we choose 
\begin{eqnarray}
{\cal N}^{[j]} = \frac{\max(\Plin^{[j]})}{\max(\Plin^{[0]})} \ . 
\end{eqnarray}
Here, ${\cal N}^{[j]}$ rescales the Planck power spectrum by the ratio of the maximum values of $\Plin^{[j]}$ and $\Plin^{[0]}$. The term $\DeltaP_{\rm lin.}^{[j]}(k)$ captures the shape difference between the rescaled $j$-th power spectrum and the Planck power spectrum. For realistic cosmologies, the contribution of $\DeltaP_{\rm lin.}^{[j]} (k)$ to the loop integrals is typically small.

As an example of the procedure outlined in the beginning of the section, we consider a second set of parameters for the $\Lambda$CDM model, labelled $j=1$. The parameters read 
\begin{align}
\begin{split}
    h^{[1]} &= 0.662 \ ,   \\
    \omega_{b}^{[1]}  &= 0.02193 \ ,  \\
    \omega_{\rm cdm}^{[1]}  &=  0.1173 \ ,  \\
    \ln 10^{10}A_s^{[1]} &=   2.909 \ , \\
    n_s^{[1]} &= 1.043 \ ,  \\
    \sum m_{\nu_i}^{[1]} &= 0 \ . 
\end{split}
\end{align}
Throughout, we evaluate this model at $z=0$.
We have chosen this second model $j=1$ to yield significant differences with respect to the shape of the Planck linear power spectrum.  We observe in the lower panel of \figref{subfig:DeltaP} that, even in this scenario, the relative contribution of $\Delta \mathcal{P}_{\rm lin.}^{[1]}(k)$ remains small: it is below $\sim 30\%$ throughout the $k$ range where loops are significant.  

\begin{figure}[t!]
    \centering
\subfloat[The Planck linear power spectrum $\Plin^{[0]}$ rescaled by ${\cal N}^{[j]}$ and the linear power spectrum $\Plin^{[j]}$ for the $j=1$ model which is expected to have quite substantial differences. The power spectrum $\Plin^{[1]}$ is evaluated at redshift $z = 0$.]{
    \includegraphics[width=0.45\textwidth]{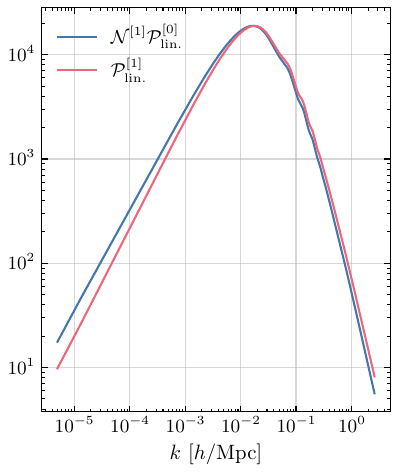}
    \label{subfig:Plin0vsPlin1_nofit}
}
\hfill
\subfloat[
The part of the linear power spectrum  $\DeltaP_{\rm lin.}^{[1]}(k)$, defined in Eq.~\eqref{eq:DeltaP_definition}, which is not accounted for by rescaling the Planck linear power spectrum. The value of  $\DeltaP_{\rm lin.}^{[1]}(k)$ contribution is plotted in the upper part. In the lower part we plot the relative contribution $\vert \DeltaP_{\rm lin.}^{[1]}(k)/ \Plin^{[1]}(k) \vert$.  We evaluate $\DeltaP_{\rm lin.}^{[1]}$ at the same redshift as the 
$j=1$ cosmology, here $z=0$. The ratio in the bottom panel is redshift independent. 
]{
    \includegraphics[width=0.45\textwidth]{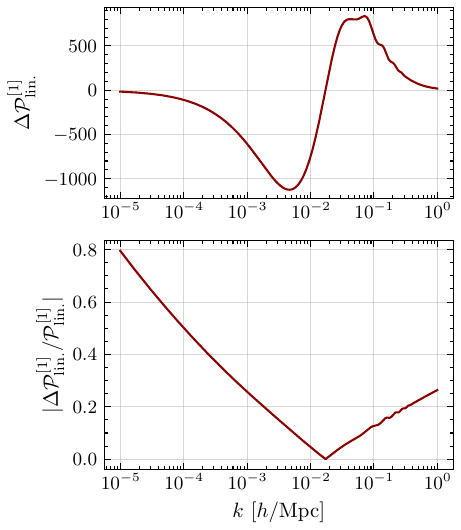}
    \label{subfig:DeltaP}
}
\caption{ \small The relative contribution of $\DeltaP_{\rm lin.}^{[1]}$
to $\Plin^{[1]}$ is small in the range of interest for our loop computation.}
    \label{fig:datasets}
\end{figure}

As it can be observed from the diagrammatic expressions, the one-loop corrections are at most quadratic in $\DeltaP_{\rm lin.}^{[j]}$ and the two-loop corrections are at most cubic in $\DeltaP_{\rm lin.}^{[j]}$.  To exploit the smallness of $\DeltaP_{\rm lin.}^{[j]}$, we expand the one-loop and two-loop corrections as:
\begin{eqnarray}
\label{eq:DPexpansion1loop}
    \Pcal_{\rm 1-loop}^{[j]} = 
    \l \mathcal{N}^{[j]} \r^2 \Pcal_{\rm 1-loop}^{\rm Planck}
    +  \mathcal{N}^{[j]} \, \Delta_{1} \Pcal_{\rm 1-loop}^{[j]}
    + \Delta_{2} \Pcal_{\rm 1-loop}^{[j]}  \ , 
\end{eqnarray}
and
\begin{eqnarray}
\label{eq:DPexpansion2loop}
    \Pcal_{\rm 2-loop}^{[j]} = 
    \l \mathcal{N}^{[j]}\r^3 \Pcal_{\rm 2-loop}^{\rm Planck}
    + \l \mathcal{N}^{[j]} \r^2 \Delta_1 \Pcal_{\rm 2-loop}^{[j]}
    + \mathcal{N}^{[j]} \, \Delta_2 \Pcal_{\rm 2-loop}^{[j]}
    +\Delta_3 \Pcal_{\rm 2-loop}^{[j]} \ .
\end{eqnarray}
In this expansion, $\Delta_m \Pcal_{r\rm-loop}^{[j]}$ represents the $r-$loop correction when $\DeltaP_{\rm lin.}^{[j]}$ from Eq.~\eqref{eq:DeltaP_definition} is substituted into the integrand, retaining terms with $m$ factors of $\DeltaP_{\rm lin.}^{[j]}$. In other words, $\Delta_m \Pcal_{r\rm-loop}^{[j]}$ is proportional to $\left(\DeltaP_{\rm lin.}^{[j]}\right)^m$. 

Based on this structure, we expect $\Pcal_{\rm 1-loop}^{{\rm Planck}} $ and $\Pcal_{\rm 2-loop}^{{\rm Planck}}$ to account for approximately $60\%$ of the correction. We test this hypothesis with the full amplitudes at one and two loops. The results are shown in Figs.~\ref{fig:P1loopexpansion} and \ref{fig:P2loopexpansion}. The large spikes near $k \sim 0.09 \, \um$ at one-loop and $k \sim 0.4 \, \um$ at two-loop are due to the loop corrections vanishing rather than the contributions becoming large. Evidently, 
the sum of the leading and next-to-leading order in $\DeltaP_{\rm lin.}$
dominate. 
 The ratios in Figs.~\ref{fig:P1loopexpansion} and \ref{fig:P2loopexpansion} are redshift independent. Changing redshift rescales the power spectrum of model $j$ by a $k$-independent factor; since $\mathcal{N}^{[j]}$ is fixed by the maxima, it rescales by the same factor. Thus, the common rescaling cancels in the ratios.

To achieve an accuracy better than $1\%$ in $\Pnnlo^{[j]}$, it is necessary to calculate terms up to $\Delta_2 \Pcal_{r\rm-loop}^{[j]}$, but the last term in \eqn{eq:DPexpansion2loop}, $\Delta_3 \Pcal_{\rm 2-loop}^{[j]}$, can be ignored. As we shall see in the next section, this term is also the most computationally challenging.  An expansion analogous to the one in Eqs.~\eqref{eq:DPexpansion1loop} and \eqref{eq:DPexpansion2loop} can be performed for the two-loop counterterms. This is carried out in App.~\ref{sec:2-loop_counterterms}, Eqs.~\eqref{eq:DPexpansion_I42}--\eqref{eq:DPexpansion_I15_2}.

\begin{figure}[t!]
\centering
  \includegraphics[]{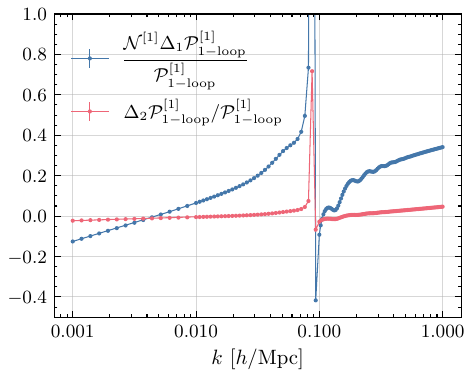}
    \caption{
\small   The ratios $\frac{\l \mathcal{N}^{[1]} \r \Delta_1 \Pcal_{\rm 1-loop}^{[1]}}{ \Pcal_{\rm 1-loop}^{[1]}}$ and 
    {$\frac{\Delta_2 \Pcal_{\rm 1-loop}^{[1]}}{ \Pcal_{\rm 1-loop}^{[1]}}$. We see that the $\Delta_2 \mathcal{P}^{[1]}_{1-\rm loop}$ contributes only a small correction relative to the full one-loop result across all $k$'s. The large spike near $k\sim0.1$ $h/$Mpc is due to the $\mathcal{P}^{[1]}_{1-\rm loop}$ correction vanishing.}
    }
    \label{fig:P1loopexpansion}
\end{figure}

\begin{figure}[t!]
    \centering
    \includegraphics[]{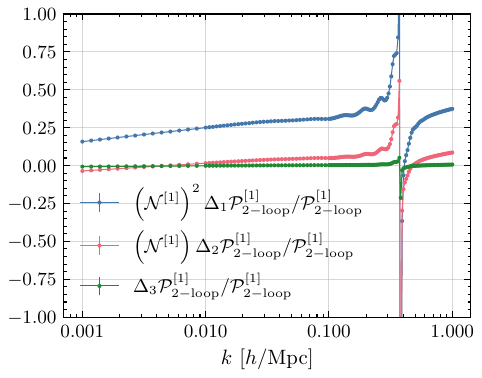}
    \caption{ \small 
    The ratios {   $\frac{\l \mathcal{N}^{[j]}\r^2 \Delta_1 \Pcal_{\rm 2-loop}^{[1]}}{ \Pcal_{\rm 2-loop}^{[1]}}$}, 
    {$\frac{ \l \mathcal{N}^{[j]}\r  \Delta_2 \Pcal_{\rm 2-loop}^{[1]}}{ \Pcal_{\rm 2-loop}^{[1]}}$}
    and     {    $\frac{\Delta_3 \Pcal_{\rm 2-loop}^{[1]} }{\Pcal_{\rm 2-loop}^{[1]}}$}. We see that the $\Delta_2 \mathcal{P}^{[1]}_{2-\rm loop}$ and $\Delta_3 \mathcal{P}^{[1]}_{2-\rm loop}$ terms contribute only a small correction relative to the full two-loop integral across all $k$'s. In particular, we confirm that $\Delta_3 \mathcal{P}^{[1]}_{2-\rm loop}$ is negligible over most of the displayed range. The sharp spike near $k\sim0.4$ $h/$Mpc arises from the vanishing of $\mathcal{P}^{[1]}_{2-\rm loop}$. }
    \label{fig:P2loopexpansion}
\end{figure}

\section{Decoupling the dependence on cosmological models and integrations}
\label{sec:Pfit}

The method explained in the previous section allows us to calculate loop corrections to any $j \neq 0$ cosmology {as a perturbation around} the Planck ($j=0$) cosmology, expanding in $\Delta \Pcal^{[j]}_{\rm lin.}$.
We can now follow the approach described in \cite{Simonovic:2017mhp,Anastasiou:2022udy} to decouple the problem of computing loop corrections from the choice of cosmology by means of an {emulation of the linear power spectrum,  fitting it to predefined sets of functions which are agnostic to the cosmological model. A novelty in this article}  is that we only fit $\Delta \Pcal_{\rm lin.}^{[j]}$ in terms of the cosmology independent functions, {rather than the full linear power spectrum $\Pcal_{\rm lin}^{[j]}$ as was done in Refs.~\cite{Simonovic:2017mhp,Anastasiou:2022udy}}. The advantage of this line of action is that the error due to the fitting procedure will only affect the terms {$\Delta_m \Pcal_{r-{\rm loop}}^{[j]}$} with $m > 0$, which give a subleading contribution to the integrals.

The results of the analysis in the previous section confirm that the highest-order contribution in $\Delta \Pcal_{\rm lin.}^{[j]}$, namely $\Delta_3 \Pcal_{2-{\rm loop}}^{[j]}$, is negligible. We also know that the $\Delta_2 \Pcal_{r-{\rm loop}}^{[j]}$ contribution is significantly smaller than $\Delta_1 \Pcal_{r-{\rm loop}}^{[j]}$, although the $\Delta_2 \Pcal_{r-{\rm loop}}^{[j]}$ contribution must be retained to achieve our goal of sub-percent precision. However, while it is important to compute the larger $\Delta_1 \Pcal_{r-{\rm loop}}^{[j]}$ with a very small uncertainty relative to its magnitude, we can afford to compute the subleading $\Delta_2 \Pcal_{r-{\rm loop}}^{[j]}$ with a larger relative uncertainty. Consequently, we use a basis of only 16 fitting functions for $\Delta_2 \Pcal_{r-{\rm loop}}^{[j]}$, 
\begin{align}
   \DeltaP^{[j]}_{\rm lin.}(k) \to & 
   \DeltaP_{\rm lin.,fit, 2}^{[j]} (k) =  \frac{a_0^{[j], (2)}}{1 + \frac{k^2}{k_{\UV,0}^{2}}}+
    \sum_{i=1}^{3} a_i^{[j], (2)} \, f(k^2, k^{2}_{\peak,1}, k^{2}_{\UV, 1}, 0, i)\, \nonumber \\
    &+\sum_{i=0}^{3} \Big(b_i^{[j], (2)} \, f(k^2, k^{2}_{\peak,2}, k^{2}_{\UV, 2}, 1, i+1)  +
    c_i^{[j], (2)} \, f(k^2, k^{2}_{\peak,3}, k^{2}_{\UV, 3}, 0, i+2)  \nonumber \\
    & \quad \quad +d_i^{[j], (2)} \, f(k^2, k^{2}_{\peak,4}, k^{2}_{\UV, 4}, 0, i+1) \Big)  \ , 
\label{eq:Pfit_2}    
\end{align}
where we defined
\begin{equation}
    f(k^2, k^2_{\rm peak}, k^2_{\rm UV}, i, j) = \frac{(k^2/k_0^2)^i}{\left( 1 + \frac{(k^2 - k_{{\rm peak}}^{2})^2}{k^{4}_{\rm UV} }\right)^j} \ , 
\end{equation}
$k_0 = 1 / 20 \, h / \text{Mpc}$, and the sets of parameters $\{ a_i^{[j], (2)},  b_i^{[j], (2)},  c_i^{[j], (2)},  d_i^{[j], (2)} \}$ with $i = 0, \dots, 3$ are determined by fitting $\Delta \Pcal^{[j]}$. We also have:
\begin{align*}
		& k^2_{\rm UV, 0} = 1 \times 10^{-4} \ h^2/{\rm Mpc}^2\ , \\
		& k^2_{\rm peak, 1} = -3.4 \times 10^{-2} \ h^2/{\rm Mpc}^2 \ ,\quad k^2_{\rm UV, 1} = 6.9 \times 10^{-2} \ h^2/{\rm Mpc}^2 \ , \\
		& k^2_{\rm peak, 2} =-1 \times 10^{-3}  \ h^2/{\rm Mpc}^2\ ,\quad k^2_{\rm UV, 2} = 8.2 \times 10^{-3} \ h^2/{\rm Mpc}^2 \ , \\
		& k^2_{\rm peak, 3} = -7.6 \times 10^{-5}  \ h^2/{\rm Mpc}^2\ ,\quad k^2_{\rm UV, 3} = 1.3 \times 10^{-3}   \ h^2/{\rm Mpc}^2 \ ,\\
		& k^2_{\rm peak, 4} = -1.56 \times 10^{-5} \ h^2/{\rm Mpc}^2 \ ,\quad k^2_{\rm UV, 4} = 1.35 \times 10^{-5} \ h^2/{\rm Mpc}^2\ .
\end{align*}	
When $\Delta \Pcal_{\rm lin.}^{[j]}$ enters the more significant $\Delta_1 \Pcal_{1-{\rm loop}}^{[j]}$ and $\Delta_1 \Pcal_{2-{\rm loop}}^{[j]}$, we emulate it more precisely using the following basis of 24 fitting functions:
\begin{align}
\begin{split}
\Delta \Pcal_{\rm lin.}^{[j]}(k) \to & \DeltaP_{\rm lin.,fit, 1}^{[j]}(k) =  \DeltaP_{\rm lin.,fit, 2}^{[j]}(k) + \sum_{i = 1}^8 e_i^{[j], (1)} \, g(k, k^{(i)}_{\rm mean}, k_{\rm width}^2, i)  \ , \\
\end{split}
\label{eq:Pfit_1}    
\end{align}
where,
\begin{align}
    g(k, k_{\rm mean}, k_{\rm width}^2, i) = \frac{1}{\frac{(k - k_{\rm mean})^2}{k_{\rm width}^2} + 1} \ , 
\end{align}
with $k_{\rm width}^2 = 1 \times 10^{-4} \ h^2/{\rm Mpc}^2$ and
\begin{align}
\begin{split}
    k^{(0)}_{\rm mean} = 0.0848 \ h/{\rm Mpc} \ , \\
    k^{(1)}_{\rm mean} = 0.141 \ h/{\rm Mpc}\ , \\
k^{(2)}_{\rm mean} = 0.113 \ h/{\rm Mpc} \ , \\
k^{(3)}_{\rm mean} = 0.175 \ h/{\rm Mpc}\ , \\
k^{(4)}_{\rm mean} = 0.205 \ h/{\rm Mpc}\ , \\
k^{(5)}_{\rm mean} = 0.236 \ h/{\rm Mpc} \ , \\
k^{(6)}_{\rm mean} = 0.269 \ h/{\rm Mpc}\ , \\
k^{(7)}_{\rm mean} = 0.155 \ h/{\rm Mpc}\ .
\end{split}
\end{align}
We will indicate explicitly in our later formulae when one of these two fitting functions $\Delta \Pcal^{[j]}_{{\rm lin.,fit}, m}$ is used to compute loop corrections instead of the exact $\Delta \Pcal_{\rm lin.}^{[j]}$, with the  subscript ``fit.''

The two emulations, $\Delta \Pcal^{[j]}_{{\rm lin., fit}, 1}$ and $\Delta \Pcal^{[j]}_{{\rm lin.,fit}, 2}$, as well as their relative difference to the exact 
 $\Delta \Pcal_{\rm lin.}^{[j]}$, are shown in \figref{fig:DPvsPfit} for the cosmological model $j=1$. Clearly, $\Delta \Pcal_{\rm lin.,fit,1}^{[1]}$ fits $\Delta \Pcal_{\rm lin.}^{[1]}$ better than $\Delta \Pcal_{\rm lin.,fit,2}^{[1]}$, and thus the contribution {$\Delta_1 \Pcal_{r-{\rm loop}}^{[1]}$} will be evaluated more accurately. {As we discussed}, the lower accuracy of {$\Delta_2 \Pcal_{r-{\rm loop}}^{[1]}$} will have a negligible impact on the full power spectrum as it is {subleading relatively to $\Delta_1 \Pcal_{r-{\rm loop}}^{[1]}$}.

\begin{figure}[t!]
\centering
    \includegraphics[]{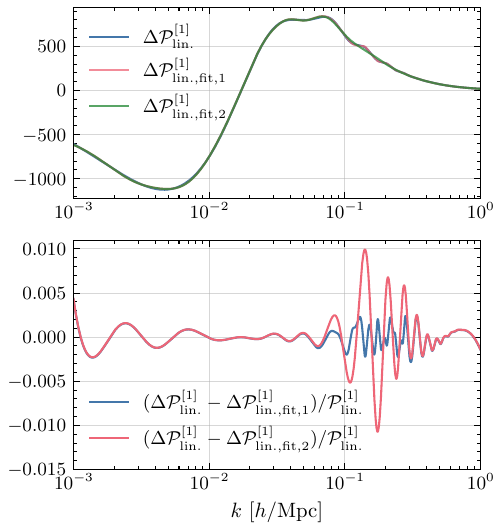}

\caption{\small Comparison between the exact and fitted differences in the linear power spectra.
On the top panel, difference between the linear power-spectra of the two cosmologies, $\DeltaP_{\rm lin.}^{[1]}$ (blue), and fitted approximations $\DeltaP_{\rm lin., fit, 1}^{[1]}$ (red) and  $\DeltaP_{\rm lin., fit, 2}^{[1]}$ (green). 
Below, relative residual between the interpolated $\DeltaP_{\rm lin.}^{[1]}$, at redshift $z=0$,  and the fitted function $\DeltaP_{\rm lin., fit, 1}^{[1]}$ (blue) and $\DeltaP_{\rm lin., fit, 2}^{[1]}$ (red). While the residual is always small, we see that $\Delta \mathcal P_{\rm lin., fit, 1}$ is more accurate than $\Delta \mathcal P_{\rm lin., fit, 2}$.
}
\label{fig:DPvsPfit}
\end{figure}

{Our numerical method provides significant freedom in selecting fitting functions. While the default set presented here allows for a faithful emulation of the linear power spectrum and an efficient numerical integration, alternative functions may offer opportunities for further optimization.}

\subsection{$\Delta \Pcal_{\rm lin.}^{[j]}(k)$ outside integration kernels}

Replacing the exact $\DeltaP_{\rm lin.}^{[j]}$ with the emulated $\DeltaP_{\rm lin.,  fit}^{[j]}$ in the loop corrections $\Delta_m \Pcal_{r\rm{-loop}}$ indroduces a theoretical uncertainty.  
We observe that when $\DeltaP_{\rm lin., fit}^{[j]}$ depends on a loop momentum and it gets convoluted in the loop integration, the uncertainty is small. 
The reason for this is that the difference between the fitted and exact functions, $\DeltaP_{\rm lin.,fit}^{[j]}(q) - \DeltaP^{[j]}_{\rm lin.}(q)$, is generally an oscillating function of the loop momentum $q$. Upon integration, the positive and negative contributions tend to cancel to a large degree.

Conversely, the replacement $\DeltaP_{\rm lin.}^{[j]}(k) \to \DeltaP_{\rm lin., fit}^{[j]}(k)$ leads to significantly higher errors when its argument is the external wavenumber $k$, which is independent of the loop momenta. In this case, $\DeltaP_{\rm lin., fit}(k)$ is a factor outside the loop integral, and the beneficial cancellation of positive and negative fitting deviations no longer occurs. The relative error of the approximation can therefore become large, particularly in regions where the loop corrections $\Delta_m \mathcal{P}_{r\rm{-loop}}$ themselves are small. It is crucial to note that this is a systematic error inherent to the fit's accuracy, not a statistical one from the Monte Carlo integration, and can only be mitigated by improving the emulation.

A straightforward solution is to use the exact $\DeltaP_{\rm lin.}^{[j]}(k)$ when its argument is independent of loop momenta. To achieve this, we decompose each component of the loop corrections, $\Delta_m \mathcal{P}_{r\text{-loop}}$, into a part proportional to $\DeltaP_{\rm lin.}^{[j]}(k)$ and a remainder,
\begin{equation}
\label{eq:delta_m_final_splitting}
\Delta_m \mathcal{P}^{[j]}_{r\rm{-loop}} (k) = \Delta_{m,0} \mathcal{P}^{[j]}_{r\rm{-loop}} (k) + \DeltaP_{\rm lin.}^{[j]} (k) \Delta_{m,1}   \mathcal{P}^{[j]}_{r\rm{-loop}}  (k) \ , 
\end{equation}
where $\Delta_{m,0} \mathcal{P}^{[j]}_{r\rm{-loop}}  (k) $ contains all terms independent of $\DeltaP_{\rm lin.}^{ [j]} (k) $, and $\Delta_{m,1} \mathcal{P}^{[j]}_{r\rm{-loop}} (k) $ contains the rest.\footnote{\eqn{eq:delta_m_final_splitting} is the general expression for any loop order, since there can only ever be up to one factor of $\DeltaP_{\rm lin.}^{[j]}  (k)$, where $k$ is the external wavenumber.  To see this, consider contracting $\delta^{(n)}$ and $\delta^{(2 r +2-n)}$ to form an $r$-loop term.  To get a factor of  $\DeltaP_{\rm lin.}^{[j]}  (k)$, we must contract only one $\delta^{(1)}$ from $\delta^{(n)}$ and one $\delta^{(1)}$ from $\delta^{(2 r +2-n)}$; if we contract more legs between $\delta^{(n)}$ and $\delta^{(2 r +2-n)}$, this would form a loop and would not have a factor of $\DeltaP_{\rm lin.}^{[j]}  (k)$.  } The replacement of $\DeltaP_{\rm lin.}^{[j]}$ with $\DeltaP^{[j]}_{\rm lin., fit}$ is then performed only within the integrals defining $\Delta_{m,\sigma} \mathcal{P}^{[j]}_{r\rm{-loop}}$ (for $\sigma = 0,1$), i.e. the explicit $\DeltaP_{\rm lin.}^{ [j]}  (k) $ in \eqn{eq:delta_m_final_splitting} is kept as the exact function in \eqn{eq:DeltaP_definition}.
This finer decomposition yields cosmology-independent integrals with a very small theoretical error from the emulation, as will be discussed in the next section.

The integrals on the right-hand side of Eq.~\eqref{eq:delta_m_final_splitting} exhibit infrared cancellations, as described in \secref{sec:IR}. These cancellations are realized perfectly only if the $\DeltaP_{\rm lin.}^{[j]}(k)$ factor in the second term and the $\DeltaP_{\rm lin.}^{[j]}$ functions within the loop integrands have identical functional forms. Using an emulated function, $\DeltaP_{{\rm lin., fit},m}^{[j]}$, in the loop integrands disrupts these delicate cancellations.  However, this problem is effectively mitigated. The IR-safe integrands from \secref{sec:IR} confine any potential infrared enhancements to the origins of the loop momenta ($q=0$ and/or $p=0$). Infrared enhancements, including those induced by the emulation of the linear power spectrum, are fully integrable and their impact is manifestly suppressed by generating loop momenta in spherical coordinates. The numerical effects of these residual enhancements can then be controlled by increasing the Monte Carlo integration statistics.

\subsection{Numerical integration of cosmology independent integrals\label{sec:cosmology-independent-integrals}}

Both expansions for $\Delta \Pcal_{\rm lin.}^{[j]}$ in \eqn{eq:Pfit_2} and \eqn{eq:Pfit_1} can generally be written as 
\begin{equation}
    \Delta \Pcal_{{\rm lin., fit}, m}^{[j]}(k) = \sum_{i=1}^{N_m} \alpha_{i}^{[j], (m)} \, f_{i}^{(m)}(k) \  , 
\end{equation}
where $N_m$ is the number of basis functions used for the fit labeled by $m$, $\alpha_{i}^{[j],(m)}$ are cosmology-dependent fit coefficients, and $\{f_{i}^{(m)}(k)\}$ denotes the chosen set of basis functions.  As mentioned near \eqn{eq:Pfit_2} and \eqn{eq:Pfit_1}, in this work, we find that it is sufficient to use $\DeltaP_{\rm lin.,fit, 1}^{[j]}$ for $\Delta_{1,\sigma}\Pcal^{[j]}_{1\text{-loop}}$ and $\Delta_{1,\sigma}\Pcal^{[j]}_{2\text{-loop}}$, while we use $\DeltaP_{\rm lin., fit, 2}^{[j]}$ for $\Delta_{2,\sigma}\Pcal^{[j]}_{1\text{-loop}}$ and $\Delta_{2,\sigma}\Pcal^{[j]}_{2\text{-loop}}$.  Substituting $\DeltaP_{\rm lin.}^{[j]} \to \Delta \Pcal_{{\rm lin., fit}, m}^{[j]}$ into the definition of $\Delta_{m,\sigma}\Pcal^{[j]}_{r\text{-loop}}$ \eqn{eq:delta_m_final_splitting} for $m = 1,2$ gives
\begin{equation}
\label{eq:loop-tensor-integrals}
    \Delta_{m, \sigma} \Pcal_{r\rm-loop, fit}^{[j]}(k) = \frac{1}{(2\pi)^{3r}} \sum_{i_1, \cdots, i_{m-\sigma}  = 1}^{N_m} \alpha_{i_1}^{[j], (m)} \cdots \alpha_{i_ m-\sigma}^{[j], (m)}  T_{r-{\rm loop},(m,\sigma)}^{i_1 \cdots i_{m-\sigma}}(k) \ ,
\end{equation}
where $\sigma = 0,1$. In the expression above, we identify $T_{r-{\rm loop},(m,\sigma)}^{i_1 \cdots i_{m-\sigma}}(k)$ as our set of cosmology independent\footnote{Of course, these integrals still depend on the base $[0]$ cosmology, which we choose to be the Planck cosmology \eqn{eq:Planck_cosmological_parameters}.  We use the phrase ``cosmology independent'' to mean that they do not change when we scan over cosmological models away from the one of Planck.} integrals with $N_m$ basis functions. The explicit form of these integrals is reported in App.~\ref{sec:Tensor_integrals}. 

Equation~\eqref{eq:loop-tensor-integrals} makes the separation of cosmology dependence explicit: for fixed $(r,m,\sigma)$, all computational cost is contained in the tensors $T^{i_1\cdots i_{m-\sigma}}_{r\text{-loop},(m,\sigma)}(k)$, which depend only on $k$, while the cosmology enters exclusively through the coefficients $\alpha_i^{[j], (m)}$. We therefore precompute these tensors once on the reference $[0]$ cosmology, and obtain $\Delta_{m,\sigma}\mathcal{P}^{[j]}_{r\text{-loop,fit}}(k)$ for any target cosmology $j$ by the contraction in Eq.~\eqref{eq:loop-tensor-integrals}. Combining $\Delta_{m,0}\mathcal{P}^{[j]}_{r\text{-loop,fit}}$ and $\Delta_{m,1}\mathcal{P}^{[j]}_{r\text{-loop,fit}}$ according to Eq.~\eqref{eq:delta_m_final_splitting} yields $\Delta_m\mathcal{P}^{[j]}_{r\text{-loop}}(k)$, which can then be inserted into Eqs.~\eqref{eq:DPexpansion1loop} and \eqref{eq:DPexpansion2loop}.  As discussed in \secref{sec:perturbative_expansion}, in order to obtain our desired precision, we only need to include up to the terms with $m = 2$.  Therefore, we obtain the emulated one- and two-loop corrections for any cosmology $j$ as 
\begin{align}\label{eq:Pr-loop_fit_j}
\mathcal{P}^{[j]}_{r\text{-loop,fit}}(k)
&= \left(\mathcal{N}^{[j]}\right)^{r+1}\,
  \mathcal{P}^{\mathrm{Planck}}_{r\text{-loop}}(k) \nonumber\\
&\quad +\left(\mathcal{N}^{[j]}\right)^{r}
\Bigl(
  \Delta_{1,0}\,\mathcal{P}^{[j]}_{r\text{-loop,fit}}(k)
  + \Delta\mathcal{P}^{[j]}(k)\,\Delta_{1,1}\,\mathcal{P}^{[j]}_{r\text{-loop,fit}}(k)
\Bigr) \nonumber\\
&\quad
 + \left(\mathcal{N}^{[j]}\right)^{r-1}
\Bigl(
  \Delta_{2,0}\,\mathcal{P}^{[j]}_{r\text{-loop,fit}}(k)
  + \Delta\mathcal{P}^{[j]}(k)\,\Delta_{2,1}\,\mathcal{P}^{[j]}_{r\text{-loop,fit}}(k)
\Bigr)\  .
\end{align}
We emphasize that the $m=1$ contributions, $\Delta_{1,0}\mathcal{P}^{[j]}_{r\text{-loop,fit}}$ and $\Delta_{1,1}\mathcal{P}^{[j]}_{r\text{-loop,fit}}$, are evaluated using the more accurate fit $\DeltaP_{{\rm lin.,fit},1}^{[j]}$ (Eq.~\eqref{eq:Pfit_1}), whereas the $m=2$ contributions, $\Delta_{2,0}\mathcal{P}^{[j]}_{r\text{-loop,fit}}$ and $\Delta_{2,1}\mathcal{P}^{[j]}_{r\text{-loop,fit}}$, use the less accurate but more economical fit $\DeltaP_{{\rm lin.,fit},2}^{[j]}$ defined in Eq.~\eqref{eq:Pfit_2}.  With the emulated one- and two-loop contributions in hand, the emulated NNLO matter power spectrum follows by restoring the standard growth-factor time dependence, as done in Eq.~\eqref{eq:Pnnlo},
\begin{equation} \label{eq:PNNLO_fit_j}
    \Pcal_{{\rm NNLO,fit}}^{[j]}(k,a)=
    \mathcal{D}(a)^2\,\Plin^{[j]}(k)
    + \mathcal{D}(a)^4\,\Pcal^{[j]}_{1{\rm-loop,fit}}(k)
    + \mathcal{D}(a)^6\,\Pcal^{[j]}_{2{\rm-loop,fit}}(k)\ . 
\end{equation}

We numerically integrate $T_{r-{\rm loop},(m,\sigma)}^{i_1 \cdots i_{m-\sigma}}(k)$ for 140 values of $k$ in the range $[10^{-3}, 1]$ $\um$ with the algorithm referenced in Sec.~\ref{sec:NumericalMethod}. The number of Monte Carlo evaluations (and computational time) used varies significantly between integrals due to different rates of convergence. The range of Monte Carlo samples used for the one- and two-loop integrations is reported in tab.~\ref{tab:MCpoints}.

\begin{table}[h]
    \centering
    \begin{tabular}{r|c|c}
                  & At least & Up to \\
                  \hline
         One-loop & $10^7$ & $10^9$ \\
         Two-loop & $10^7$ & $10^{11}$
    \end{tabular}
    \caption{ \small Range of Monte Carlo points used at one and two loop. Lower numbers of Monte Carlo points in these ranges are used depending on how fast the integral converges.}
    \label{tab:MCpoints}
\end{table}

Finally, we construct a set of cosmology-independent integrals for the two-loop counterterms as well. These are defined in \appref{sec:2-loop_counterterms}, in Eqs.~\eqref{eq:DPexpansion_I15_fit}--\eqref{eq:DPexpansion_I42_fit}. The corresponding integrals are evaluated using the same numerical procedure as for the one-loop contributions.

We have now achieved the goal of decoupling the cosmological dependence from the loop integrations. As we shall see in the next section, this procedure will allow us to compute the predictions of the EFTofLSS for dark matter up to two loops with a small theoretical error compared to the experimental error of current generation surveys, such as the DESI surveys \cite{DESI:2016fyo}. 

\section{Validation and results}
\label{sec:Results}
In this section, we present tests and results of our novel numerical approach. In particular, we will be comparing two ways enabled by our method for computing the loop corrections to the power spectrum of the $j=1$ cosmology:
\begin{itemize}
    \item The first way is our new fitting procedure (described in the previous section), which uses the fitted difference of linear power spectra $\DeltaP_{\rm lin.,fit}^{[1]}$ and gives expressions for the loop-corrections in \eqn{eq:Pr-loop_fit_j}.  Quantities discussed in this section using this procedure are labeled with the subscript or superscript `fit.'
    \item The second way is a direct numerical integration that uses the exact $\Plin^{[1]}$ and the IR- and UV-regulated integrands in \eqn{eq:Pr-loop_jmodel} and \eqn{uvirintdefs}.  Quantities discussed in this section using this procedure do not have an extra subscript or superscript.
\end{itemize}
In all plots, we show the errors coming from Monte Carlo integration.

The results for the one- and two-loop corrections at redshift $z=0$ are shown in \figref{subfig:1-loop_results} and \figref{subfig:2-loop_results}, while the full power spectrum \textit{up to} two-loops, $\Pcal_{\rm NNLO}$, evaluated at the same redshift, is presented in \figref{fig:PfitvsPlin_LO+NLO+NLLO}. 
The error spike in $\Pcal_{\rm 2-loop}$ at $k\sim 0.3 \, \um$ {(bottom panel of \figref{subfig:2-loop_results})} arises from the two-loop integral approaching zero. Thus, this does not significantly impact the total power spectrum $\Pcal_{\rm NNLO}$, as seen in \figref{fig:PfitvsPlin_LO+NLO+NLLO}, where no corresponding precision loss is observed. In \figref{fig:1&2-loop_results} and \figref{fig:PfitvsPlin_LO+NLO+NLLO}, we set all the counterterms to zero, as the precision of the associated one-loop diagrams can be inferred by the shown one-loop diagrams.  As we can see {in \figref{fig:PfitvsPlin_LO+NLO+NLLO}}, $\Pcal_{\rm NNLO}^{[1]}$ at redshift $z = 0$ has an error below $5$ per mill for ${k \leq 0.6 \, \um}$. However, a more realistic application would be at redshift $z = 1.23$, where we have an error below $0.6$ per mill. 

To further assess the accuracy of our approach, as our theoretical error we define the quantity $|\Pcal_{\rm NNLO} - \Pcal_{\rm NNLO, fit}|$.  We compare this quantity to the expected experimental uncertainty of the DESI surveys~\cite{DESI:2016fyo}, as estimated using the methodology described in \cite[Sec. 2]{Braganca:2023pcp}.   At redshift $z = 1.23$, we find that our theoretical error, due to the Monte-Carlo integration and the emulation of the linear power spectrum in the loop integrands, remains below $15\%$ of the estimated DESI uncertainty $\sigma_{\rm DESI}$ (with binning $\Delta k=0.005\, \um$), as shown in \figref{fig:sigma_DESI_comparison}.
To match the binning used in $\sigma_{\rm DESI}$~\cite[Sec. 2]{Braganca:2023pcp}, we employed a linear interpolation for our calculated values of $\Pcal_{\rm NNLO}(k)$ and $\Pcal_{\rm NNLO, fit}(k)$.

We also display the range of theoretical error affecting the computation of the one-loop-type integrals $I_{42}^{i, [1]}$, $I_{15}^{i, [1]}$ and $I_{33}^{[1]}$, with the index $i$ denoting the corresponding EFT coefficient. These integrals, which enter the two-loop EFT counterterms, are defined in \appref{sec:2-loop_counterterms} in Eqs.~\eqref{eq:P42_counterterm_integral},~\eqref{eq:P15_counterterm_integral} and \eqref{eq:P33_counterterm_integral}. Since there are 12 independent integrals in total, we do not show each individually. Instead, we plot the pointwise extremal envelope, {\it i.e.}, the range between the maximum and minimum values over all components $i$ of the relative residuals $(I^{i, [1]}_{15} - I^{i, [1]}_{15, \rm fit}) / I^{i, [1]}_{15}$ in \figref{subfig:I15_plots} and $(I^{i, [1]}_{42} - I^{i, [1]}_{42, \rm fit}) / I^{i, [1]}_{42}$ in \figref{subfig:I42_plots}.  The relative residual for a representative component, $i=1$, is also shown as a black line. We then show $(I^{[1]}_{33} - I^{[1]}_{33, \rm fit}) / I^{[1]}_{33}$ in \figref{subfig:I33_plots}.

We conclude by highlighting three important aspects of our method:
\begin{itemize}
    \item The choice of basis functions in \eqref{eq:Pfit_2} and \eqref{eq:Pfit_1} could be further optimized. Although these choices {may facilitate future comparisons with emerging analytical methods}, a key advantage of {our} purely numerical approach is that alternative function bases can be explored freely. 
    In fact, if one is required to compute the integrals of order $(\Delta \Pcal_{\rm lin.})^3$ to achieve greater precision, reducing the number of fitting functions may be beneficial.
    \item The chosen test case, cosmology $j = 1$, represents a particularly challenging scenario where $\DeltaP_{\rm lin.}$ is larger than in realistic cases. In practical applications, we expect $\DeltaP_{\rm lin.}$ to be smaller, leading to improved accuracy.
    \item Our current framework expands around a single reference cosmology (labeled Planck), but this can be extended.
        A finite set of reference cosmologies could be selected, each with exactly computed loop corrections. Then, the loop tensors in Eq.~\eqref{eq:loop-tensor-integrals} could be computed for each reference cosmology and used to approximate loop corrections for any cosmology with a sufficiently small $\DeltaP_{\rm lin.}$. 
\end{itemize}

\begin{figure}[H]
    \centering
    \subfloat[\label{subfig:1-loop_results}]
    {
        \includegraphics[width=0.9\textwidth]{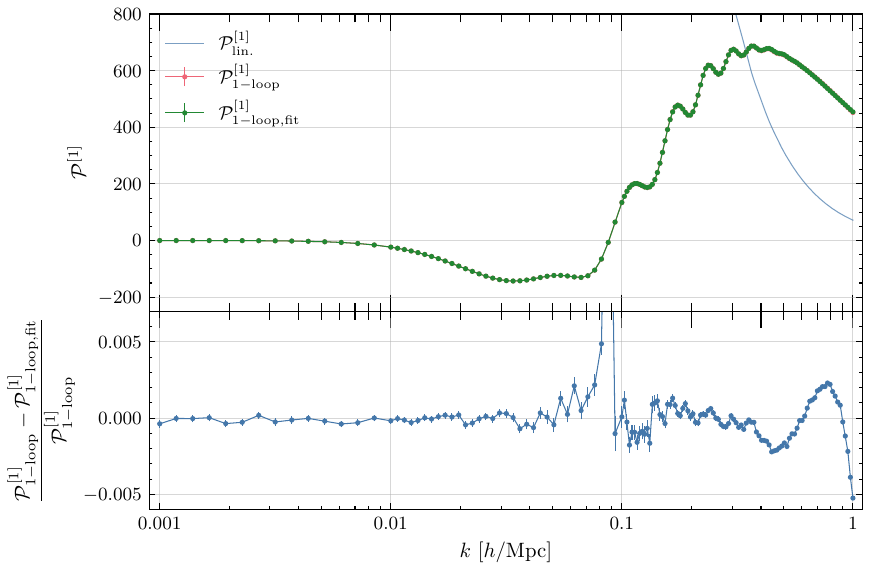}
    }
    \par 
    \subfloat[\label{subfig:2-loop_results}]{
        \includegraphics[width=0.9\textwidth]{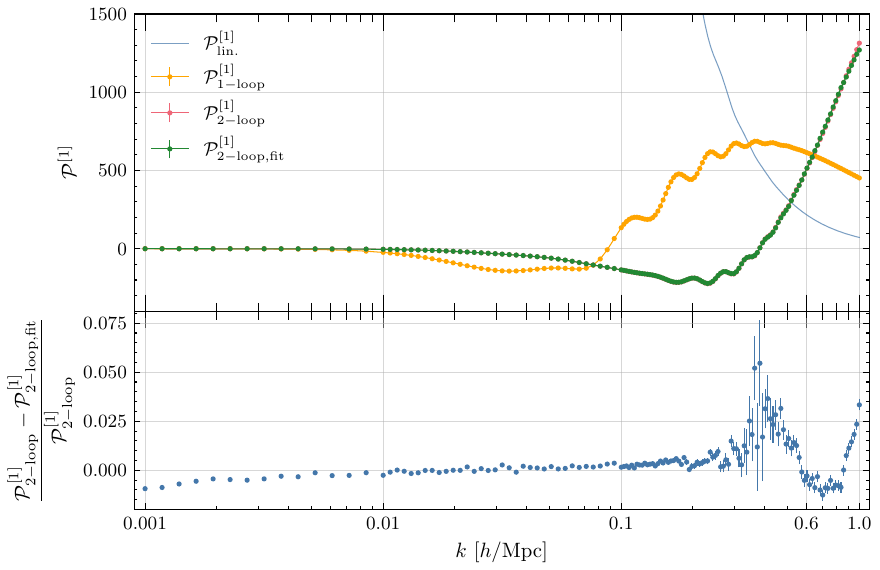}
       
    }    
    \caption{\small  Top panels: One-loop (\ref{subfig:1-loop_results}) and two-loop (\ref{subfig:2-loop_results}) corrections at redshift $z=0$ computed using the exact linear power spectrum $\Plin^{[1]}$ (red) and the fitted $\DeltaP^{[1]}_{\rm lin. fit}$ (green). The exact and fitted predictions are almost visually indistinguishable over the full $k$-range shown.  
    Bottom panels: Relative residuals between the exact and fitted results for each case.
 We stress that $\Pcal^{[1]}_{2\rm -loop, fit}$ has been computed up to second order in $\DeltaP_{\rm lin.}$. Error bars in the bottom panels are from Monte Carlo integration. 
    }
    \label{fig:1&2-loop_results}
\end{figure}

\begin{figure}[H]
    \centering
    \subfloat[]
    {
        \includegraphics[]{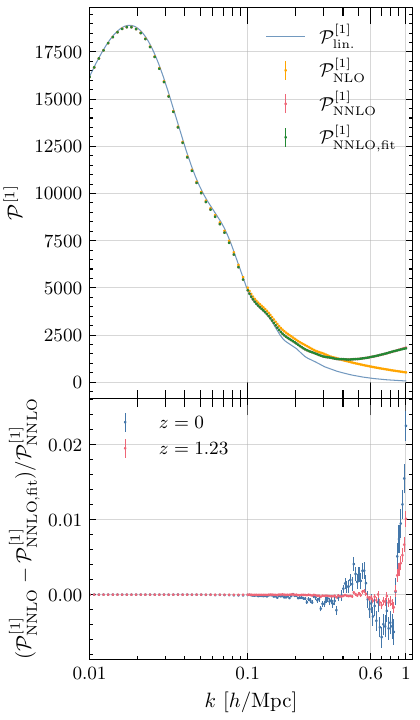}
        \label{subfig:PNNLO_fit_full}
    }    
    \subfloat[]{
        \includegraphics[]{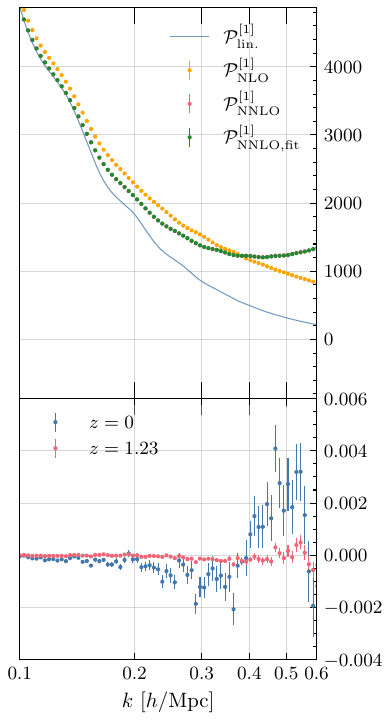}
        \label{subfig:PNNLO_fit_focus}
    }    
    \caption{\small  Above: Full power spectrum for cosmology $j = 1$ at redshift $z=0$, at LO (blue), exact numeric computation of NLO (orange), exact numeric computation of NNLO  (red), and NNLO computed with cosmology independent fitting functions (green). Below: Relative residual between the next-to-next leading order power spectrum computed with the exact $\Plin^{[j]}$ and the fitting procedure at redshift $z = 0$ (blue) and at $z=1.23$ (red). {In all these plots we set the counterterms of the EFTofLSS to zero.} Error bars in the bottom panels are from Monte Carlo integration.}
    \label{fig:PfitvsPlin_LO+NLO+NLLO}
\end{figure}

\begin{figure}
    \centering
    \includegraphics[]{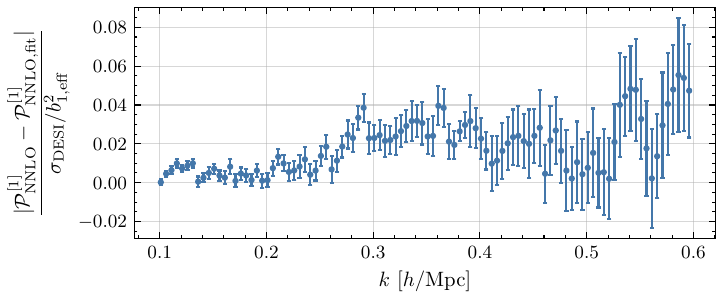}
    \caption{ \small   Comparison of the theoretical uncertainty {$|\Pnnlo^{[1]} - \mathcal{P}^{[1]}_{\rm NNLO, fit}|$} at $z = 1.23$ and $\sigma_{\rm DESI}$ (see Footnote~\ref{covfn}) for the $j = 1$ cosmology.  We see that this error is negligible with respect to the DESI error bars, where we typically allow a $\sigma_{\rm DESI} /3$ theoretical error.  Error bars are from Monte Carlo integration. }
    \label{fig:sigma_DESI_comparison}
\end{figure}

\begin{figure}[H]
    \centering
    \subfloat[\label{subfig:I15_plots}]
    {
        \includegraphics[width=0.48\textwidth]{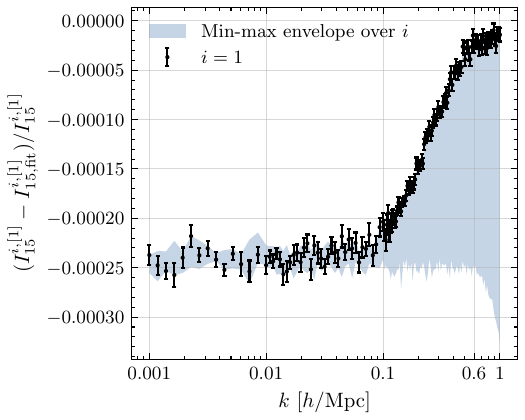}
    }
    \hfill
    \subfloat[\label{subfig:I42_plots}]{
        \includegraphics[width=0.48\textwidth]{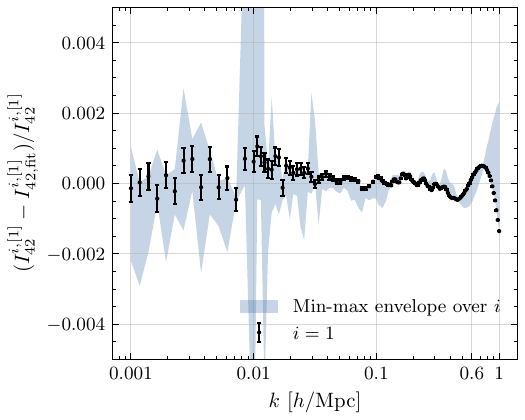}
       
    }    
    \\
    \centering
    \subfloat[\label{subfig:I33_plots}]
    {
        \includegraphics[width=0.7\textwidth]{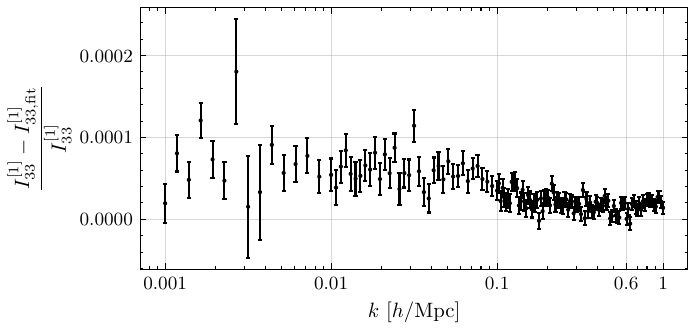}
    }
    \caption{ \small 
        {
        Relative residuals between the fitted and exact values of the one-loop integrals $I^{i, [1]}_{15}$ (\ref{subfig:I15_plots}), $I^{i, [1]}_{42}$ (\ref{subfig:I42_plots}) and $I^{[1]}_{33}$ (\ref{subfig:I33_plots}) for the cosmology $j=1$. The blue bands show the min-max envelope of the relative residuals over all components $i$ (7 for $I^{i, [1]}_{15}$, 4 for $I^{i, [1]}_{42}$), while the black data points show the residual for~$i = 1$.  Error bars are from Monte Carlo integration. All these ratios are redshift independent.} 
    }
    \label{fig:EFT_counterterms}
\end{figure}

\section{Conclusions}
\label{sec:conclusion}

{In this paper we have {developed} a formalism to compute the two-loop dark-matter power spectrum in the EFTofLSS in a way that is fast and therefore allows to explore how predictions from points of the {vast  space} of cosmological parameters fit the data, as {is normally done} in cosmological analysis. 

The formalism is based on a few simple ideas. {We establish a numerical method for computing loop integrals in Euclidean space with high accuracy. This is achieved by creating smooth loop integrands; we remove enhancements with systematic UV-subtractions and use diagrammatic rearrangements that enable IR cancellations to occur locally within the integration domain. With this efficient numerical method, we compute the exact loop predictions for a realistic `reference' cosmology; in our case we have chosen the one that best fits the Planck data.
Next, we fit the difference between each linear power spectrum and the reference one, $\Delta \mathcal{P}_{\rm lin.}$, to a cosmology-independent basis of functions. 
Then, we evaluate the loop integrals for each combination of the basis functions that emerge as kernels in the integrands of the one and two-loop power spectrum. We call this  set of universal (cosmology independent) integrals  $T_{r-{\rm loop},(m,\sigma)}^{i_1 \cdots i_{m-\sigma}}$, where the indices, 
indicate which  basis functions are present in their integrand. The prediction for generic values of cosmological parameters is then simply obtained as a linear combination of the  $T_{r-{\rm loop},(m,\sigma)}^{i_1 \cdots i_{m-\sigma}}$ integrals with coefficients} derived from the coordinates $i_1,i_2,\ldots$ which map $\Delta \Pcal_{\rm lin.}$ to our basis functions (in the spirit of Refs~\cite{Simonovic:2017mhp,Anastasiou:2022udy}). By expanding in the smallness of  $\Delta \Pcal_{\rm lin.}$, we can limit the number of integrals $T_{r-{\rm loop}}$ with $r=1,2$. The expansion  allows  us a rather large dimensionality in the set of basis functions without producing an unbearable computational cost to produce the set  $T_{r-{\rm loop}}$ or a very large memory cost for loading them. In this way, we check we achieve a fast and accurate evaluation at a relatively small computation cost.
}

We have applied these ideas to the two-loop power spectrum of dark matter, where we compute the two-loop diagrams and the various one-loop and tree-level diagrams necessary to make the answer insensitive to uncontrolled short distance physics. These ingredients are essentially all that are needed to compute the galaxy lensing power spectrum, which is an observable quantity. In total, we find eight EFT counterterms, all of which are needed to fully cancel the UV limits of the loops. 

{While our prediction is essentially all that is needed to analyze the lensing signal in galaxy surveys, this formalism is general, and one can envision extensions to two-loop power spectra of other tracers, such as galaxies in redshift space, or to higher $n-$point functions or even higher loop orders. We leave these explorations for future work.}

{We provide the set of integrals $ T_{r-{\rm loop},(m,\sigma)}^{i_1 \cdots i_{m-\sigma}}$ together with the arXiv version of this article. A detailed description of the data format, column contents, and import instructions can be found in the attached \texttt{README} file.
}

\section*{Acknowledgments}

{{We thank the organizers and participants of the ``Galaxies meet QCD'' conference where part of this work was initiated.  M.L. thanks Alex Edison for insightful discussions related to this work.  C.A. would like to thank Bernhard Mistlberger for collaboration at early stages of this work.} 
We acknowledge the use of the Euler cluster at ETH for much of the numerical computations. {C.A. is supported by the SNF grant  10001706.} L.S. is supported by the SNSF grant 200021 213120. }

\appendix
\section{Non-local-in-time expansion}

%
\subsection{A method for computation of non-local-in-time kernels} \label{kernelsderiv}

In this Appendix, we give an explicit general method to compute all non-local-in-time kernels.  For convenience, we can explicitly rewrite \eqn{eq:final_stress} as 
\begin{align}
\begin{split} \label{generaltauij}
\tau^{ij}_{(n)} ( \xvec , t ) & = \sum_{m = 1}^n \sum_{\{\mathcal{O}^{ij}_m\}} \int^t \, d t_1 \cdots d t_m\, H ( t_1 ) \cdots H(t_m )  \, c_{\mathcal{O}_m } ( t , t_1 , \dots , t_m ) \\
 & \hspace{1in} \times  \left[ \mathcal{O}^{ij}_{m} \big|_{\xfl}  ( \xvec , t ; t_1 , \dots , t_m )  \right]^{(n)}  \ .
\end{split}
\end{align}
The set $\{ \mathcal{O}^{ij}_m \}$ is the set of all symmetric tensors built out of $r^{ij}$ and $p^{ij}$ that have $m$ factors, and the notation $\mathcal{O}^{ij}_{m} \big|_{\xfl}$ will be explained below.

Each $\mathcal{O}_m^{ij}$ is the product of $m$ factors of $r^{ij}$ and $p^{ij}$, i.e.
\be
\mathcal{O}_m^{ij} ( \xvec , t ) = t^{i_1 j_1}_{\mathcal{O}_m , 1} ( \xvec , t )  \cdots t^{i_m j_m}_{\mathcal{O}_m , m} ( \xvec , t ) \, \Delta^{ij}_{\mathcal{O}_m} [ i_1 , j_1 , \dots , i_m , j_m ] \ ,
\ee
where $t^{i_k j_k}_{\mathcal{O}_m , k}$ is either $r^{ij}$ or $p^{ij}$, and $\Delta^{ij}_{\mathcal{O}_m} [ i_1 , j_1 , \dots , i_m , j_m ] $ is the combination of Kroneker delta functions needed to build the relevant $\mathcal{O}_m^{ij}$.  For example, for $\mathcal{O}_2^{ij} = r^{i k } p^{k j}$, we have $t_{\mathcal{O}_2, 1}^{i_1 j_1} = r^{i_1 j_1}$, $t_{\mathcal{O}_2, 2}^{i_2 j_2} = p^{i_2 j_2}$, and  $\Delta^{ij}_{\mathcal{O}_2} [ i_1 , j_1 , i_2 , j_2 ] = \delta_K^{i i_1} \delta_K^{j_1 i_2} \delta_K^{j_2 j } $.  For the fields that have to be symmetrized over the indices, this can be included in the $\Delta_{\mathcal{O}_m}^{ij}$.    Given this, we use the following notation
\begin{align}
\begin{split}
& \mathcal{O}^{ij}_{m} \big|_{\xfl}  ( \xvec , t ; t_1 , \dots , t_m ) \equiv t^{i_1 j_1}_{\mathcal{O}_m , 1} ( \xfl ( \xvec , t , t_1) , t_1 )  \cdots t^{i_m j_m}_{\mathcal{O}_m , m} ( \xfl ( \xvec , t , t_m)  , t_m ) \\
&\hspace{2in}  \times  \Delta^{ij}_{\mathcal{O}_m} [ i_1 , j_1 , \dots , i_m , j_m ] \ ,
\end{split}
\end{align}
i.e. each factor is evaluated at a different time on the fluid element.  

Now, we can expand up to $n$-th order by trivially writing
\begin{align}
 \left[ \mathcal{O}^{ij}_{m} \big|_{\xfl}  ( \xvec , t ; t_1 , \dots , t_m )  \right]^{(n)}   & = \Delta^{ij}_{\mathcal{O}_m} [ i_1 , j_1 , \dots , i_m , j_m ]   \sum_{\{ n_1 , \dots , n_m \}_n}  \\
 & \times  [ t^{i_1 j_1}_{\mathcal{O}_m , 1} ( \xfl ( \xvec , t , t_1) , t_1 ) ]^{(n_1)}  \cdots [ t^{i_m j_m}_{\mathcal{O}_m , m} ( \xfl ( \xvec , t , t_m)  , t_m ) ]^{(n_m)} \ , \nonumber
\end{align}
where $\{ n_1 , \dots , n_m \}_n$ is the set of all $(n_1 , \dots , n_m)$ such that $n_1 + \cdots + n_m = n$ with $n_a \geq 1$ for $a = 1 , \dots , m$. Using the notation of~\cite{Donath:2023sav}, each individual term on the right-hand side can be written as
\be
 [ t_{\mathcal{O}_m,a}^{i_a j_a}( \xfl ( \xvec , t , t_a)  , t_a ) ]^{(n_a)} =  \sum_{\alpha_a = 1}^{n_a} \left( \frac{D(t_a)}{D(t)} \right)^{\alpha_a}  [ \mathbb{C}_{ (\mathcal{O}_m , a)  , \alpha_a}^{ i_a j_a}]^{(n_a)} ( \xvec , t )  \ , 
\ee
since by definition each of the $t_{\mathcal{O}_m,a}$, being $r$ and $p$, has $m = 1$.  Plugging this into \eqn{generaltauij}, we have 
\begin{align}
\begin{split}
\tau^{ij}_{(n)} ( \xvec , t ) & = \sum_{m = 1}^n \sum_{\{\mathcal{O}^{ij}_m\}} \int^t \, d t_1 \cdots d t_m H ( t_1 ) \cdots H(t_m )  \, c_{\mathcal{O}_m } ( t , t_1 , \dots , t_m )   \Delta^{ij}_{\mathcal{O}_m} [ i_1 , j_1 , \dots , i_m , j_m ]   \\
& \times \sum_{\{ n_1 , \dots , n_m \}_n}  \prod_{a = 1}^m \left[ \sum_{\alpha_a = 1}^{n_a} \left( \frac{D(t_a)}{D(t)} \right)^{\alpha_a}  [ \mathbb{C}_{ (\mathcal{O}_m , a)  , \alpha_a}^{ i_a j_a}]^{(n_a)} ( \xvec , t )   \right]  \ ,
\end{split}
\end{align}
which we can rewrite as 
\begin{align}
\begin{split} \label{timeints}
\tau^{ij}_{(n)} ( \xvec , t ) & = \sum_{m = 1}^n \sum_{\{\mathcal{O}^{ij}_m\}} \int^t \, d t_1 \cdots d t_m H ( t_1 ) \cdots H(t_m )  \, c_{\mathcal{O}_m } ( t , t_1 , \dots , t_m )   \Delta^{ij}_{\mathcal{O}_m} [ i_1 , j_1 , \dots , i_m , j_m ]   \\
& \times \sum_{\{ n_1 , \dots , n_m \}_n} \sum_{\alpha_1 = 1}^{n_1} \dots  \sum_{\alpha_m = 1}^{n_m}  \left[  \prod_{a = 1}^m  \left( \frac{D(t_a)}{D(t)} \right)^{\alpha_a}  [ \mathbb{C}_{ (\mathcal{O}_m , a)  , \alpha_a}^{ i_a j_a}]^{(n_a)} ( \xvec , t )   \right]  \ .
\end{split}
\end{align}
Next, we define
\begin{align}
\begin{split} \label{gentimefns}
c_{\mathcal{O}_m; \alpha_1 , \dots , \alpha_m } (t ) \equiv  \int^t \, d t_1 \cdots d t_m H ( t_1 ) \cdots H(t_m )  \, c_{\mathcal{O}_m } ( t , t_1 , \dots , t_m )    \prod_{a = 1}^m  \left( \frac{D(t_a)}{D(t)} \right)^{\alpha_a}  \ , 
\end{split}
\end{align}
so that \eqn{timeints} becomes 
\begin{align}
\begin{split} 
\tau^{ij}_{(n)} ( \xvec , t ) & = \sum_{m = 1}^n \sum_{\{\mathcal{O}^{ij}_m\}}   \sum_{\{ n_1 , \dots , n_m \}_n} \sum_{\alpha_1 = 1}^{n_1} \dots  \sum_{\alpha_m = 1}^{n_m}    \\
& \times c_{\mathcal{O}_m; \alpha_1 , \dots , \alpha_m } (t )  \left(  \Delta^{ij}_{\mathcal{O}_m} [ i_1 , j_1 , \dots , i_m , j_m ]   \prod_{a = 1}^m   [ \mathbb{C}_{ (\mathcal{O}_m , a)  , \alpha_a}^{ i_a j_a}]^{(n_a)} ( \xvec , t )    \right) \ .
\end{split}
\end{align}

Since the independent coefficients $c_{\mathcal{O}_m; \alpha_1 , \dots , \alpha_m } (t )$ do not depend on the $n_a$, we would like to swap the sum over the $n_a$ with the sum over the $\alpha_a$.  For this, we use  
\be
  \sum_{\{ n_1 , \dots , n_m \}_n} \sum_{\alpha_1 = 1}^{n_1} \dots  \sum_{\alpha_m = 1}^{n_m}  =  \sum_{\{\alpha_1, \dots , \alpha_m\}_{\leq n} } \sum_{\{n_1 , \dots , n_m\}^\alpha_n }
\ee
where $\{\alpha_1, \dots , \alpha_m\}_{\leq n} $ is the set of all $(\alpha_1 , \dots , \alpha_m)$ with $\alpha_1 + \cdots + \alpha_m \leq n$, and $\{n_1 , \dots , n_m\}^\alpha_n $ is the set of all $(n_1 , \dots , n_m )$ such that $n_1 + \cdots + n_m = n$ and $n_a \geq \alpha_a$ for $a = 1 , \dots , m$.  Finally, we have
\begin{align}
\begin{split}  \label{finalfulltauij}
\tau^{ij}_{(n)} ( \xvec , t ) & = \sum_{m = 1}^n \sum_{\{\mathcal{O}^{ij}_m\}} \sum_{\{\alpha_1, \dots , \alpha_m\}_{\leq n} } c_{\mathcal{O}_m; \alpha_1 , \dots , \alpha_m } (t )  [ \mathbb{C}^{ij}_{\mathcal{O}_m; \alpha_1 , \dots , \alpha_m} ]^{(n)} ( \xvec , t) \ ,
\end{split}
\end{align}
where 
\be
[\mathbb{C}^{ij}_{\mathcal{O}_m; \alpha_1 , \dots , \alpha_m}]^{(n)}  ( \xvec , t) \equiv  \sum_{\{n_1 , \dots , n_m\}^\alpha_n }     \Delta^{ij}_{\mathcal{O}_m} [ i_1 , j_1 , \dots , i_m , j_m ]   \prod_{a = 1}^m   [ \mathbb{C}_{ (\mathcal{O}_m , a)  , \alpha_a}^{ i_a j_a}]^{(n_a)} ( \xvec , t )    \ , 
\ee
so that the initial set of potentially independent functions is one for each $\mathcal{O}_m^{ij}$ and each tuple $(\alpha_1 , \dots , \alpha_m)$ with $\alpha_1 + \cdots + \alpha_m \leq n$.

At third order, we have found that it is possible to make the following choice for the selection of the basis of independent functions from the above construction.  This will define the final set of basis functions that we use in \eqn{tauijexpansion}.  We consider a subset of the functions in \eqn{finalfulltauij} by setting  
\be \label{singletimelimit}
c_{\mathcal{O}_m} ( t , t_1 , \dots , t_m ) =  c_{\mathcal{O}_m} ( t , t_1 ) \frac{\delta_D ( t _1 - t_2)}{H(t_2)} \frac{\delta_D ( t _1 - t_3)}{H(t_3)} \dots \frac{\delta_D ( t _1 - t_m)}{H(t_m)}  \ ,
\ee
in \eqn{generaltauij}.  As physical intuition, this corresponds to the single-time-insertion construction of \cite{Donath:2023sav}.  Plugging \eqn{singletimelimit} into \eqn{gentimefns}, the time-dependent coefficients become
\be
c_{\mathcal{O}_m ; \alpha_1 , \dots , \alpha_m} ( t ) = \int^t dt' H(t') \, c_{\mathcal{O}_m} ( t , t' ) \left( \frac{D(t')}{D(t)} \right)^{\alpha_1 + \cdots + \alpha_m} \ ,
\ee
so we see that the coefficient only depends on the sum $\alpha_1 + \cdots + \alpha_m$.  Because of this, it is useful to define
\begin{align}
    c_{\mathcal{O}_m, \alpha } ( t)  = \int^t dt' H(t')\, c_{\mathcal{O}_m}(t,t') \left( \frac{D(t')}{D(t)} \right)^{\alpha +  m - 1} \  .
\end{align}
Given this, we can simplify the sum over the $\alpha_i$ in \eqn{finalfulltauij} by noting that 
\begin{align}
\begin{split}
&\hspace{-1in}  \sum_{\{\alpha_1, \dots , \alpha_m\}_{\leq n} } c_{\mathcal{O}_m; \alpha_1 , \dots , \alpha_m } (t )  [ \mathbb{C}^{ij}_{\mathcal{O}_m; \alpha_1 , \dots , \alpha_m} ]^{(n)} ( \xvec , t)   \\
& = \sum_{\alpha_{\rm tot}=m}^n \sum_{\{\alpha_1, \dots , \alpha_m\}_{\alpha_{\rm tot}} } c_{\mathcal{O}_m; \alpha_1 , \dots , \alpha_m } (t )  [ \mathbb{C}^{ij}_{\mathcal{O}_m; \alpha_1 , \dots , \alpha_m}]^{(n)}  ( \xvec , t)  \\
& = \sum_{\alpha_{\rm tot}=m}^n  \sum_{\{\alpha_1, \dots , \alpha_m\}_{\alpha_{\rm tot}} } c_{\mathcal{O}_m, \alpha_{\rm tot}-m+1} ( t )   [\mathbb{C}^{ij}_{\mathcal{O}_m; \alpha_1 , \dots , \alpha_m}]^{(n)}  ( \xvec , t)  \\
& = \sum_{\alpha = 1}^{n - m + 1} c_{\mathcal{O}_m , \alpha} (t)\sum_{\{\alpha_1, \dots , \alpha_m\}_{\alpha + m -1} } [ \mathbb{C}^{ij}_{\mathcal{O}_m; \alpha_1 , \dots , \alpha_m}]^{(n)}  ( \xvec , t)   \ .
\end{split}
\end{align}
Plugging this back into \eqn{finalfulltauij}, we find
\begin{align}
\begin{split}  \label{finalfulltauijsimp}
\tau^{ij}_{(n)} ( \xvec , t ) \Big|_{1\text{I}}& = \sum_{m = 1}^n \sum_{\{\mathcal{O}^{ij}_m\}}   \sum_{\alpha = 1}^{n - m + 1} c_{\mathcal{O}_m , \alpha} (t)  [\mathbb{C}^{ij}_{\mathcal{O}_m , \alpha}]^{(n)} ( \xvec , t)  \ ,
\end{split}
\end{align}
where the subscript $1\text{I}$ stands for one time insertion, and 
\be
[ \mathbb{C}^{ij}_{\mathcal{O}_m , \alpha} ]^{(n)} ( \xvec , t)\equiv \sum_{\{\alpha_1, \dots , \alpha_m\}_{\alpha + m -1} } [ \mathbb{C}^{ij}_{\mathcal{O}_m; \alpha_1 , \dots , \alpha_m} ]^{(n)} ( \xvec , t)  \ . 
\ee
At third order, we find that the set of functions $[ \mathbb{C}^{ij}_{\mathcal{O}_m , \alpha} ]^{(n)}$ defines the same basis as the set of functions $ [ \mathbb{C}^{ij}_{\mathcal{O}_m; \alpha_1 , \dots , \alpha_m} ]^{(n)} $, and so for simplicity, we use the $[ \mathbb{C}^{ij}_{\mathcal{O}_m , \alpha} ]^{(n)}$  functions.

%
%

\subsection{Explicit expressions for counterterms\label{app:counterterm expressions}}

For the kernels $\mathbb{C}^{ij(n)}_{\mathcal{O}_m , \alpha}$ and the coefficients $c_{\mathcal{O}_m, \alpha}$, the labels for $\mathcal{O}_m$ correspond to the following matrix structures
\be
\normalsize
  \begin{tabular}{|c|c|} 
  \hline
  $\mathcal{O}_m$ 	 	&	 \text{matrix}	  \\
  \hline 
 $r$ 	 	&	 $r_{ij}$	 \\
  \hline
   $\delta$ 	 	&	 $\delta^K_{ij} \delta $	 \\
  \hline
   $r^2$ 	 	&	 $r_{ik}r_{kj} $	 \\
   \hline
   $r \delta$ 	 	&	 $r_{ij} \delta $	 \\
   \hline
   $\delta^2$ 	 	&	 $\delta^K_{ij} \delta^2 $	 \\
   \hline
      $r^3$ 	 	&	 $r_{ik}r_{km}r_{mj}$	 \\
   \hline
      $(r^2) r$ 	 	&	 $r_{km}r_{mk} r_{ij}$	 \\
   \hline
       $\delta r^2$ 	 	&	 $\delta r_{ik}r_{kj}$	 \\
   \hline
       $\delta^2 r$ 	 	&	 $\delta^2 r_{ij} $	 \\
   \hline
          $\delta^3$ 	 	&	 $\delta^3 \delta^K_{ij} $	 \\
   \hline
             $\partial^2 \delta $ 	 	&	 $ \delta^K_{ij}  \partial^2 \delta $	 \\
   \hline
   \end{tabular}  \ .
\ee
The functions $\mathbb{C}^{ij(n)}_{\mathcal{O}_m, \alpha}$ are explicitly given by
\begin{align}
    \mathbb{C}^{ij(1)}_{r,1} &= \frac{\partial_i\partial_j\tilde{\delta}^{(1)}}{\partial^2}\  ,\\
    \mathbb{C}^{ij(1)}_{\delta,1} &= \delta^K_{ij}\tilde{\delta}^{(1)}\  , \nonumber\\
    \mathbb{C}^{ij(2)}_{r,1} &= \frac{\partial_i\partial_j\partial_k\td^{(1)}}{\partial^2}\frac{\partial_k\td^{(1)}}{\partial^2}\  , \nonumber\\
    \mathbb{C}^{ij(2)}_{\delta,1} &= \delta_{ij}\partial_k\td^{(1)}\frac{\partial_k\td^{(1)}}{\partial^2}\  , \nonumber\\
    \mathbb{C}^{ij(2)}_{r,2} &= \frac{\partial_i\partial_j\td^{(2)}}{\partial^2} - \frac{\partial_i\partial_j\partial_k\td^{(1)}}{\partial^2}\frac{\partial_k\td^{(1)}}{\partial^2}\  , \nonumber\\
    \mathbb{C}^{ij(2)}_{\delta,2} &= \delta^K_{ij}\left[\td^{(2)} - \partial_k\td^{(1)}\frac{\partial_k\td^{(1)}}{\partial^2}\right]\  , \nonumber\\
    \mathbb{C}^{ij(2)}_{r\delta,1} &= \frac{\partial_i\partial_j\td^{(1)}}{\partial^2}\td^{(1)}\  , \nonumber\\
    \mathbb{C}^{ij(2)}_{r^2,1} &= \frac{\partial_i\partial_k\td^{(1)}}{\partial^2}\frac{\partial_j\partial_k\td^{(1)}}{\partial^2}\  , \nonumber\\
    \mathbb{C}^{ij(2)}_{\delta^2,1} &= \delta^K_{ij}\td^{(1)}\td^{(1)}\  , \nonumber\\
    \mathbb{C}^{ij(3)}_{r,1} &= \frac{1}{2}\left[\frac{\partial_i\partial_j\partial_k\td^{(1)}}{\partial^2}\frac{\partial_k\tilde{\theta}^{(2)}}{\partial^2} + \frac{\partial_m\partial_n\partial_i\partial_j\td^{(1)}}{\partial^2}\frac{\partial_m\td^{(1)}}{\partial^2}\frac{\partial_n\td^{(1)}}{\partial^2} + \frac{\partial_m\partial_i\partial_j\td^{(1)}}{\partial^2}\frac{\partial_m\partial_n\td^{(1)}}{\partial^2}\frac{\partial_n\td^{(1)}}{\partial^2}\right]\  , \nonumber\\
    \mathbb{C}^{ij(3)}_{r,2} &=   \frac{\partial_i\partial_j\partial_k\td^{(2)}}{\partial^2}\frac{\partial_k\td^{(1)}}{\partial^2} - \frac{\partial_m\partial_n\partial_i\partial_j\td^{(1)}}{\partial^2}\frac{\partial_m\td^{(1)}}{\partial^2}\frac{\partial_n\td^{(1)}}{\partial^2} - \frac{\partial_m\partial_i\partial_j\td^{(1)}}{\partial^2}\frac{\partial_m\partial_n\td^{(1)}}{\partial^2}\frac{\partial_n\td^{(1)}}{\partial^2}\  , \nonumber\\
    \mathbb{C}^{ij(3)}_{r,3} &= \frac{\partial_i\partial_j\td^{(3)}}{\partial^2} - \frac{\partial_i\partial_j\partial_k\td^{(2)}}{\partial^2}\frac{\partial_k\td^{(1)}}{\partial^2} - \frac{1}{2}\frac{\partial_i\partial_j\partial_k\td^{(1)}}{\partial^2}\frac{\partial_k\tilde{\theta}^{(2)}}{\partial^2} + \frac{1}{2}\frac{\partial_m\partial_n\partial_i\partial_j\td^{(1)}}{\partial^2}\frac{\partial_m\td^{(1)}}{\partial^2}\frac{\partial_n\td^{(1)}}{\partial^2} \nonumber \\
    & + \frac{1}{2}\frac{\partial_m\partial_i\partial_j\td^{(1)}}{\partial^2}\frac{\partial_m\partial_n\td^{(1)}}{\partial^2}\frac{\partial_n\td^{(1)}}{\partial^2}\  , \nonumber \\
    \mathbb{C}^{ij(3)}_{\delta,1} &=  \frac{\delta^K_{ij}}{2}\left[\partial_k \td^{(1) }\frac{\partial_k  \tilde{\theta}^{(2)}}{\partial^2} + \partial_m\partial_n\td^{(1)}\frac{\partial_m\td^{(1)}}{\partial^2}\frac{\partial_n\td^{(1)}}{\partial^2} + \partial_m\td^{(1)}\frac{\partial_m\partial_n\td^{(1)}}{\partial^2}\frac{\partial_n\td^{(1)}}{\partial^2} \right]\  , \nonumber \\
    \mathbb{C}^{ij(3)}_{\delta,2} &= \delta^K_{ij}\left[\partial_k\td^{(2)}\frac{\partial_k\td^{(1)}}{\partial^2} - \partial_m\partial_n\td^{(1)}\frac{\partial_m\td^{(1)}}{\partial^2}\frac{\partial_n\td^{(1)}}{\partial^2} -  \partial_m\td^{(1)}\frac{\partial_m\partial_n\td^{(1)}}{\partial^2}\frac{\partial_n\td^{(1)}}{\partial^2}\right]\  ,\nonumber \\
    \mathbb{C}^{ij(3)}_{\delta,3} &= \frac{\delta^K_{ij}}{2}\left[2\td^{(3)} -2\partial_k\td^{(2)}\frac{\partial_k\td^{(1)}}{\partial^2}-\partial_k\td^{(1)}\frac{\partial_k\tilde{\theta}^{(2)}}{\partial^2} + \partial_m\partial_n\td^{(1)}\frac{\partial_m\td^{(1)}}{\partial^2}\frac{\partial_n\td^{(1)}}{\partial^2}\right. \nonumber \\
    &\left. + \partial_m\td^{(1)}\frac{\partial_m\partial_n\td^{(1)}}{\partial^2}\frac{\partial_n\td^{(1)}}{\partial^2} \right]\  ,\nonumber \\
    \mathbb{C}^{ij(3)}_{r^2,1} &=  \frac{\partial_i\partial_m\td^{(1)}}{\partial^2}\frac{\partial_n\partial_m\partial_j\td^{(1)}}{\partial^2}\frac{\partial_n\td^{(1)}}{\partial^2} + {(i \leftrightarrow j)} \  ,\nonumber \\
    \mathbb{C}^{ij(3)}_{r^2,2} &=  \left( \frac{\partial_i\partial_k\td^{(2)}}{\partial^2}\frac{\partial_k\partial_j\td^{(1)}}{\partial^2} + {(i \leftrightarrow j)} \right) - \mathbb{C}^{ij(3)}_{r^2,1} \  ,\nonumber \\
    \mathbb{C}^{ij(3)}_{r\delta,1} &= \td^{(1)}\frac{\partial_i\partial_j\partial_k\td^{(1)}}{\partial^2}\frac{\partial_k\td^{(1)}}{\partial^2} + \frac{\partial_i\partial_j\td^{(1)}}{\partial^2}\partial_k\td^{(1)}\frac{\partial_k\td^{(1)}}{\partial^2}\  ,\nonumber \\
    \mathbb{C}^{ij(3)}_{r\delta,2} &= \frac{\partial_i\partial_j\td^{(2)}}{\partial^2}\td^{(1)} + \frac{\partial_i\partial_j\td^{(1)}}{\partial^2}\td^{(2)}  - \frac{\partial_i\partial_j\partial_k\td^{(1)}}{\partial^2}\frac{\partial_k\td^{(1)}}{\partial^2}\td^{(1)} - \frac{\partial_i\partial_j\td^{(1)}}{\partial^2}\partial_k\td^{(1)}\frac{\partial_k\td^{(1)}}{\partial^2}\  ,\nonumber \\
    \mathbb{C}^{ij(3)}_{\delta^2,1} &= 2\delta^K_{ij}\td^{(1)}\partial_k\td^{(1)}\frac{\partial_k\td^{(1)}}{\partial^2}\  ,\nonumber \\
    \mathbb{C}^{ij(3)}_{\delta^2,2} &= 2\delta^K_{ij}\left[\td^{(2)}\td^{(1)} -\td^{(1)}\partial_k\td^{(1)}\frac{\partial_k\td^{(1)}}{\partial^2} \right]\  , \nonumber \\
    \mathbb{C}^{ij(3)}_{r^3,1} &= \frac{\partial_i\partial_m\td^{(1)}}{\partial^2}\frac{\partial_m\partial_n\td^{(1)}}{\partial^2}\frac{\partial_n\partial_j\td^{(1)}}{\partial^2}\  , \nonumber \\
    \mathbb{C}^{ij(3)}_{r^2\delta,1} &= \td^{(1)}\frac{\partial_i\partial_k\td^{(1)}}{\partial^2}\frac{\partial_k\partial_j\td^{(1)}}{\partial^2}\  , \nonumber \\
    \mathbb{C}^{ij(3)}_{r\delta^2,1} &= \td^{(1)}\td^{(1)}\frac{\partial_i\partial_j\td^{(1)}}{\partial^2}\  , \nonumber \\
    \mathbb{C}^{ij(3)}_{\delta^3,1} &= \delta^K_{ij}\td^{(1)}\td^{(1)}\td^{(1)}\  , \nonumber \\
    \mathbb{C}^{ij(3)}_{(r^{2})r,1} &= \frac{\partial_m\partial_n\td^{(1)}}{\partial^2}\frac{\partial_m\partial_n\td^{(1)}}{\partial^2}\frac{\partial_i\partial_j\td^{(1)}}{\partial^2}\  , \nonumber \\
    \mathbb{C}^{ij(1)}_{\partial^2 \delta,1} &= \delta^K_{ij}\frac{\partial^2\td^{(1)}}{\knl^2}\  .\nonumber
\end{align}

%
%

\subsection{Degeneracy equations for the contribution to the power spectrum}\label{sec:counterDegens}
We here provide the set of degeneracy equations to find the independent counterterms that contribute to the power spectrum {through the integrand of the counterterm diagrams.  At the integrand level, we have}
\begin{align}
\mathbb{K}^{(3)}_{\delta ^2,2}&=\frac{775 \mathbb{K}^{(3)}_{\delta ,3}}{187}+\frac{60 \mathbb{K}^{(3)}_{r^3,1}}{187}+\frac{200 \mathbb{K}^{(3)}_{r,3}}{187}-\frac{401 \mathbb{K}^{(3)}_{r \delta^2,1}}{187}-\frac{130 \mathbb{K}^{(3)}_{r\delta ,2}}{187}\  , \\ 
\mathbb{K}^{(3)}_{(r^{2})r,1}&=\frac{1085 \mathbb{K}^{(3)}_{\delta ,3}}{374}+\frac{416 \mathbb{K}^{(3)}_{r^3,1}}{187}+\frac{140 \mathbb{K}^{(3)}_{r,3}}{187}-\frac{1459 \mathbb{K}^{(3)}_{r\delta^2,1}}{374}-\frac{91 \mathbb{K}^{(3)}_{r\delta ,2}}{187}\  , \\ 
\mathbb{K}^{(3)}_{r^2,1}&=-\frac{7 \mathbb{K}^{(3)}_{\delta ^2,1}}{4}+\frac{7 \mathbb{K}^{(3)}_{\delta ,2}}{4}+\mathbb{K}^{(3)}_{r\delta ,1}\  , \\ 
\mathbb{K}^{(3)}_{r^2,2}&=-\frac{1225 \mathbb{K}^{(3)}_{\delta ,3}}{374}-\frac{156 \mathbb{K}^{(3)}_{r^3,1}}{187}-\frac{520 \mathbb{K}^{(3)}_{r,3}}{187}+\frac{1225 \mathbb{K}^{(3)}_{r\delta^2,1}}{374}+\frac{525 \mathbb{K}^{(3)}_{r\delta ,2}}{187}\  ,\\
\mathbb{K}^{(3)}_{r,2}&=\frac{3 \mathbb{K}^{(3)}_{\delta ^2,1}}{4}-\frac{3 \mathbb{K}^{(3)}_{\delta ,2}}{4}+\frac{705 \mathbb{K}^{(3)}_{\delta ,3}}{374}+\frac{160 \mathbb{K}^{(3)}_{r^3,1}}{187} \\
& \hspace{1in} -\frac{90 \mathbb{K}^{(3)}_{r,3}}{187}-\frac{705 \mathbb{K}^{(3)}_{r\delta^2,1}}{374}+\mathbb{K}^{(3)}_{r\delta ,1}-\frac{35 \mathbb{K}^{(3)}_{r\delta ,2}}{187}\  , \\ 
\mathbb{K}^{(3)}_{r,1}&=-\frac{3 \mathbb{K}^{(3)}_{\delta ^2,1}}{4}+\mathbb{K}^{(3)}_{\delta ,1}+\frac{7 \mathbb{K}^{(3)}_{\delta ,2}}{4}-\frac{331 \mathbb{K}^{(3)}_{\delta ,3}}{374} \\
& \hspace{1in} -\frac{160 \mathbb{K}^{(3)}_{r^3,1}}{187}-\frac{97 \mathbb{K}^{(3)}_{r,3}}{187}+\frac{705 \mathbb{K}^{(3)}_{r\delta^2,1}}{374}-\mathbb{K}^{(3)}_{r\delta,1}+\frac{35 \mathbb{K}^{(3)}_{r\delta ,2}}{187}\  , \\ 
\mathbb{K}^{(3)}_{\delta ^3,1}&=\frac{1085 \mathbb{K}^{(3)}_{\delta ,3}}{374}+\frac{42 \mathbb{K}^{(3)}_{r^3,1}}{187}+\frac{140 \mathbb{K}^{(3)}_{r,3}}{187}-\frac{711 \mathbb{K}^{(3)}_{r\delta^2,1}}{374}-\frac{91 \mathbb{K}^{(3)}_{r\delta,2}}{187}\  , \\ 
\mathbb{K}^{(3)}_{\delta  r^2,1}&=-\frac{1085 \mathbb{K}^{(3)}_{\delta ,3}}{748}-\frac{21 \mathbb{K}^{(3)}_{r^3,1}}{187}-\frac{70 \mathbb{K}^{(3)}_{r,3}}{187}+\frac{1833 \mathbb{K}^{(3)}_{r\delta^2,1}}{748}+\frac{91 \mathbb{K}^{(3)}_{r\delta ,2}}{374}\  .
\end{align}
This allows us to write 
\be
\int_{\qvec} \sum_{\mathcal{O}_m , \alpha } c_{\mathcal{O}_m , \alpha }\mathbb{K}^{(3)}_{\mathcal{O}_m , \alpha }(\vk, \vq) =  \int_{\qvec} \sum_{\Gamma } c_{\Gamma}\mathbb{K}^{(3)}_{\Gamma}(\vk, \vq) \ ,  \label{degeneq1temp}
\ee
for $\Gamma =  \{ ({\delta ,1}), ({\delta ,2}), ({\delta ,3}), (\delta  r,1), ({\delta  r,2}) ,({\delta ^2,1}), ({\delta ^2 r,1} ), (r^3,1), (r,3) \}  $.  Additionally, after angular integration in the $\mathcal{P}_{15}^{\rm ct}$ diagram, we also have the following degeneracy equations
\begin{align}
\begin{split}
& \int d^2 \hat q \, \Bigg(
- \frac{20}{13} \, \mathbb{K}^{(3)}_{r,3}(\vec{k},\vec{q})
+ \frac{33513}{2366} \, \mathbb{K}^{(3)}_{r^3,1}(\vec{k},\vec{q})
- \frac{155}{26} \, \mathbb{K}^{(3)}_{\delta,3}(\vec{k},\vec{q})
+ \mathbb{K}^{(3)}_{r\delta,2}(\vec{k},\vec{q})
\Bigg) = 0 \ ,
\\
& \int d^2 \hat q \, \Bigg(
- \frac{25}{13} \, \mathbb{K}^{(3)}_{r^3,1}(\vec{k},\vec{q})
+ \mathbb{K}^{(3)}_{r\delta^2,1}(\vec{k},\vec{q})
\Bigg) = 0 \ ,
\end{split}
\end{align}
which allows us to remove $ \mathbb{K}^{(3)}_{r^3,1}$ and $\mathbb{K}^{(3)}_{r,3}$ from our list of operators, leaving us with the set of operators in \eqn{eq:counterKernels}, for a total of seven independent operators in the expression \eqn{eq:EFT_ct_2L_sum}.  Along with the higher derivative operator proportional to $c_{\partial^2 \delta , 1}$, this gives a total of eight non-stochastic EFT counterterms.

%
%

\section{Expressions for $F_{{\rm ct}, n}$} 
\label{sec:counterKernels}

Writing the list of EFT coefficients that appear in $F_{\rm ct, 3}$  as 
\be
\label{eq:ct_coeff_F3ct}
\alpha_i = \{ c_{\delta ,1}, c_{\delta ,2}, c_{\delta ,3}, c_{\delta  r,1}, c_{\delta  r,2},c_{\delta ^2,1}, c_{\delta ^2 r,1}  \}  \ , 
\ee
we can write expression for $F_{\rm ct, 3}$ as
\be
\label{eq:F3ct_decomposition}
6 \knl^2 F_{\rm ct, 3}( \kvec , \qvec , - \qvec ) = \sum_{i = 1}^{7}\alpha_i e^i_{\rm ct , 3} ( \kvec , \qvec) \ ,
\ee
where 
\begin{align}
\begin{split}
e^1_{\rm ct, 3} = &  -\left[\frac{-7 k^8 \left(143 (\vk \cdot \vq)^2-9 q^4\right)-k^6 \left(4039 (\vk \cdot \vq)^2 q^2+6 q^6\right)}{9009 k^2 q^4  |\vk-\vq|^{2}|\vk + \vq|^{2}}\right. \\
&+\frac{k^4 \left(3976 (\vk \cdot \vq)^4-4135 (\vk \cdot \vq)^2 q^4-69 q^8\right)+k^2 \left(7144 (\vk \cdot \vq)^4 q^2-2813 (\vk \cdot \vq)^2 q^6\right)}{9009 k^2 q^4  |\vk-\vq|^{2}|\vk + \vq|^{2}}  \\
&+\left.\frac{880 (\vk \cdot \vq)^4 q^4}{9009 k^2 q^4  |\vk-\vq|^{2}|\vk + \vq|^{2}} \right] \ ,   \\
e^2_{\rm ct,3}  & =  \frac{4 \left(75 k^4 q^4+85 k^2 (\vk \cdot \vq)^2 q^2+22 (\vk \cdot \vq)^4\right)}{3003 k^2 q^4} \ ,   \\
e^3_{\rm ct, 3} & = -\left[\frac{35 k^6 q^4+k^4 \left(32 (\vk \cdot \vq)^2 q^2+74 q^6\right)+k^2 \left(-4 (\vk \cdot \vq)^4-100 (\vk \cdot \vq)^2 q^4+39 q^8\right)}{273 q^4 |\vk-\vq|^{2}|\vk + \vq|^{2}} \right. \\
&+\left.\frac{-100 (\vk \cdot \vq)^4 q^2+24 (\vk \cdot \vq)^2 q^6}{273 q^4 |\vk-\vq|^{2}|\vk + \vq|^{2}} \right] \ ,  \\
e^4_{\rm ct , 3} = &  2 \left[\frac{45 k^8 q^4+k^6 \left(113 (\vk \cdot \vq)^2 q^2+90 q^6\right)+k^4 \left(-2 (\vk \cdot \vq)^4+51 (\vk \cdot \vq)^2 q^4+45 q^8\right)}{1287 k^2 q^4 |\vk-\vq|^{2}|\vk + \vq|^{2}}\right. \\
&\left.+\frac{k^2 \left(100 (\vk \cdot \vq)^2 q^6-453 (\vk \cdot \vq)^4 q^2\right)+11 (\vk \cdot \vq)^4 q^4}{1287 k^2 q^4 |\vk-\vq|^{2}|\vk + \vq|^{2}}\right]  \ ,  \\
e^5_{\rm ct , 3} = & -  \left[\frac{17 k^8 q^4+3 k^6 \left(7 (\vk \cdot \vq)^2 q^2+8 q^6\right)+k^4 \left(4 (\vk \cdot \vq)^4-6 (\vk \cdot \vq)^2 q^4+7 q^8\right)}{91 k^2 q^4 |\vk-\vq|^{2}|\vk + \vq|^{2}}\right.\\
&\left.+\frac{k^2 \left(41 (\vk \cdot \vq)^2 q^6-86 (\vk \cdot \vq)^4 q^2\right)-2 (\vk \cdot \vq)^4 \left(8 (\vk \cdot \vq)^2+3 q^4\right)}{91 k^2 q^4 |\vk-\vq|^{2}|\vk + \vq|^{2}} \right] \ , \\
e^6_{\rm ct , 3} = &  \left[\frac{20 k^2}{143} + \frac{4 (\vk \cdot \vq)^2}{39 q^2} \right] \ ,   \\
e^7_{\rm ct, 3} = & \frac{-1}{13} \left[ k^2 + 2 \frac{(\kvec \cdot \qvec)^2}{q^2} \right] \ . 
\end{split}
\end{align}

Writing the list of EFT coefficients that appear in $F_{\rm ct, 2}$  as 
\be
\label{eq:ct_coeff_F2ct}
\beta_i = \{ c_{\delta , 1} , c_{r \delta , 1} , c_{\delta^2 , 1} , c_{\delta , 2}  \}  \ , 
\ee
we can write expression for $F_{\rm ct, 2}$ as
\be
\label{eq:F2ct_decomposition}
4 \knl^2 F_{\rm ct, 2}( \kvec - \qvec ,  \qvec ) = \sum_{i = 1}^{4}\beta_i e^i_{\rm ct , 2} ( \kvec , \qvec) \ ,
\ee
with  
\begin{align}
\begin{split}
e^1_{\rm ct , 2} = &\frac{-24 (\kvec \cdot \qvec)^3 + 10 k^2 q^2 (k^2 + 2 q^2) + 8 (\kvec \cdot \qvec)^2 (7 k^2 + 3 q^2) - 2 (\kvec \cdot \qvec) ( 11 k^4 + 32 k^2 q^2) }{99 q^2 | \kvec - \qvec |^2}   \ ,  \\
e^2_{\rm ct , 2} = & \frac{8 (\kvec \cdot \qvec)^3 + 8 (\kvec \cdot \qvec) k^2 q^2 - 4 k^4 q^2 - 4 (\kvec \cdot \qvec)^2 ( k^2 + 2 q^2)}{33 q^2 |\kvec - \qvec|^2}  \ ,  \\
e^3_{\rm ct, 2} = & - \frac{8 k^2}{33}  \ , \\
e^4_{\rm ct, 2} = & - \frac{8 k^2 ( 2 (\kvec \cdot \qvec)^2 -14 (\kvec \cdot \qvec) q^2 + 5 k^2 q^2 + 7 q^4 )}{231 q^2 |\kvec - \qvec|^2} \ . 
\end{split}
\end{align}
The expression for $F_{\partial^{2}\rm ct,1}$ is given as,
\begin{align}\label{eq:Fct1}
F_{\partial^{2}\rm ct,1}(\vk) &= \frac{k^{4}}{26\knl^{4}}c_{\partial^{2}\delta,1}\  .
\end{align}

\section{Tensor reduction of UV approximations at two loops}
\label{sec:TRUVapproximations}

In this section we give the explicit expression of the double-UV contribution of $\Pcal_{15}$ after tensor reduction. This can be written as the sum of two terms:
\begin{align}
    \eqs[0.15]{Figures/Diagrams/2_loop/P15_UV_2.pdf} =& \int_{\qvec, \pvec}  \mathcal{R}^{(0)}_{(q, p) \to \infty}(p_{15}( \kvec , \qvec , \pvec) ) + \int_{\qvec, \pvec}  \mathcal{R}^{(2)}_{(q, p) \to \infty}(p_{15} ( \kvec , \qvec , \pvec)) \ ,
\end{align}
where we defined 
\begin{align}
    \int_{\qvec, \pvec } \mathcal{R}^{(i)}_{(q, p) \to \infty}(p_{15}( \kvec , \qvec , \pvec)) =& 30\Plin(k) \int_{\qvec, \pvec }   \Plin(q) \Plin(p) \, f_{\rm screen}(q^2) f_{\rm screen}(p^2) \times \nonumber \\ 
    & \times \left( \Tcal^{(i)}_{(q, p) \to \infty}[\Ffive] + \right. \nonumber\\
    & \left. \qquad - \Tcal^{(i)}_{(q, p) \to \infty} \left[ \left( \Tcal^{(0)}_{q \to \infty} [\Ffive] + \right. \right. \right. \nonumber \\
    & \left. \left. \left. \qquad \qquad \qquad + \Tcal^{(0)}_{p \to \infty} [\Ffive] \right) \right] \right) \ ,
\end{align}
with $i= 0,2$.
The explicit expressions for these UV-approximations after tensor reduction are:
\begin{align}
    \int_{\qvec , \pvec }  \mathcal{R}_{(q,p)\to\infty}^{(0)}(p_{15} ( \kvec , \qvec , \pvec)) = & - 30k^2 \Plin(k) \times \nonumber \\ 
    & \, \times \left\{ \int_{\qvec} \int_{\pvec} \frac{\Plin(q)}{q^2} \frac{\Plin(p)}{p^2} \fscr{q^2} \fscr{p^2} \times \nonumber \right. \\
    & \qquad \times \frac{(p^2 + q^2) ( (\vq \cdot \vp)^2 - p^2 \, q^2)^2}{429975 (p^2)^2 (q^2)^2 (|\vq + \vp|^2)^2 (|\vq - \vp|^2)^2}  \times \nonumber \\ 
    & \qquad \times \bigg[ -1362 (p^2)^2 (\vq\cdot\vp)^2 + 2136 (\vq\cdot\vp)^4 + 2345 (p^2)^3 q^2 + \nonumber \\
    & \qquad \qquad - 8792 p^2 (\vq\cdot\vp)^2 q^2 + 4690 (p^2)^2 (q^2)^2 + \nonumber \\
    & \qquad \qquad - 1362 (\vq\cdot\vp)^2 (q^2)^2 + 2345 p^2 (q^2)^3\bigg] + \nonumber \\
    \label{eq:P15_TR_1}
    & \, \qquad \left. +  \frac{120424}{45147375} \int_{\qvec} \frac{\Plin(q)}{q^2}\fscr{q^2} \int_{\pvec} \frac{\Plin(p)}{p^2} \fscr{p^2} \right\}, \\
\int_{\qvec , \pvec }  \Rcal_{(q,p)\to\infty}^{(2)}(p_{15}( \kvec , \qvec , \pvec)) =& - 30k^4\Plin(k) \times \nonumber \\
& \times \left\{ \int_{\qvec} \int_{\pvec} \frac{\Plin(q)}{q^2} \frac{\Plin(p)}{p^2} \fscr{q^2} \fscr{p^2} \times \right. \nonumber \\
    & \qquad \times \frac{1}{264864600 (p^2)^3 (q^2)^3 (|\vq - \vp|^2)^3 (|\vq + \vp|^2)^3} \times \nonumber \\ 
    & \qquad \times \bigg[ 143401 (p^2)^9 (q^2)^3 + \nonumber \\
    & \qquad \quad + (p^2)^8 \left(784346 (\vp \cdot \vq)^2 (q^2)^2 + \right. \nonumber \\ 
    & \qquad \quad \quad \left. + 903528 (q^2)^4 \right) + \nonumber \\
    & \qquad \quad - 3 (p^2)^7 \left( 1418692 (\vq \cdot \vp)^4 q^2 + 591116 (\vq\cdot\vp)^2 (q^2)^3 + \right. \nonumber \\
    & \qquad \quad \quad \left. - 774501 (q^2)^5 \right) + \nonumber \\
    & \qquad \quad + 4 p^2 (\vq \cdot \vp)^4 q^2 \left( 6878464 (\vq \cdot \vp)^6 + \right. \nonumber \\
    & \qquad \quad \quad - 11872120 (\vq \cdot \vp)^4 (q^2)^2 + 5810922 (\vq \cdot \vp)^2 (q^2)^4 + \nonumber \\ 
    & \qquad \quad \quad \left. - 1064019 (q^2)^6 \right) + \nonumber \\
    & \qquad \quad + 16 (\vq \cdot \vp)^6 \left( 185856 (\vq\cdot\vp)^6 + 569200 (\qp)^4 (q^2)^2 + \right. \nonumber \\  
    & \qquad \quad \quad \left. - 540776 (\qp)^2 (q^2)^4 + 151363 (q^2)^6 \right) + \nonumber \\
    & \qquad \quad + 2 (p^2)^6 \left(1210904 (\qp)^6 - 4141163 (\qp)^4 (q^2)^2 + \right. \nonumber \\  
    & \qquad \quad \quad \left. - 6389949 (\qp)^2 (q^2)^4 + 1563376 (q^2)^6 \right) + \nonumber \\  
    & \qquad \quad + (p^2)^5 \left( 23243688 (\qp)^6 q^2 + 11493924 (\qp)^4 (q^2)^3 + \right. \nonumber  \\ 
    & \qquad \quad \quad \left. - 20444408 (\qp)^2 (q^2)^5 + 2323503 (q^2)^7 \right) + \nonumber \\
    & \qquad \quad + (p^2)^4 \left(-8652416 (\qp)^8 + 23745936 (\qp)^6 (q^2)^2 + \right. \nonumber \\ 
    & \qquad \quad \quad + 32180540 (\qp)^4 (q^2)^4 - 12779898 (\qp)^2 (q^2)^6 + \nonumber \\  
    & \qquad \quad \quad \left. + 903528 (q^2)^8 \right) + \nonumber \\
    & \qquad \quad + (p^2)^2 \left( 9107200 (\qp)^{10} - 24365792 (\qp)^8 (q^2)^2 + \right. \nonumber \\  
    & \qquad \quad \quad + 23745936 (\qp)^6 (q^2)^4 - 8282326 (\qp)^4 (q^2)^6  + \nonumber \\  
    & \qquad \quad \quad \left. + 784346 (\qp)^2 (q^2)^8 \right) + \nonumber \\  
    & \qquad \quad + (p^2)^3 \left(-47488480 (\qp)^8 q^2 - 2854224 (\qp)^6 (q^2)^3 + \right. \nonumber \\  
    & \qquad \quad \quad + 11493924 (\qp)^4 (q^2)^5 - 1773348 (\qp)^2 (q^2)^7 + \nonumber \\
    \label{eq:P15_TR_2}
    & \qquad \quad \quad \left. + 143401 (q^2)^9 \right) \bigg] + \nonumber \\
    & \quad \left. - \frac{10541261}{27810783000} \int_{\qvec} \frac{\Plin(q)}{q^2} \fscr{q^2} \int_{\pvec} \frac{\Plin(p)}{p^2} \fscr{p^2} \right\}.
\end{align}

\section{Two-loop counterterms}
\label{sec:2-loop_counterterms}

In this section, we describe the computation of the two-loop EFT counterterms in the power spectrum, defined in Eq.~\eqref{eq:2LEFTCT_diagrams}. Our goal is to numerically evaluate the associated one-loop integrals and express them in terms of a basis of cosmology-independent integrals. To enable numerical integration, we use the decompositions of $F_{\rm ct, 3} (\vk, \vq, -\vq)$ and $F_{\rm ct, 2} (\vk -\vq, \vq)$ given in Eqs.~\eqref{eq:F3ct_decomposition} and~\eqref{eq:F2ct_decomposition}, which separate the momentum dependence from the EFT coefficients. The corresponding sets of EFT parameters are shown in Eqs.~\eqref{eq:ct_coeff_F3ct} and~\eqref{eq:ct_coeff_F2ct}.

As discussed in \secref{sec:check_renormalization}, the $\Pcal_{15}^{\rm ct}$ counterterm diagram is UV divergent. The UV divergence needs to be subtracted locally, {\it i.e.}, at the integrand level, to enable the numerical evaluation of the integrals. Using the same procedure described in \secref{sec:UV_1loop}, we expand the counterterm kernels as
\begin{equation}
    e^{ i}_{{\rm ct}, 3}(\vk, \frac{\vq}{\delta}, - \frac{\vq}{\delta}) = \delta^0 {\mathcal T}^{(0)}_{q \to \infty} \left[ e^{ i}_{{\rm ct}, 3}(\vk, - \vq, \vq) \right] + \delta^2 {\mathcal T}^{(2)}_{q \to \infty} \left[ e^{ i}_{{\rm ct}, 3}(\vk, - \vq, \vq) \right] + \mathcal{O} (\delta^3) \ , 
\end{equation}
where the $\mathcal T^{(0)}_{q \to \infty}$ and $\mathcal T^{(2)}_{q \to \infty}$ are the first and third term in the Taylor expansions around $\delta \to 0$. For all the kernels the second term, $\mathcal T^{(1)}_{q \to \infty}$, vanishes.

According to the power-counting defined in \secref{sec:UV_1loop}, both terms must be removed to reduce the UV sensitivity of the integral. We therefore define the UV-subtracted kernels $e^{ i}_{{\rm ct}, 3}$ as
\begin{multline}
    e^{i, \rm UV-reg., \, loc.}_{{\rm ct}, 3}(\vk, - \vq, \vq) = e^{ i}_{{\rm ct}, 3}(\vk, - \vq, \vq) +  \\  
    - \left( {\mathcal T}^{(0)}_{q \to \infty} \left[ e^{ i}_{{\rm ct}, 3}(\vk, - \vq, \vq) \right] + {\mathcal T}^{(2)}_{q \to \infty} \left[ e^{ i}_{{\rm ct}, 3}(\vk, - \vq, \vq) \right] \right) \fscr{q^2}.
    \label{eq:ct3_uv_local}
\end{multline}
In the expression above, the label ``loc.'' emphasizes that the UV subtraction is applied locally, in contrast to Eq.~\eqref{eq:ct3_uv}, where the angular components of the subtracted terms are integrated analytically. 
The subtracted UV modes,
\begin{equation}
    e^{ i, \rm UV}_{{\rm ct}, 3}(\vk, - \vq, \vq) = \left( {\mathcal T}^{(0)}_{q \to \infty} \left[ e^{ i}_{{\rm ct}, 3}(\vk, - \vq, \vq) \right] + {\mathcal T}^{(2)}_{q \to \infty} \left[ e^{ i}_{{\rm ct}, 3}(\vk, - \vq, \vq) \right] \right) \fscr{q^2} \ ,
\end{equation}
are degenerate, after angular integration, with the EFT counterterms as discussed in \secref{sec:check_renormalization}.

With this definition, we define the UV-regularated {counterterm} diagram $\Pcal_{15}^{\rm ct, UV-reg.}$ as
\begin{align}
\begin{split}
    \Pcal_{15}^{\rm ct, UV-reg.}(k) & = 2 \Biggl( \Plin(k) F_{\partial^2 {\rm ct}, 1} + \\
    & +  3 \sum_{i = 1}^{7} \frac{\alpha_i}{6 k_{\rm NL}^2} \int_{\qvec} e^{ i, \rm UV-reg., \, loc.}_{{\rm ct}, 3}(\vk, - \vq, \vq) \Plin (k) \Plin (q) \Biggr)  \ , 
\end{split}
\end{align}
where $7$ is the number of independent EFT coefficients in the two-loop counterterm kernel $F_{{\rm ct}, 3}$.

The following two-loop counterterm contributions to the power spectrum can be written as 
\begin{eqnarray}
    \Pcal_{42}^{\rm ct}(k) &=& \frac{ 1}{k_{\rm NL}^2} \sum_{i = 1}^{4} \beta_i I_{42}^{i}(k) \ , \\
    \Pcal_{15}^{\rm ct, UV-reg.}(k) - 2\Plin(k) F_{\partial^2 {\rm ct}, 1} &=& \frac{1}{k_{\rm NL}^2} \sum_{i = 1}^{7} \alpha_i I_{15}^{i}(k) \  ,
\end{eqnarray}
where the integrals $I_{15}^{i}(k)$ and $I_{42}^{i}(k)$ are defined as,
\begin{eqnarray}
    I_{42}^{i}(k) &=& \int_{\qvec} e_{{\rm ct}, 2}^{ i}(\vk - \vq, \vq) F_2(\vk - \vq, \vq) \Plin(q) \Plin(|\kvec - \qvec|)\  ,\label{eq:P42_counterterm_integral} \\
    I_{15}^{i}(k) &=& \Plin(k)\int_{\qvec} e_{{\rm ct}, 3}^{ i, {\rm UV-reg., \, loc.}}(\vk, \vq, -\vq) \Plin(q) \ . \label{eq:P15_counterterm_integral}
\end{eqnarray}
Note that since $e_{{\rm ct},2}^{i}$ and $e_{{\rm ct},3}^{i}$ do not depend on $k_{\rm NL}$, neither do $I_{42}^{i}(k)$ and $I_{15}^{i}(k)$.  We remind the reader that $\mathcal{P}_{42}^{\rm ct}$ does not need UV regulation as it has a negative degree of UV divergence.

From Eq.~\eqref{eq:P33renorm} we see that $\Pcal^{\rm ct}_{33}$ combines with $\Pcal_{33}^{\rm (I)}$ into finite quantities. Among those terms, the only contribution that requires numerical integration is:
\begin{equation}
    F_{{\rm ct},1}^{\rm fin.}(\vk) \mathcal{P}_{13}^{\rm UV-reg.}(k) = -c_{\delta,1} \frac{ k^2}{9k_{\rm NL}^2} \Pcal_{13}^{\rm UV\text{-}reg}(k) = \frac{c_{\delta,1}}{k_{\rm NL}^2} I_{33}(k) \ , 
\end{equation}
where $\Pcal_{13}^{\rm UV-reg.}(k) = \int_{\qvec} p_{13}^{\rm UV-reg.}(\vk, \vq)$ is the integrated form of the one-loop UV-regulated diagram defined in Eq.~\eqref{eq:p13-subtract}.  
In the last equality we introduced
\begin{equation}
    \label{eq:P33_counterterm_integral}
    I_{33}(k) = - \frac{k^2}{9}\, \Pcal_{13}^{\rm UV-reg.}(k) \ . 
\end{equation}
Since $\Pcal_{13}^{\rm UV-reg.}(k)$ is both UV and IR finite, $I_{33}(k)$ is directly suitable for numerical evaluation.

In Eqs.~\eqref{eq:P42_counterterm_integral}, \eqref{eq:P15_counterterm_integral} and \eqref{eq:P33_counterterm_integral} there are 12 cosmology-dependent one-loop integrals in total. This number exceeds the 7 EFT coefficients that contribute to loop integrals because the definitions of the $F_{{\rm ct},1}$, $F_{{\rm ct},2}$ and $F_{{\rm ct},3}$ kernels lead to a splitting of contributions into multiple integrals.  The explicit expressions for the kernels used in these computations are provided in the supplementary material attached to the arXiv version of this article.

\subsection{Infrared cancellations\label{subapp:2-loop_count_IR}}

We now apply the procedure described in \secref{sec:IR} to the one-loop integrals defined in Eqs.~\eqref{eq:P42_counterterm_integral} and~\eqref{eq:P15_counterterm_integral}. The integrands $I_{15}^i$ have a single IR singularity at $\vq = 0$ for $i=1$, which is cancelled by the integration measure. For all other values of $i$, the integrand is regular. In contrast, the integral $I^{i}_{\rm ct, 42}$ for $i=1$ share the same IR structure as $\Pcal_{22}$, with two singularities at $\vq = 0$ and $\vq = \vk$, while remaining regular for $i \neq 1$.

To enhance the numerical stability, we choose to apply the partitioning procedure to the product of kernels $e_{{\rm ct}, 2}^{ i} (\vk - \vq, \vq) F_2(\vk - \vq, \vq)$ for all values of $i$, not only for those with true IR singularities. Although the integrands for $i\neq1$ are regular, the kernels still contain singular structures that are cancelled only by the linear scaling of $\Plin$ in the IR. These cancellations may lead to numerical instabilities. Importantly, the partitioning leaves the value of the integral unchanged; it rearranges the integrand to improve numerical convergence.

Thus, we define the partitioned kernels entering in \eqn{eq:P42_counterterm_integral}
\begin{align}
\begin{split}
\label{eq:F2_ct_multi}
 e_{{\rm ct}, 2}^{ i}(\vk - \vq, \vq) F_2(\vk - \vq, \vq) \ \  & \to\ \    F^{{\rm part.}, i}_{{\rm ct}, 22}(\vq, \vk - \vq) \\
 & \quad \quad \equiv \frac{2(|\vk - \vq|^2)^{j_i}}{(q^2)^{j_i} + (|\vk - \vq|^2)^{j_i}} e_{{\rm ct}, 2}^{ i} (\vk - \vq, \vq) F_2(\vk - \vq, \vq) \ , 
\end{split}
\end{align}
where the exponent $j_i \in \left\{ 1, 2 \right\}$ depends on the index $i$ and is chosen to appropriately regulate the denominators of the kernel. 

\subsection{Perturbative expansion around a standard cosmology model}

We can apply the expansion in Eq.~\eqref{eq:DPexpansion1loop} to the one-loop integrals $I^{i, [j]}_{42}$ and $I^{i, [j]}_{15}$, where $i$ indexes the counterterm and $j$ the cosmology. 
For $I^{i, [j]}_{42}$, this yields
\begin{eqnarray}
\label{eq:DPexpansion_I42}
    I_{42}^{i, [j]} = 
    \l \mathcal{N}^{[j]} \r^2 I_{42}^{i, [0]}
    +  \mathcal{N}^{[j]} \, \Delta_{1} I_{42}^{i, [j]}
    + \Delta_{2} I_{42}^{i, [j]} \ . 
\end{eqnarray}

The expansion of $I^{i, [j]}_{15}$ can be rearranged as
\begin{align}
    I_{15}^{i, [j]} &= \l \mathcal{N}^{[j]} \Plin^{[0]} (k) + \DeltaP_{\rm lin.}^{[j]} (k) \r   \\
    \label{eq:DPexpansion_I15}
        & \qquad \times \l \mathcal{N}^{[j]} \int_{\qvec} e^{ i, {\rm UV-reg., loc.}}_{{\rm ct}, 3}(\vk, \vq, -\vq) \Plin^{[0]} (q) + \int_{\qvec} e^{ i, {\rm UV-reg., loc.}}_{{\rm ct}, 3}(\vk, \vq, -\vq) \DeltaP_{\rm lin.}^{[j]} (q) \r  \ .  \nonumber
\end{align}

Thus, to minimize the number of integrals that must be evaluated, we compute $I_{15}^{i, [j]}$ in terms of the integrals
\begin{eqnarray}\label{eq:DPexpansion_I15_2}
    I_{15}^{i, [0]} = \int_{\qvec} e^{ i, {\rm UV-reg., loc.}}_{{\rm ct}, 3}(\vk, \vq, -\vq) \Pcal^{[0]}_{\rm lin.} (q) \ , \\
    \Delta_1 I_{15}^{i, [j]} = \int_{\qvec} e^{ i, {\rm UV-reg., loc.}}_{{\rm ct}, 3}(\vk, \vq, -\vq) \DeltaP_{\rm lin.}^{[j]} (q) \ . 
\end{eqnarray}

A similar spitting can be achieved for $I_{33}$. We define
\begin{eqnarray} 
    \label{eq:DPexpansion_I33_1}
    I_{33}^{[0]} = - \frac{k^2}{9}\int_{\qvec} 6 F^{\rm UV-reg.}_{3}(\vk, \vq, -\vq) \Pcal^{[0]}_{\rm lin.} (q)  \ , \\
    \Delta_1 I_{33}^{[j]} = - \frac{k^2}{9} \int_{\qvec} 6 F^{\rm UV-reg.}_{3}(\vk, \vq, -\vq) \DeltaP_{\rm lin.}^{[j]} (q)  \ , 
\end{eqnarray}
With these definitions we can write
\begin{eqnarray}
    \label{eq:DPexpansion_I33}
    I_{33}^{[j]} = \l \mathcal{N}^{[j]} \Plin^{[0]} (k) + \DeltaP^{[j]}_{\rm lin.} (k) \r  \l I_{33}^{[0]} + \Delta_1 I_{33}^{[j]} \r \ . 
\end{eqnarray}

\subsection{Decoupling the dependence on the cosmological models}

We adopt the same approach described in \secref{sec:cosmology-independent-integrals} to compute the contributions from $\DeltaP_{\rm lin.}$ to the one-loop integrals $I_{33}^{[j]}$, $I_{42}^{i,[j]}$, and $I_{15}^{i,[j]}$ using a cosmology-independent set of integrals. Due to the rearrangement of the $\DeltaP_{\rm lin.}$ expansion for $I_{15}^{i,[j]}$ and $I_{33}^{[j]}$ in Eqs.~\eqref{eq:DPexpansion_I15_2} - \eqref{eq:DPexpansion_I33}, only terms involving $\DeltaP_{\rm lin.,fit,1}$ contribute to these integrals.

Using a notation similar to the one used for the one- and two-loop corrections above, we can write 
\begin{eqnarray}\label{eq:DPexpansion_I15_fit}
    \Delta_1 I^{[j]}_{33, \rm fit} (k) = \frac{1}{(2\pi)^3} \sum_{n = 1}^{24} \alpha_{n}^{[j],(1)} \, T_{{\rm ct}, 33}^{n}(k)  \ , \\
    \Delta_1 I^{i, [j]}_{15, \rm fit} (k) = \frac{1}{(2\pi)^3} \sum_{n = 1}^{24} \alpha_{n}^{[j],(1)} \, T_{{\rm ct}, 15}^{i n}(k) \ , \\
    \Delta_1 I^{i, [j]}_{42, \rm fit} (k) = \frac{\mathcal{N}^{[j]} }{(2\pi)^3} \sum_{n = 1}^{24} \alpha_{n}^{[j],(1)}  \, T_{{\rm ct}, 42, 1}^{i n}(k) \ , \\ \label{eq:DPexpansion_I42_fit}
    \Delta_2 I^{i, [j]}_{42, \rm fit} (k) = \frac{1}{(2\pi)^3} \sum_{n_1, n_2 = 1}^{16} \alpha_{n_1}^{[j],(2)} \alpha_{n_2}^{[j],(2)} \,  T_{{\rm ct}, 42, 2}^{i n_1 n_2}(k) \ ,
   \end{eqnarray}
where the expressions for the symbols $ T_{{\rm ct}, 33}^{n} $, $T_{{\rm ct}, 15}^{in}$, $T_{{\rm ct}, 42, 1}^{in}$, and $T_{{\rm ct}, 42, 2}^{i n_1 n_2}$, are reported in Eqs.~\eqref{eq:T33ct_fit}-\eqref{eq:T42_2_ct_fit}.

\section{Loop tensor integrals expression}
\label{sec:Tensor_integrals}
Here we write the explicit formulas for the loop tensor integrands defined in Eq.~\eqref{eq:loop-tensor-integrals}. These can be derived by plugging Eqs.~\eqref{eq:1L_lin_IRUV_integrand} and \eqref{eq:2L_lin_IRUV_integrand} into \eqref{eq:delta_m_final_splitting}, and, at one-loop, read,
\begin{align}
    T_{{\rm 1-loop}, (1,0)}^{i} =& \int d^3q \Bigg[ 6 F^{ \rm UV-reg.}_{3}(\vk, -\vq, \vq) \Plin^{[0]}(k) f_{i}^{(1)}(q) + \nonumber \\
                &+ 2 F^{\rm part.}_{22}(\vk, \vq - \vk) \l f_{i}^{(1)}(q) \Plin^{[0]}(|\vk - \vq|) + \Plin^{[0]}(q) f_{i}^{(1)}(|\vk - \vq|) \r \Bigg] \  ,\\
    T_{{\rm 1-loop}, (1,1)} =& \int d^3q \; 6 F_{3}^{ \rm UV-reg.}(\vk, -\vq, \vq) \Plin^{[0]}(q) \  , \\
    T_{{\rm 1-loop}, (2,0)}^{i j} =& \int d^3q \; 2  F^{\rm part.}_{22}(\vk, \vq - \vk)\ f_{i}^{(2)}(q) f_{j}^{(2)}(|\vk - \vq|) \  ,\\
    T_{{\rm 1-loop}, (2,1)}^{i} =& \int d^3q \; 6 F^{ \rm UV-reg.}_{3}(\vk, -\vq, \vq) f_{i}^{(2)}(q) \  .
\end{align}

At two-loops we have
\begin{align}
    T_{{\rm 2-loop}, (1,0)}^{i} =& \int d^3q \int d^3 p \Bigg[ \Bigg( \left( 9 F_{3}^{ \rm UV-reg.}(\vk, -\vq, \vq)  F_{3}^{\rm UV-reg.}(\vk, -\vp, \vp) + \right. \nonumber\\
    & \qquad \left. + 30F_{5}^{ \rm UV-reg.}(\vk, -\vq, \vq, -\vp, \vp) \right) \Plin^{[0]}(k) + \nonumber \\
    & \,\, + 6 F_{33}^{\rm (II), part.}(\vq, \vp, \vk) \, \Plin^{[0]}(|\vk-\vq -\vp|)  + 24 F_{42}^{\rm part.}(\vq, \vp, \vk) \Plin^{[0]}(|\vk-\vq|) \Bigg) \times \nonumber \\
    & \qquad \qquad \qquad \qquad \times \l \Plin^{[0]}(q) f_{i}^{(1)} (p) + f_{i}^{(1)} (q) \Plin^{[0]}(p) \r + \nonumber\\
    & + \left( 6 F_{33}^{\rm (II), part.}(\vq, \vp, \vk) f_{i}^{(1)} (|\vk - \vq - \vp|) + 24 F_{42}^{\rm part.}(\vq, \vp, \vk) f_{i}^{(1)} (|\vk - \vq|) \right) \times \nonumber \\
    & \qquad \qquad \qquad \qquad \times \Plin^{[0]}(q) \Plin^{[0]}(p) \Bigg]  \ ,  \\
    T_{{\rm 2-loop}, (1,1)} =& \int d^3 q \int d^3 p \Bigg[ 9 F_{3}^{ \rm UV-reg.}(\vk, -\vq, \vq)  F_{3}^{\rm UV-reg.}(\vk, -\vp, \vp) \Plin^{[0]}(q)  \Plin^{[0]}(p) + \nonumber \\
    & \qquad + 30 F_{ 5}^{ \rm UV-reg.}(\vk, -\vq, \vq, -\vp, \vp) \Plin^{[0]}(q)  \Plin^{[0]}(p) \Bigg] \ ,  \\
    T_{{\rm 2-loop}, (2,0)}^{i j} =& \int d^3 q \int d^3 p \Bigg[ \Bigg( \left( 9 F_{3}^{ \rm UV-reg.}(\vk, -\vq, \vq)  F_{3}^{\rm UV-reg.}(\vk, -\vp, \vp) + \right. \nonumber\\
    & \qquad \left. + 30F_{5}^{ \rm UV-reg.}(\vk, -\vq, \vq, -\vp, \vp) \right) \Plin^{[0]}(k) + \nonumber \\
    & + 6 F_{33}^{\rm (II), part.}(\vq, \vp, \vk) \Plin^{[0]}(|\vk-\vq -\vp|)  + 24 F_{42}^{\rm part.} (\vq, \vp, \vk) \Plin^{[0]}(|\vk-\vq|) \Bigg) \times \nonumber \\
    & \qquad \qquad \qquad \qquad \qquad \qquad \qquad \qquad \times f_{i}^{(2)} (q)  f_{j}^{(2)} (p)  \nonumber\\
    & + \left( 6 F_{33}^{\rm (II), part.}(\vq, \vp, \vk) f_{i}^{(2)} (|\vk - \vq - \vp|) + 24 F_{42}^{\rm part.} (\vq, \vp, \vk) f_{i}^{(2)} (|\vk - \vq|) \right) \times \nonumber \\
    &  \qquad \qquad \qquad \qquad \qquad \qquad \times \l \Plin^{[0]}(q) f_{j}^{(2)} (p) + f_{j}^{(2)} (q) \Plin^{[0]}(p) \r \Bigg] \ ,  \\
    T_{{\rm 2-loop}, (2,1)}^{i} =& \int d^3 q \int d^3 p \Bigg( 9 F_{3}^{ \rm UV-reg.}(\vk, -\vq, \vq)  F_{3}^{\rm UV-reg.}(\vk, -\vp, \vp) + \nonumber\\
    & \; + 30 F_{5}^{ \rm UV-reg.}(\vk, -\vq, \vq, -\vp, \vp) \Bigg)  \l \Plin^{[0]}(q) f_{i}^{(2)} (p) + f_{i}^{(2)} (q) \Plin^{[0]}(p) \r \ . 
\end{align}

For the one-loop integrals that enter the expression of the two-loop counterterms, we have
\begin{align}
    T_{{\rm ct}, 33}^{i} =& - \frac{k^2}{9} \int d^3 q \, 6 F_{3}^{\rm UV-reg.}(\vk, -\vq, \vq) f_{i}^{(1)}(q)  \ , \label{eq:T33ct_fit} \\
    T_{{\rm ct}, 15}^{ij} =& \int d^3 q \, e_{{\rm ct}, 3}^{ i, \rm UV-reg., loc.}(\vk, -\vq, \vq) f_{j}^{(1)}(q)  \ , \label{eq:T15ct_fit} \\
    T_{{\rm ct}, 42, 1}^{i j} =& \int d^3 q \, F_{{\rm ct}, 22}^{{\rm part.}, i}(\vk -\vq, \vq) \l f_{j}^{(1)}(q) \Plin^{[0]}(|\vk - \vq|) +  \Plin^{[0]}(q) f_{j}^{(1)}(|\vk - \vq|) \r  \ , \label{eq:T42_1_ct_fit}\\
    T_{{\rm ct}, 42, 2}^{i j k} =& \int d^3 q \, F_{{\rm ct}, 22}^{{\rm part.}, i}(\vk -\vq, \vq) f_{j}^{(2)}(q) f_{k}^{(2)}(|\vk - \vq|) \  . \label{eq:T42_2_ct_fit}
\end{align}

\bibliographystyle{JHEP}
\bibliography{biblio}

\end{document}